\documentclass[twocolumn,preprintnumbers,amsmath,amssymb,aps,prb,longbibliography]{revtex4-1}
\usepackage{graphicx}
\begin{document}

\title{Skyrmion Pinball and Directed Motion on Obstacle Arrays}
 
\author{N.P. Vizarim$^{1,2}$, C.J.O. Reichhardt$^{1}$,
  P.A. Venegas$^{3}$, and
  C. Reichhardt$^{1}$}
\affiliation{$^{1}$Theoretical Division and Center for Nonlinear Studies,
  Los Alamos National Laboratory, Los Alamos, New Mexico 87545, USA}
\affiliation{$^{2}$ POSMAT - Programa de P{\' o}s-Gradua{\c c}{\~ a}o em Ci{\^ e}ncia e Tecnologia de Materiais, Faculdade de Ci\^{e}ncias, Universidade Estadual Paulista - UNESP, Bauru, SP, CP 473, 17033-360, Brazil}
\affiliation{$^{2}$ Departamento de F\'{i}sica, Faculdade de Ci\^{e}ncias, Universidade Estadual Paulista - UNESP, Bauru, SP, CP 473, 17033-360, Brazil
}

\date{\today}

\begin{abstract}
We examine skyrmions interacting with a square array of obstacles under
ac drives applied in one or two directions.  
For a single direction of ac driving,
we find that the Magnus force in conjunction with
the obstacle interactions
can create 
elliptical skyrmion orbits
of increasing size,
leading to localized phases, chaotic phases, and translating or ratcheting
orbits. 
Under two ac drives that are out of phase by $90^\circ$ and applied in
two directions, 
the skyrmions form localized commensurate orbits
that encircle an integer number of obstacles, similar 
to the electron pinball effect observed for electrons in antidot lattices.
As a function of ac amplitude, Magnus force strength, and obstacle size, we  
find that chaotic scattering regimes and directed motion
can emerge even in the absence of asymmetry
in the substrate.     
The directed motion follows different symmetry axes of the periodic
substrate, and we observe a variety of
reversed ratchet effects.
The Magnus force
in the skyrmion system
produces a significantly larger number of directed motion regimes than
are exhibited by overdamped systems.
We discuss how these results could be used to
move skyrmions in a controlled way for possible applications.  
\end{abstract}

\maketitle

\section{Introduction}
When a classical electron is translating in the presence of a magnetic field,
it undergoes cyclic motion 
with an orbit of radius $R$.
If the electron simultaneously
interacts with a
square periodic array of scatterers,
it falls into stable localized orbits for
certain values of
$R$ 
that allow it to encircle an 
integer number of scattering sites during each orbit.
When $R$ is sufficiently small,
the electron becomes locked in 
a localized orbit that fits in the space
between the obstacles,
but for larger $R$, strong localization occurs
when the electron orbit encloses
$n = 1$, 2, 4, $9$, or $m^2$ scattering sites, where $m$ is an integer
\cite{Weiss91}.
As the orbit radius $R$ varies,
the electron remains locked in a localized orbit
until it begins
to collide with the obstacles
and
enters a chaotic regime of
continuous collisions
associated with gradual diffusive motion
\cite{Weiss91,Fleischmann92}.
Samples containing an antidot lattice
are commonly known as electron pinball systems in analogy to the
pinball game in which a ball is scattered by
fixed obstacles.
The windows in which commensurate orbits appear depend on the
size and periodicity of the scattering sites
\cite{Ishizaka97,Meckler05},
and the
commensurate orbits can be identified by
the peaks they produce in the magnetic transport curves 
\cite{Weiss91,Fleischmann92,Weiss93,Ishizaka97}.

Electron pinball effects
have been studied for square
\cite{Weiss91,Fleischmann92,Weiss93}, hexagonal \cite{Meckler05,Kato12},
rectangular \cite{Onderka00,Geisler05}
and disordered dot geometries \cite{Klinkhammer08,Siboni18}. 
More recently,
similar effects were studied in graphene superlattice structures
\cite{Sandner15,Power17},
topological insulators \cite{Maier17}, and bilayer systems,  
where a pinned Wigner crystal acts as the periodic substrate for
commensurate orbits of composite fermions \cite{Deng16,Jo18}.   
Variations on electron pinball include vortex pinball
for vortices in a type-II superconductor
interacting with a square array of pinning sites under a
circulating ac drive, where
the size of the vortex orbit depends on the
amplitude and frequency of the ac drive \cite{Reichhardt02b}.  
Similar dynamics arise for colloidal particles
interacting with periodic arrays of obstacles or scattering
sites under
intrinsic or external ac driving
or combined ac and dc driving, 
when the effective ac driving is in both
the $x$ and $y$-directions and produces
circular or elliptical colloidal orbits \cite{Tierno07,Loehr16}. 

In the absence of a dc drive,
particles in a two-dimensional (2D) array under crossed ac
drives can also exhibit phase locking 
and directed motion where a net
drift of the particles appears \cite{Reichhardt03,Speer09,Platonov15}.
This is similar to a ratchet effect in which
a net dc transport is produced by an ac drive when the system
is coupled to an asymmetric substrate \cite{Reimann02};
however, in the symmetric 2D substrate arrays, 
the additional symmetry breaking arises from
the drive itself
or from the interface of two different substrate lattices
which can generate 
edge transport.
Recently, a variety of types of directed motion have
been studied for magnetic colloids
on a periodic substrate where an external magnetic field creates non-trivial 
loop motion of the colloids, which
follow translating skipping orbits along
the interfaces between two 
different substrate lattices \cite{Loehr16,Loehr18,MassanaCid19,Yazdi20}.  

Skyrmions in chiral magnets are another example of an
assembly of particle-like objects
that can interact with various types
of ordered and disordered substrates
\cite{Muhlbauer09,EverschorSitte18,Yu10,Nagaosa13}.
Magnetic skyrmions 
have been found in a variety of systems,
including 
materials in which the skyrmions are stable at room temperature
\cite{Woo16,Soumyanarayanan17}.
Skyrmions can be set into 
motion with an external current
\cite{Nagaosa13,Woo16,Schulz12,Yu12,Iwasaki13,Lin13a}, and the resulting
velocity-force relations show 
transitions from pinned to sliding states
in transport experiments measuring changes 
in the topological Hall effect \cite{Nagaosa13,Schulz12,Liang15} 
and in direct imaging  \cite{Woo16,Yu12,Montoya18}. 
It is also possible to characterize skyrmion motion
based on changes in the noise 
fluctuations as a function of drive
\cite{Diaz17,Sato19}.
Interest
in skyrmions is spurred in part by
possible applications,
since due to their size scale and mobility,
skyrmions could be used 
for memory, logic devices, and
alternative computing architectures
\cite{Tomasello14,Fert17,Prychunenko18,Song20}.  
The ability to control the direction and
distance of skyrmion motion
could open new avenues
for device creation, and as a result there have been
a number 
of proposals
to control skyrmion motion through interactions
with structured substrates such as race tracks 
\cite{Tomasello14,Fert17}, 
periodic substrate modulations
\cite{Reichhardt15a,Reichhardt15b,Reichhardt16a,Navau16}, or
ordered pinning structures
\cite{Ma16,Ma17,Stosic17,Fernandes18,Toscano19}. 
For example, one control technique is to have
the skyrmions interact with a 2D periodic substrate such as
an anti-dot array
\cite{Reichhardt15aa,Saha19,Feilhauer19,Vizarim20}. 

A dynamical feature that distinguishes skyrmions
from many other particle systems 
is
the strong non-dissipative Magnus or gyroscopic
component of the skyrmion motion
produced by
their topology  
\cite{EverschorSitte18,Nagaosa13,Vizarim20}, 
which affects both how the skyrmions move under a drive and also
how they interact with a 
substrate potential.
On a smooth landscape,  
a driven skyrmion moves at an angle with respect to the
driving direction known as the skyrmion Hall 
angle $\theta_{sk}$
\cite{EverschorSitte18,Nagaosa13,EverschorSitte14,Reichhardt15},
which increases as the
ratio of the Magnus force to the
damping term increases.
In experimental measurements,
skyrmion Hall angles ranging from just a few degrees up 
to $55^\circ$ have been observed
\cite{Jiang17,Litzius17,Woo18,Juge19,Zeissler20};
however, it is expected that much larger Hall angles could
arise in certain skyrmion-supporting materials \cite{Nagaosa13}.  
In devices such as a race track memory,
the skyrmion Hall angle strongly limits the distance the 
skyrmion can move before it reaches the edge of the sample,
and as a result, a variety of studies have focused
on methods for controlling the skyrmion Hall angle.
In the presence of pinning, the skyrmion Hall angle
develops a drive or velocity dependence,
increasing from
a value of nearly zero
just above the depinning threshold
to a saturation at the pin free value for high drives 
\cite{Reichhardt15,Jiang17,Litzius17,Woo18,Juge19,Zeissler20,Muller15,Kim17,Legrand17,Reichhardt19,Reichhardt19b,Reichhardt20}.
This behavior can result from a
side jump effect that occurs when the skyrmion interacts with the
pinning sites
which diminishes in magnitude as the skyrmion
velocity increases
\cite{Saha19,Muller15,Reichhardt20,Litzius20}.
Other effects such as skyrmion distortions can also 
alter the skyrmion Hall angle \cite{Litzius17,Litzius20}. 

Nanostructuring techniques can be used to create
periodic pinning landscapes for skyrmions in which
the pinning sites
act as repulsive or attractive scattering sites
\cite{Stosic17,Fernandes18,Feilhauer19}. 
Particle based simulations examining the motion of skyrmions in 
the presence of 2D arrays of scattering sites 
show that at low drives,
the skyrmions move in the direction of the drive, while at higher drives,
the skyrmion Hall angle is quantized and the skyrmion motion
locks to different symmetry directions of
the underlying substrate \cite{Saha19,Feilhauer19,Vizarim20},
similar 
to the dynamical symmetry locking found in overdamped systems such as
dc driven superconducting vortices \cite{Reichhardt99}
or
colloids \cite{Korda02,MacDonald03,Reichhardt12}.
In the overdamped systems, the direction of the external drive must be changed
to observe the dynamical locking, whereas
in the skyrmion system, the drive direction is fixed.
Skyrmions moving over 2D substrates under
combined dc and ac driving exhibit
various types of locking effects including Shapiro steps
and transverse phase locking \cite{Vizarim20a}.

\begin{figure}
\includegraphics[width=3.5in]{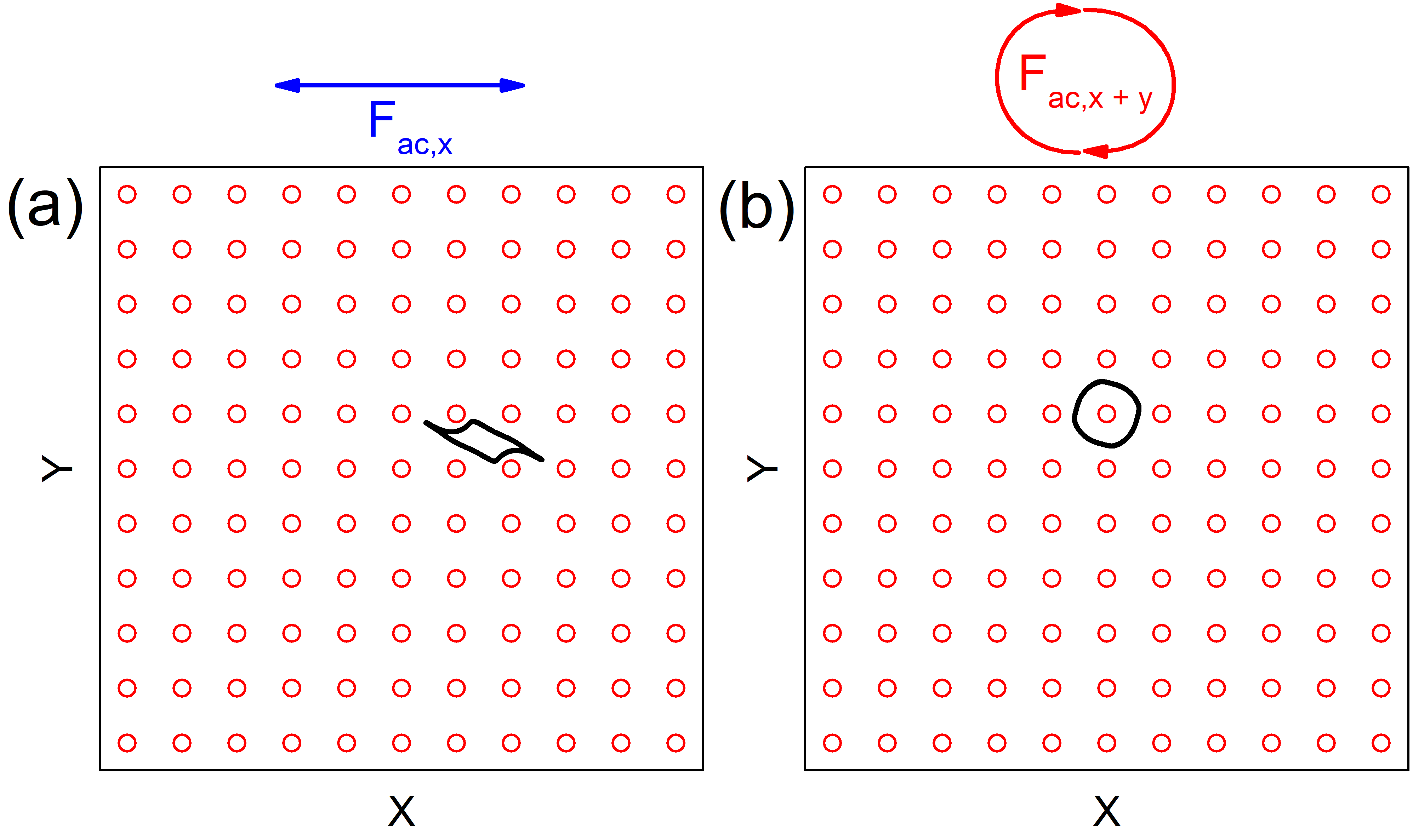}
\caption{Obstacles (red circles) and skyrmion trajectory (black line) in
a sample where the obstacles are modeled as repulsive
Gaussian scattering sites, while the skyrmion is represented as a point
particle experiencing both damping and a Magnus force.
(a) Skyrmion motion under linear ac driving $F_{ac,x}$.
(b) Skyrmion motion under circular ac driving
$F_{ac,x+y}$.
}
\label{fig:1}
\end{figure}

In this work we examine skyrmions interacting with a
two dimensional periodic array of 
obstacles or antidots and subjected to either
a linear or a circular ac drive, as shown in Fig.~\ref{fig:1}. 
Under linear ac driving, the Magnus force makes it possible
to observe
two-dimensional
skyrmion orbits
as well as directed motion in which the skyrmions translate
due to a symmetry breaking of the orbits by the Magnus force.
Under circular ac driving, an even richer
variety of dynamical phases appear, including localized periodic orbits
as well as
localized or diffusing states similar to but more varied than those found
in the electron pinball systems.
We also observe a series of ratcheting 
orbits for certain ac amplitudes and frequencies along with
reversals of the ratchet effect. 
Our results suggest that ac driving provides
a new method for controlling skyrmion motion
and achieving topological transport.  

\section{Simulation}

We consider a two-dimensional system of size 
$L\times L$ with periodic boundary conditions containing 
$N_{p}$ obstacles placed in a square array.
A single skyrmion is added to the sample and subjected to
either a linear
ac drive applied along the $x$-direction,
as shown in Fig.~\ref{fig:1}(a),
or a circular ac drive composed of
two perpendicular ac drives that are out of phase by $90^\circ$,
as shown in Fig.~\ref{fig:1}(b).  
The skyrmion dynamics are obtained
from a particle based or modified Theile equation approach
as used in previous work
\cite{Reichhardt15aa,Vizarim20a,Lin13,Brown19,Xiong19}.
The equation of motion is given by
\begin{equation}
\alpha_{d}{\bf v}_{i} + \alpha_{m}\times {\bf v}_{i} = {\bf F}^{o} + {\bf F}^{AC}  .
\end{equation}
The first term on the left
of magnitude $\alpha_d$ is the damping term,
which aligns the skyrmion velocity with
the net external forces experienced by the skyrmion.
The second term on the left is the Magnus force, 
${\alpha}_m$  
which produces a velocity component perpendicular to
the net forces on the skyrmion.
The obstacle interaction force ${\bf F}^{o}$
arises from obstacles with 
a Gaussian potential energy $U_o=C_oe^{-{\left({r_{io}}/{a_o}\right)}^2}$, where $C_o$ 
is the strength of the obstacle potential, 
$r_{io}$ is the distance between skyrmion $i$ and obstacle $o$,
and $a_o$ is the obstacle radius. 
The force between the obstacles and the skyrmion is given by
${\boldsymbol{\mathrm{F}}}^o_i=-\mathrm{\nabla }U_o=-F_or_{io}e^{-{\left({r_{io}}/{a_o}\right)}^2}{\widehat{\boldsymbol{\mathrm{r}}}}_{io}\ $, 
where $F_o=2U_o/a^2_o$.
For computational efficiency, we cut off the interaction
beyond $r_{io}=2.0$, since it becomes negligible at larger distances.
We use an obstacle density of $\rho_o=0.093/\xi^2$,
where the dimensionless unit of length is $\xi$.
The ac driving force is given by
\begin{equation} 
{\bf F}^{AC} = A\sin(2\pi\omega_{1} t){\hat {\bf x}} + B\cos(2\pi\omega_{2} t){\hat {\bf y}} .
\end{equation}
For linear driving, $B=0$, while for circular driving,
both $A$ and $B$ are nonzero.
In each simulation
we fix the frequency values $\omega_1$ and $\omega_2$
and increase the ac amplitudes $A$ and $B$ in increments
of $0.002$ every $10^5$ simulation time steps.
For each ac drive amplitude value we measure the skyrmion velocity
$\langle V_{||}\rangle=\langle {\bf v}_i \cdot {\hat {\bf x}}\rangle/(2\pi\omega_1 a)$ in the $x$ direction
and 
$\langle V_{\perp}\rangle=\langle {\bf v}_i \cdot {\hat {\bf y}}\rangle/(2\pi\omega_1 a)$ in the $y$ direction,
where the average is taken over 100 ac drive cycles.
Under our normalization, a value $\langle V_{||}\rangle=1.0$
or $\langle V_{\perp}\rangle = 1.0$ indicates that the skyrmion
is translating by one substrate lattice constant per ac drive cycle
in the $x$ or $y$ direction, respectively.

\section{Skyrmion Pinball}

\begin{figure}
  \begin{minipage}{3.5in}
    \begin{minipage}{3.5in}
      \includegraphics[width=3.5in]{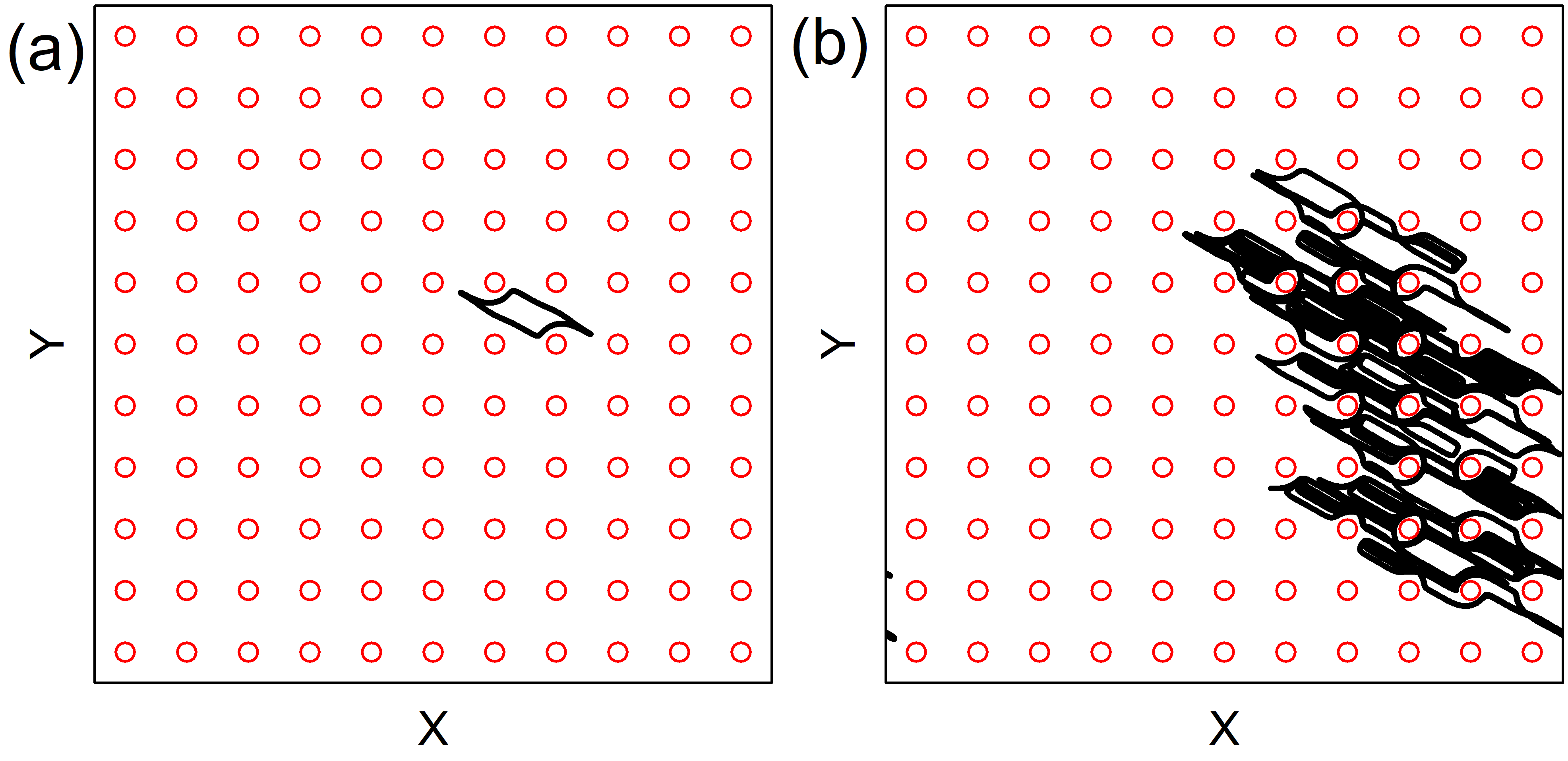}
    \end{minipage}
    \begin{minipage}{3.5in}
      \includegraphics[width=3.5in]{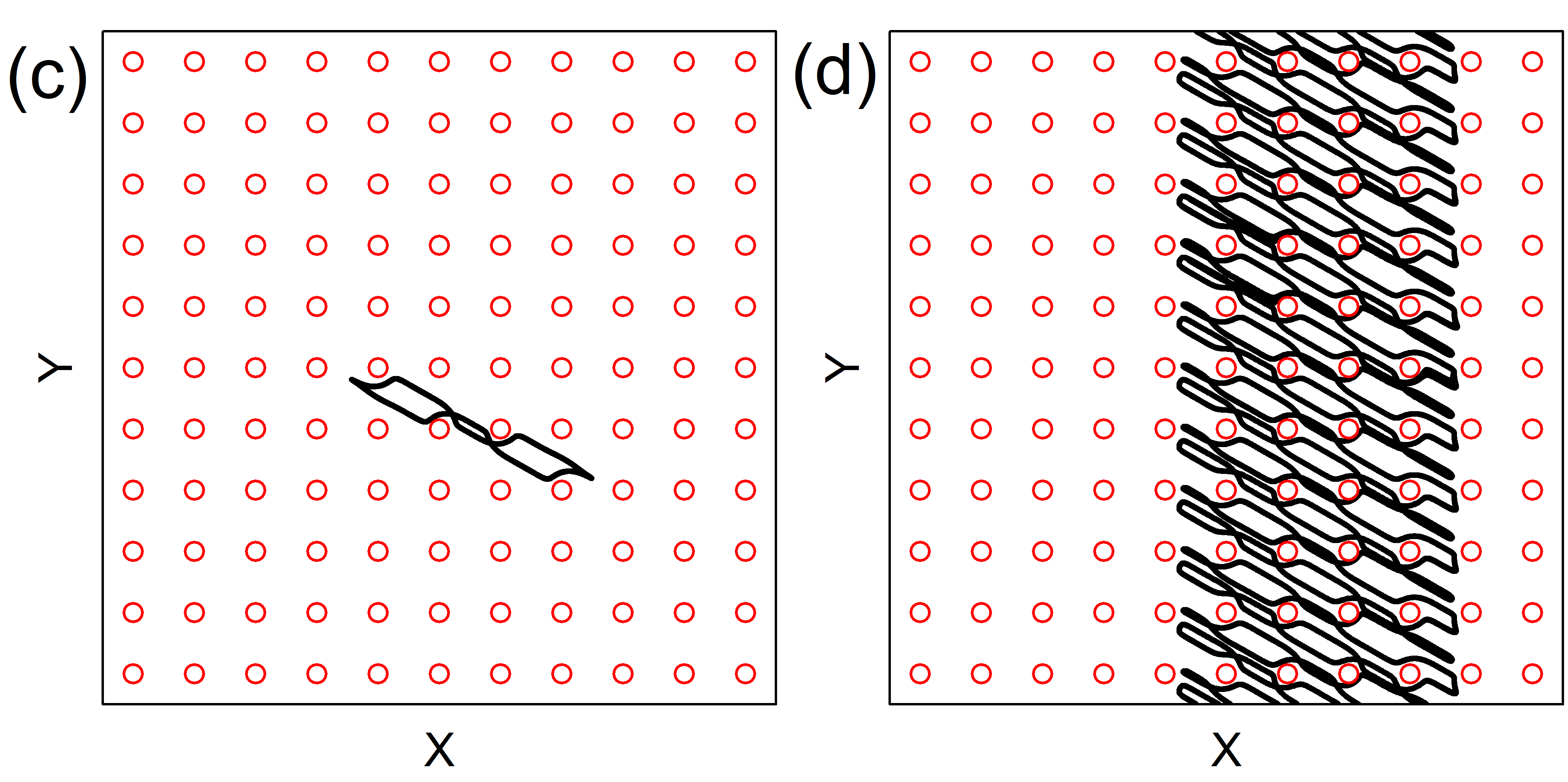}
    \end{minipage}
    \begin{minipage}{3.5in}
      \includegraphics[width=3.5in]{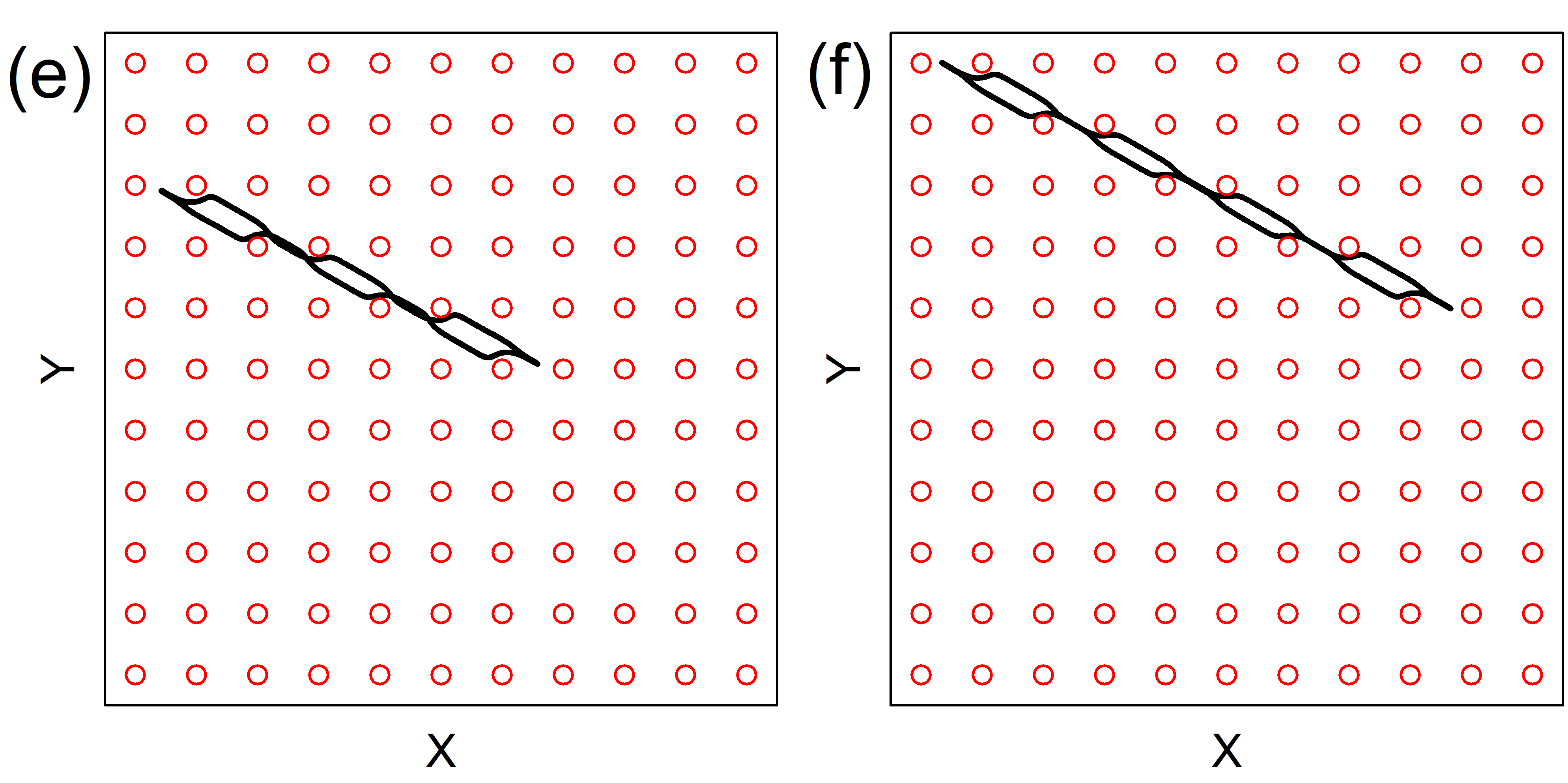}
    \end{minipage}
\end{minipage}
\caption{Obstacles (red circles) and skyrmion trajectory (black
line)
in a sample with 
$\alpha_{m}/\alpha_{d} = 0.577$ and $a_{0} = 0.65$ for
linear ac driving $F_{ac,x}$ along the $x$ direction.
(a) At $A = 0.5$, the skyrmion is oscillating between 
obstacles.
(b) At $A = 0.91$, the skyrmion motion
is delocalized.
(c) At $A = 1.0$, the skyrmion motion is localized.
(d) Directed motion at $A  = 1.22$.
The skyrmion is translating by one lattice constant
in the
negative $y$-direction every two ac drive cycles. 
(e) Localized state at $ A= 1.5$ where the skyrmion orbit spans three
plaquettes.
(f) Localized state at $A = 2.0$ where the
skyrmion orbit spans four plaquettes.  
}   
\label{fig:2}
\end{figure}

\subsection{Linear ac Drive}
We first consider the case of
a skyrmion subjected to a linear ac drive
applied along the $x$-direction, so that $B = 0$
as shown in Fig.~\ref{fig:1}(a).
In an overdamped system, a particle under such a drive would
remain localized and would
simply follow a one-dimensional trajectory
along the $x$-direction between the
obstacles.
A skyrmion with a finite Magnus term
moves in a 2D orbit, as illustrated
in Fig.~\ref{fig:2}
for a system with $\alpha_{m}/\alpha_{d} = 0.577$ and $a_{0} = 0.65$. 
At $A = 0.5$ in Fig.~\ref{fig:2}(a),
the orbiting motion is localized.
For a larger ac drive amplitude of
$A = 0.91$ in
Fig.~\ref{fig:2}(b),
the orbit size
becomes large enough that
the skyrmion
collides with multiple obstacles,
resulting in delocalization and diffusive chaotic motion but
no net drift.
In Fig.~\ref{fig:2}(c)
at $A = 1.22$,
the skyrmion has relocalized with
a larger orbit that is
similar in shape to the orbit
in Fig.~\ref{fig:2}(a) but
that spans two plaquettes during each ac drive cycle.
The chaotic phase illustrated
in Fig.~\ref{fig:2}(b)
occurs at the transition between a skyrmion orbit
that spans one plaquette and an orbit that spans two plaquettes.
In each case the orbits are oriented at an angle to the linear ac drive
as a result of the Magnus force.
In the absence of the substrate, the skyrmion would move in 
a one-dimensional orbit at an angle of
$\theta_{sk} = \arctan(\alpha_{m}/\alpha_{d})$ to the $x$ direction.      
As $A$ increases above $A=1.0$, the skyrmion remains locked
in the localized state of Fig.~\ref{fig:2}(b) until a transition
occurs between localized motion spanning two plaquettes and the localized
motion spanning three plaquettes that is illustrated in Fig.~\ref{fig:2}(e)
at $A=1.5$.
A series of such localized phases occurs each time the skyrmion
orbit spans an integer number $n$ of plaquettes, such as the
$n=4$ state shown in Fig.~\ref{fig:2}(f) at $A=2.0$.
At the transitions between different localized states,
delocalized or chaotic motion appears.
In some cases there can also be fractional localization,
as shown in Fig.~\ref{fig:2}(d) at $A = 1.22$ 
where the skyrmion repeatedly switches between localized orbits
that span two and three plaquettes.

In the overdamped limit,
particles on a 2D substrate under a linear ac sinusoidal drive
do not exhibit any kind of directed motion or ratchet effect;
however,
if two perpendicular ac drives
are applied,
the particle orbits become two-dimensional and directed
motion can occur if the ac driving breaks a symmetry
\cite{Tierno07,Loehr16,Reichhardt03,Speer09,Platonov15}. 
In the skyrmion case,
a linear ac drive in conjunction with the skyrmion-obstacle
interactions produces a 2D orbit due to the Magnus force,
as illustrated
in Fig.~\ref{fig:2}(a).
Since the orbit is chiral,
temporal symmetry is broken and a ratchet effect
can occur \cite{Reimann02}.
Under linear ac driving, we find only limited regimes where
ratcheting of the skyrmion appears; however, this result demonstrates
that even a single ac drive is sufficient to produce directed skyrmion
motion on a symmetric substrate.

\begin{figure}
\includegraphics[width=3.5in]{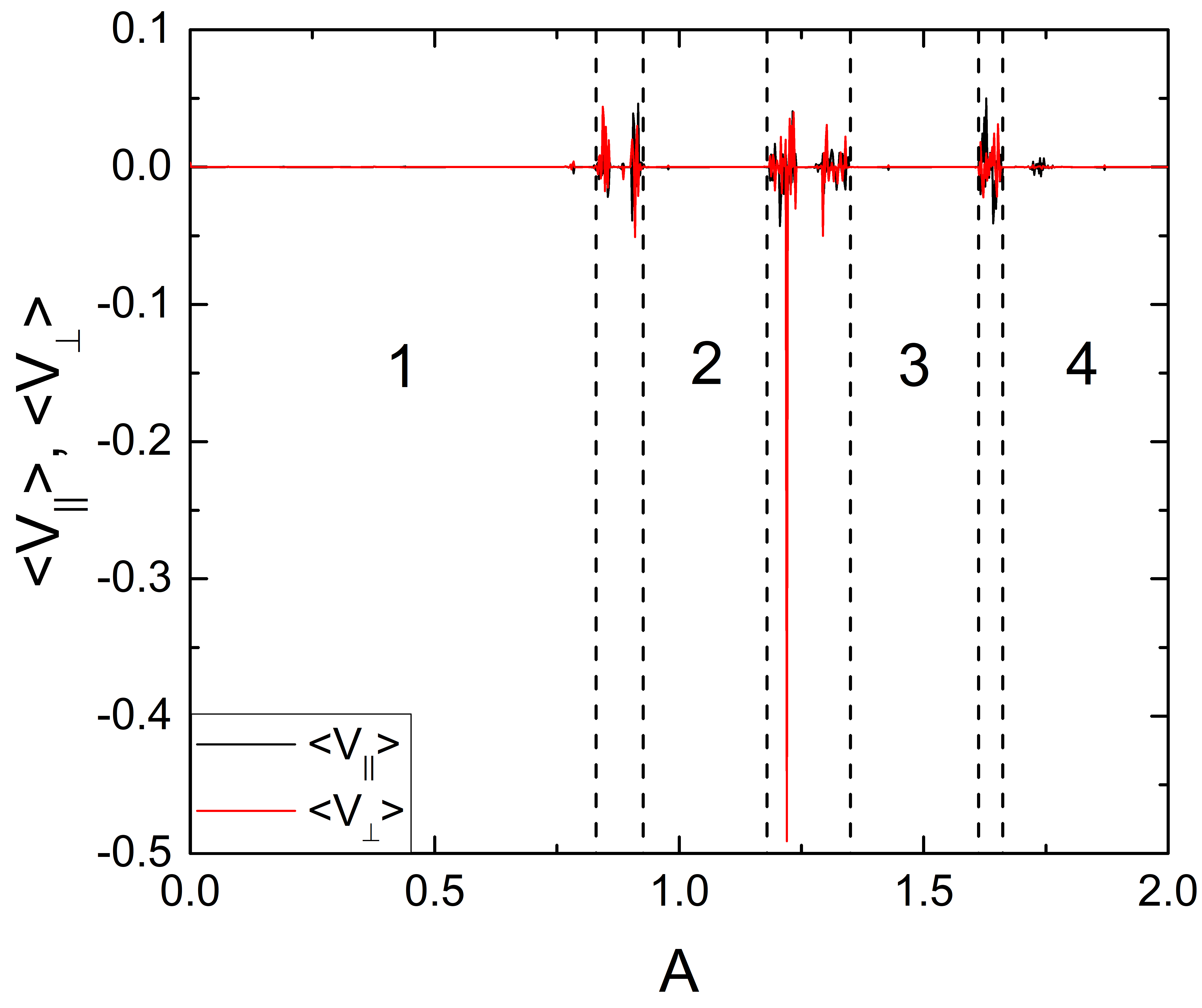}
\caption{ $\langle V_{||}\rangle$ (black) and $\langle V_{\perp}\rangle$ (red)
vs ac drive amplitude $A$ for the
system in Fig.~\ref{fig:2} with linear ac driving, $\alpha_m/\alpha_d=0.577$,
and $a_0=0.65$.
Localized states occur on the plateaus of zero velocity.
Chaotic diffusing states appear in the regions of fluctuating velocity
between the plateaus, and there is a translating state corresponding to
the negative peak in $\langle V_{\perp}\rangle$ at $A=1.22$.
Vertical dashed lines indicate the boundaries of the regions in which
the skyrmion orbit passes through $n=1$, 2, 3, or 4 plaquettes, from
left to right.
}  
\label{fig:3}
\end{figure}

In Fig.~\ref{fig:3} we plot $\langle V_{||}\rangle$ and
$\langle V_{\perp}\rangle$
versus ac drive amplitude $A$ for the
system in Fig.~\ref{fig:2} with linear ac driving.
In the localized states,
$\langle V_{||}\rangle = \langle V_{\perp}\rangle = 0$,
while in the delocalized states there are finite velocity
fluctuations.
Since the particle motion is diffusive, these fluctuations
diminish in magnitude if the velocities are averaged over
a longer period of time.
At $A = 1.22$, there is ballistic motion
in the negative $y$ direction
and we find $\langle V_{||}\rangle = 0$ and $\langle V_{\perp}\rangle=-0.5$,
indicating that the skyrmion is translating 
by one lattice constant every two ac drive cycles, as
illustrated in Fig.~\ref{fig:2}(d).
The vertical dashed lines in Fig.~\ref{fig:3}
indicate the edges of the regions in which the skyrmion motion forms
localized orbits spanning $n=1$, 2, 3, or 4 plaquettes.
We also observe smaller fractional localized phases
in which the skyrmion
spends two drive cycles in an orbit spanning
$n$ plaquettes and one drive cycle in an orbit spanning
$n+1$ plaquettes.
These fractional localized states 
occur in the boundary regions
between localized states, and are surrounded by delocalized phases.

\begin{figure}
  \begin{minipage}{3.5in}
    \begin{minipage}{3.5in}
      \includegraphics[width=3.5in]{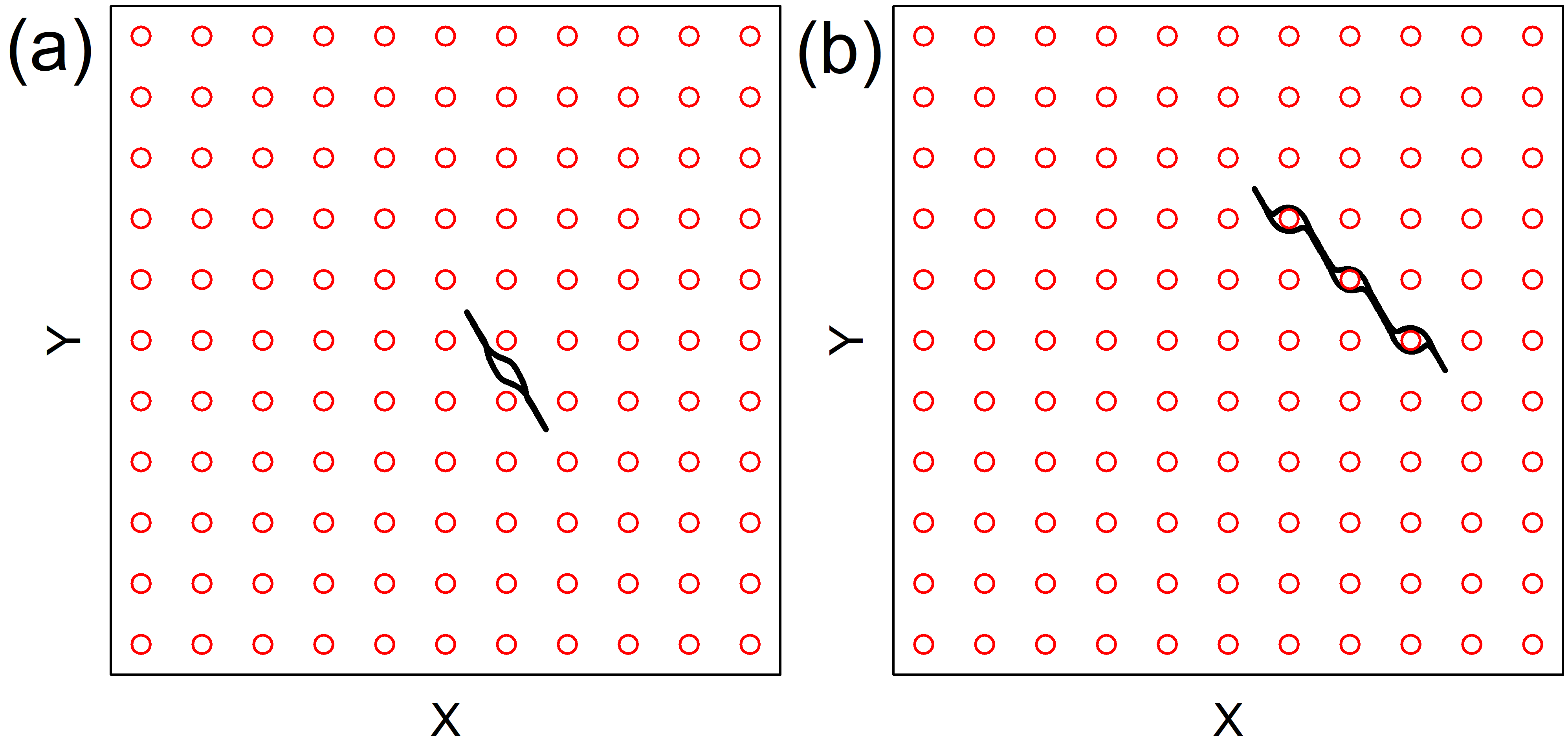}
    \end{minipage}
    \begin{minipage}{3.5in}
      \includegraphics[width=3.5in]{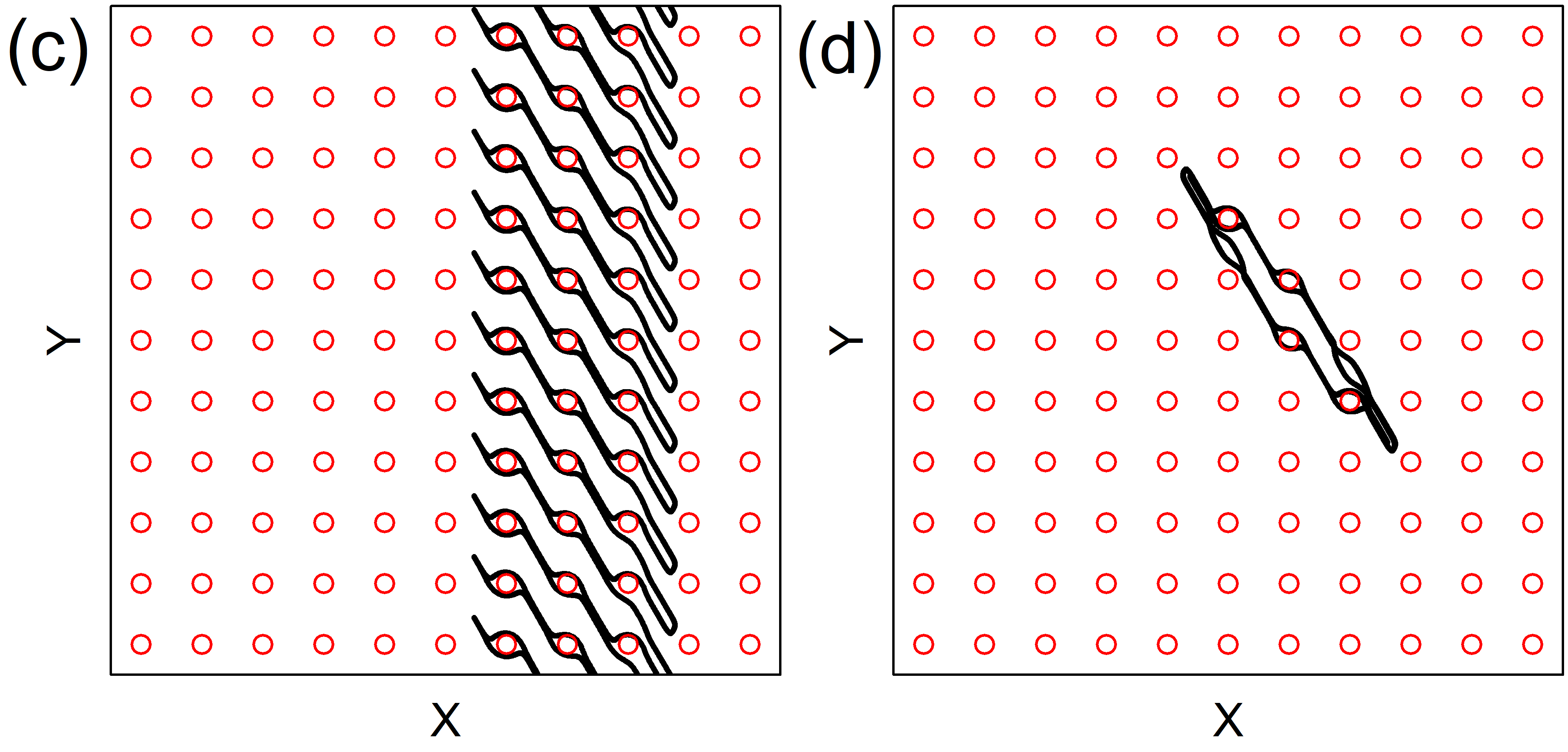}
    \end{minipage}
    \begin{minipage}{3.5in}
      \includegraphics[width=3.5in]{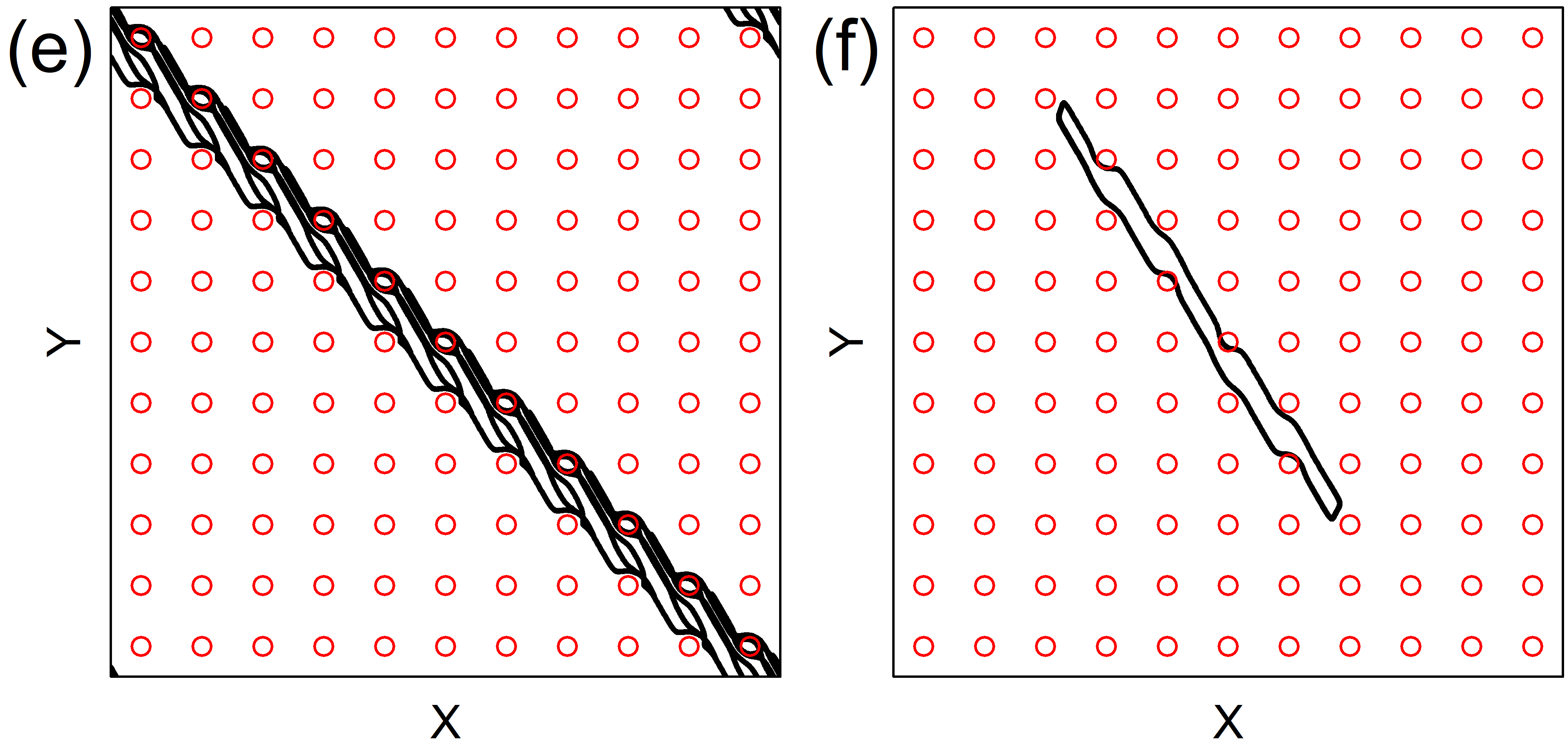}
    \end{minipage}
  \end{minipage}
\caption{
  Obstacles (red circles) and skyrmion trajectory (black line)
  in a system
  with $\alpha_m/\alpha_d=1.732$,
  $a_0=0.65$, and linear driving $F_{ac,x}$ along the $x$ direction.
  (a)
  At $A = 0.5$ the motion is localized.
  (b)
  At $A = 0.934$ the localized orbit encircles two obstacles.
  (c) At $A = 1.05$
  there is a translating orbit with motion
  in the positive $y$-direction.  
  (d)
  Localized  motion at $A  = 1.136$.
  (e)
  At
  $ A= 1.29$,
  the skyrmion translates at a
  $45^\circ$ angle.
  (f) An $n=7$ localized state at $A = 1.656$.
}  
\label{fig:4}
\end{figure}

In Fig.~\ref{fig:4}(a-f) we plot representative skyrmion trajectories
for a system with a linear ac drive and a larger
Magnus component of $\alpha_{m}/\alpha_{d} = 1.732$.  
For $A = 0.5$ and $A=0.934$ in Fig.~\ref{fig:4}(a) and (b),
the orbit is localized at $n=1$ and $n=3$ plaquettes, respectively.
The motion remains entirely within the interstitial region
between obstacles for
$A=0.5$, but when $A=0.934$, the skyrmion encircles the obstacles.
At $A = 1.05$ in Fig.~\ref{fig:4}(c),
we find
a translating orbit in
which the skyrmion moves one
lattice constant in the $+y$ direction
per ac drive cycle.
When $A = 1.136$, the orbit is localized again but has a more complex
shape, encircling four obstacles that do not fall along a one-dimensional
line, as shown in Fig.~\ref{fig:4}(d).
For $A = 1.29$ in Fig.~\ref{fig:4}(e),
there is
a translating orbit where
the skyrmion moves in the positive $y$ and negative $x$ directions
at an angle of $-45^\circ$,
translating 
by one lattice constant during every two ac drive cycles.
In Fig.~\ref{fig:4}(f) at $A = 1.656$,
there is an $n=7$ localized orbit that does not encircle any obstacles.

\begin{figure}
  \begin{minipage}{3.5in}
    \begin{minipage}{3.5in}
      \includegraphics[width=3.5in]{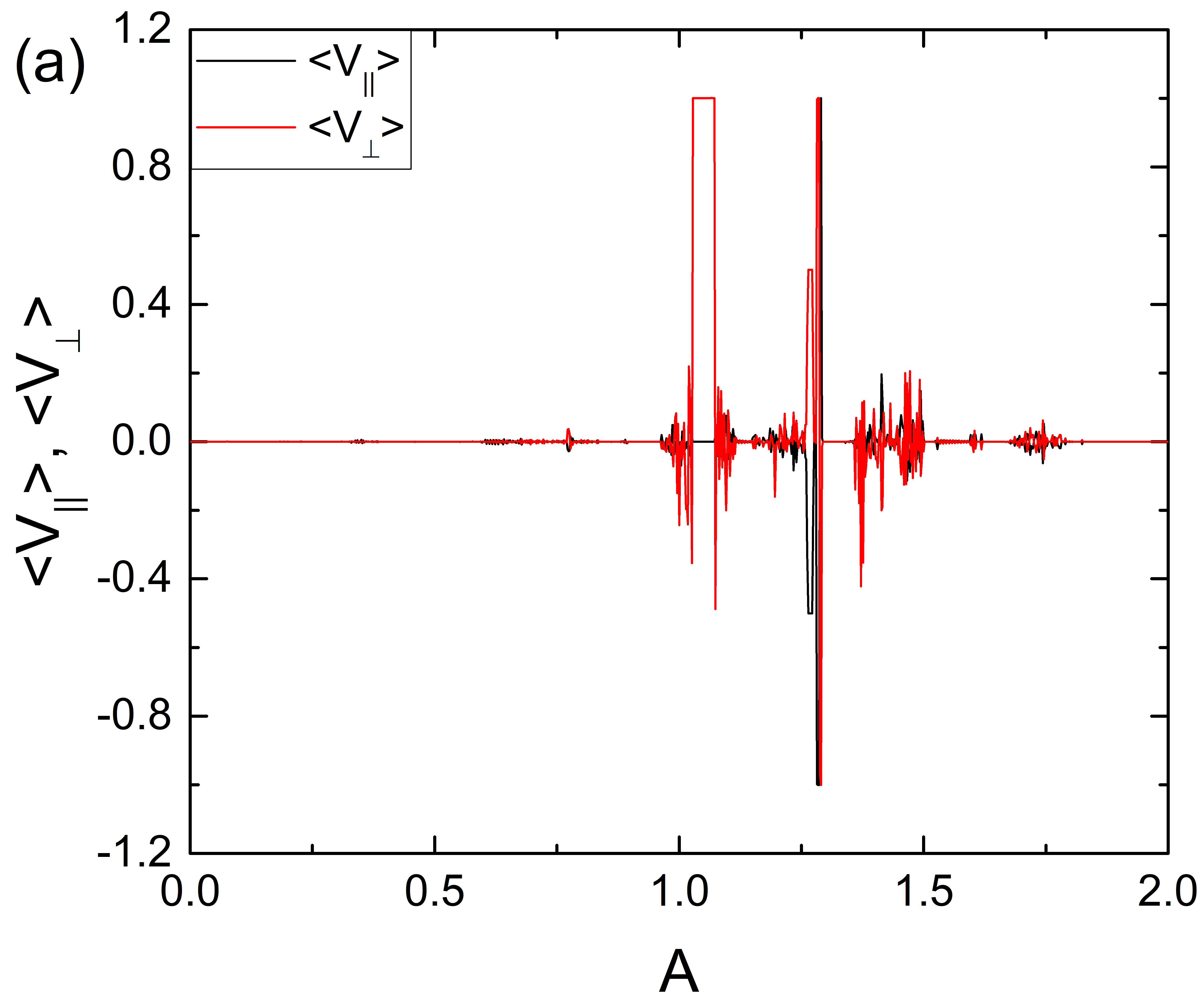}
    \end{minipage}
    \begin{minipage}{3.5in}
      \includegraphics[width=3.5in]{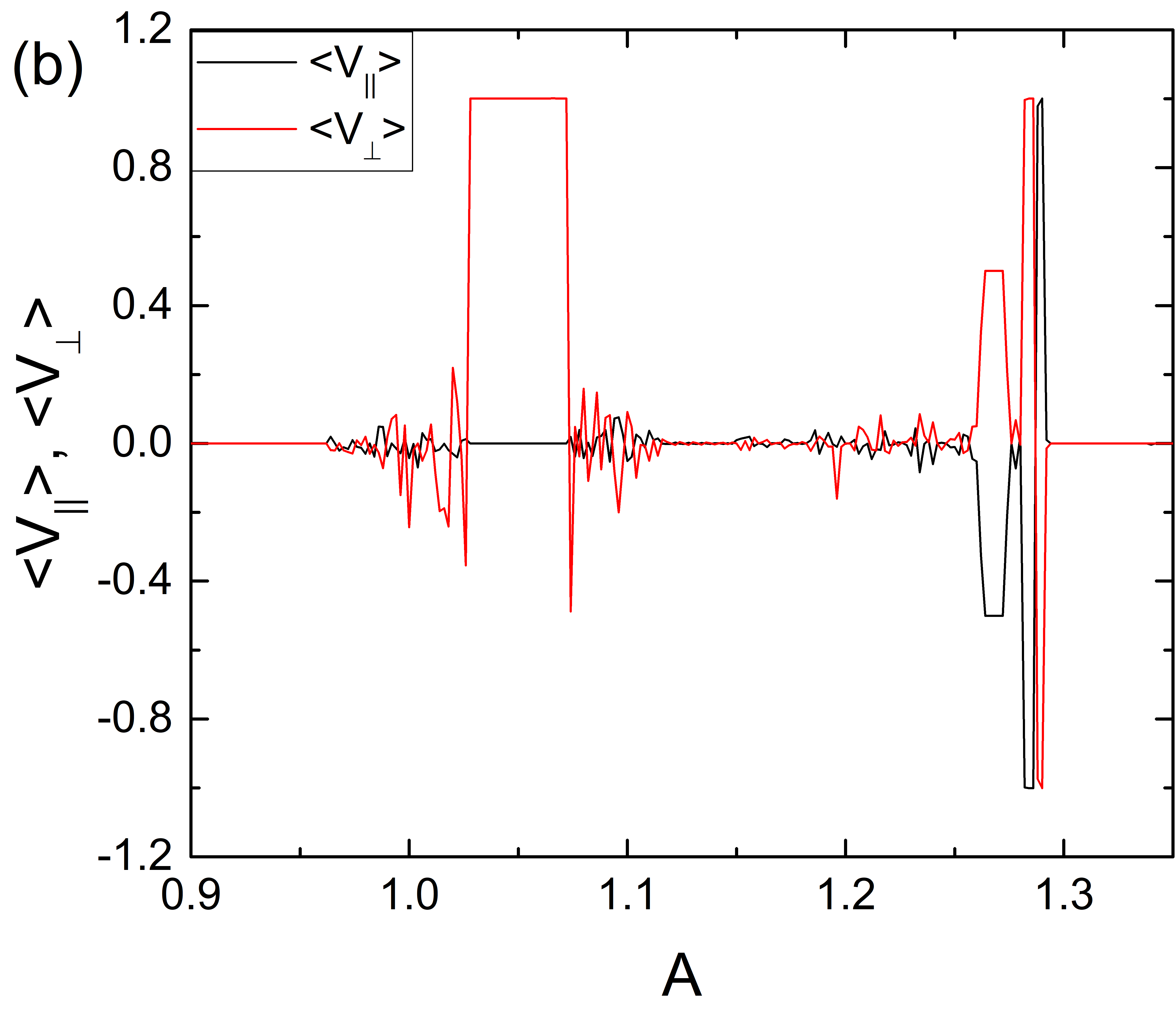}
    \end{minipage}
  \end{minipage}
\caption{(a) $\langle V_{||}\rangle$ (black) and
$\langle V_{\perp}\rangle$ versus $A$ for the
system in Fig.~\ref{fig:4} with linear ac driving,
$\alpha_{m}/\alpha_{d} = 1.732$,
and $a_0=0.65$.
(b) A blow up of panel (a) over the range
$0.9 < A < 1.35$ showing the three different regimes
in which translating orbits occur.
} 
\label{fig:5}
\end{figure}

In Fig.~\ref{fig:5}(a) we plot $\langle V_{||}\rangle$
and $\langle V_{\perp}\rangle$ versus $A$
for the system in Fig.~\ref{fig:4}
showing the different regions of localized and translating orbits.
Figure~\ref{fig:5}(b) is a blowup of Fig.~\ref{fig:5}(a)
over the range $0.9 < A < 1.35$.
Localized orbits appear when $A < 0.96$, followed by a window of
delocalized orbits for $0.96 \leq A < 1.03$.
The translating orbit illustrated
in Fig.~\ref{fig:4}(c) corresponds to the plateau of positive
$\langle V_{\perp}\rangle$ extending from $A=1.03$ to $A=1.07$.
As $A$ increases,
the system enters another regime of chaotic motion.
When $A=1.26$,
a translating state appears with motion along $-45^\circ$,
where the skyrmion moves one lattice constant in
the positive $y$ and negative $x$ directions every two ac drive cycles.
Here the velocities plateau with
$-\langle V_{||}\rangle=\langle V_{\perp}\rangle$.
This is followed by 
a small localized region, and then
at $A=1.28$ by a second regime of $-45^\circ$ translation,
where the skyrmion moves one lattice constant in
the positive $y$ and negative $x$ directions during every ac drive cycle.
At $A=1.29$ the $-45^\circ$ translation changes direction and
the skyrmion moves one lattice constant in
the positive $x$ and negative $y$ directions during every ac drive cycle,
as illustrated in Fig.~\ref{fig:4}(e).
Finally the motion becomes localized again above
$A=1.3$ as shown
in Fig.~\ref{fig:4}(f).
In general, as the Magnus term $\alpha_m$ increases,
a greater variety of distinct
localized and translating orbits appear.
The translating orbits are 
generally along either the $x$ or $y$ directions or at a
$45^\circ$ angle, since these are the most prominent symmetry directions
of the obstacle lattice.

\begin{figure}
  \begin{minipage}{3.5in}
    \begin{minipage}{3.5in}
      \includegraphics[width=3.5in]{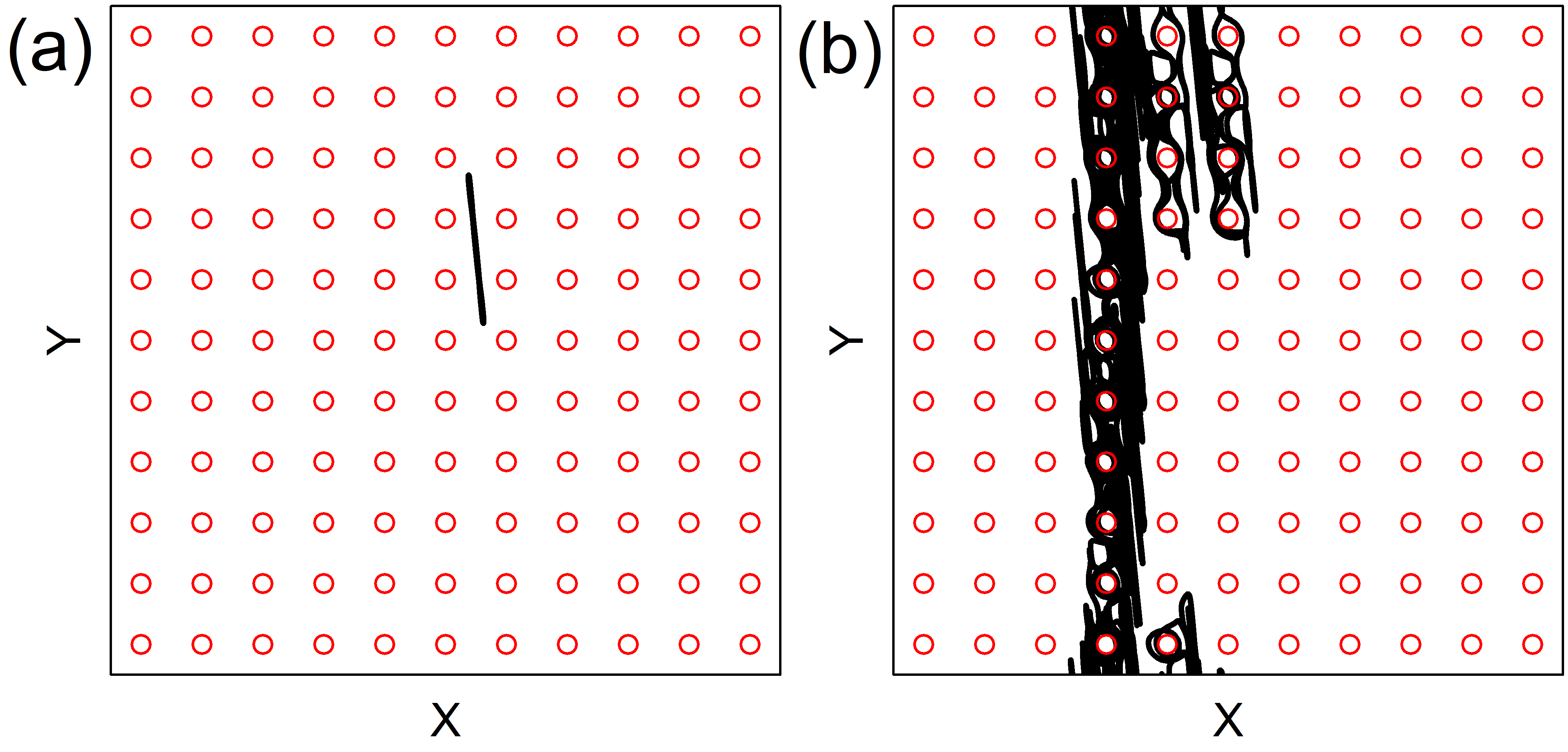}
    \end{minipage}
    \begin{minipage}{3.5in}
      \includegraphics[width=3.5in]{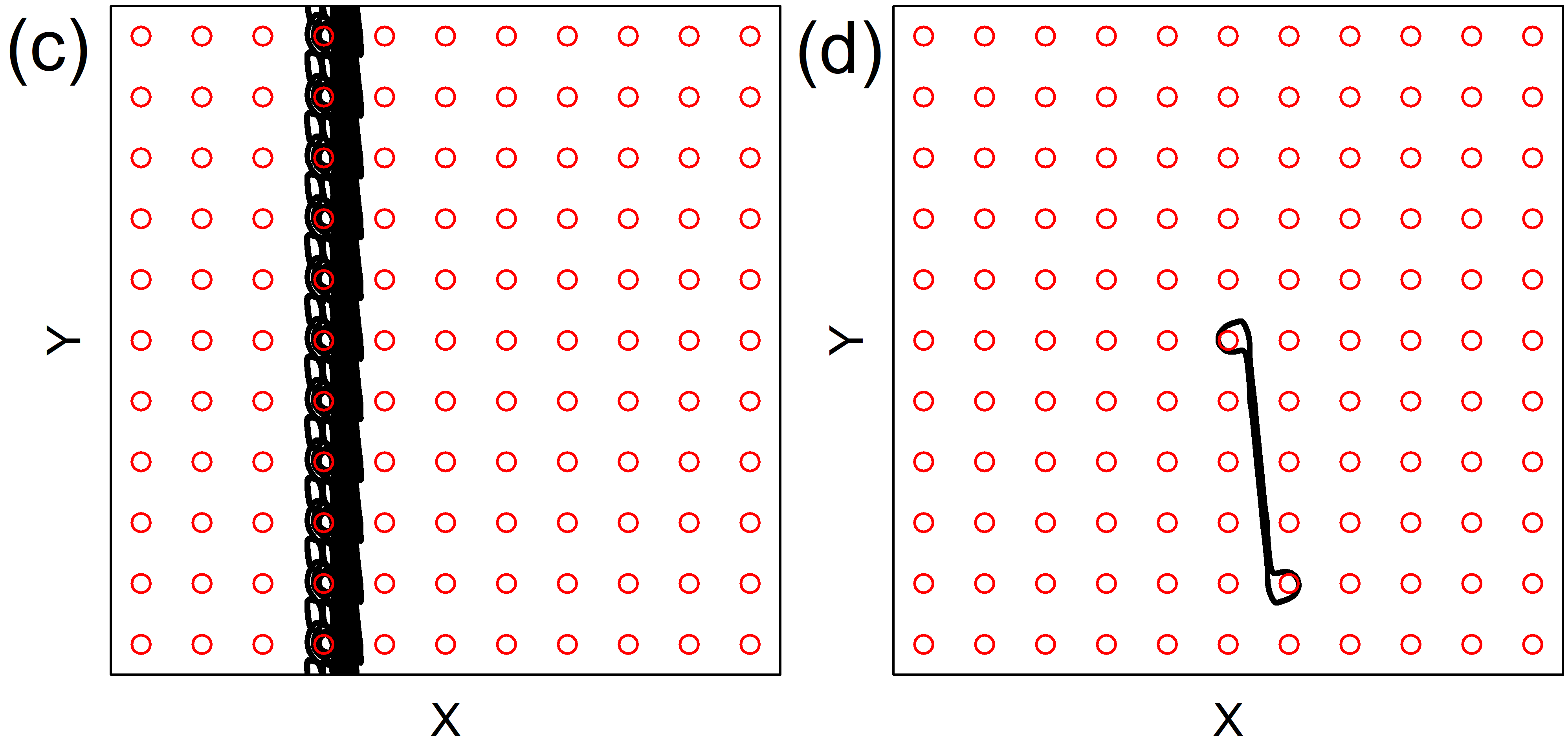}
    \end{minipage}
  \end{minipage}
\caption{Obstacles (red circles) and skyrmion trajectory (black line)
in a system with
$\alpha_{m}/\alpha_{d} = 9.962$, $a_{0} = 0.65$,
and linear driving $F_{ac,x}$ along the $x$ direction.
(a) A localized orbit at $A = 0.5$.
(b) Delocalized motion at $A = 0.576$.
(c) Translation along the $y$ direction at $A = 0.88$.
(d) A localized orbit at $A = 0.936$.
}  
\label{fig:6}
\end{figure}

\begin{figure}
\includegraphics[width=3.5in]{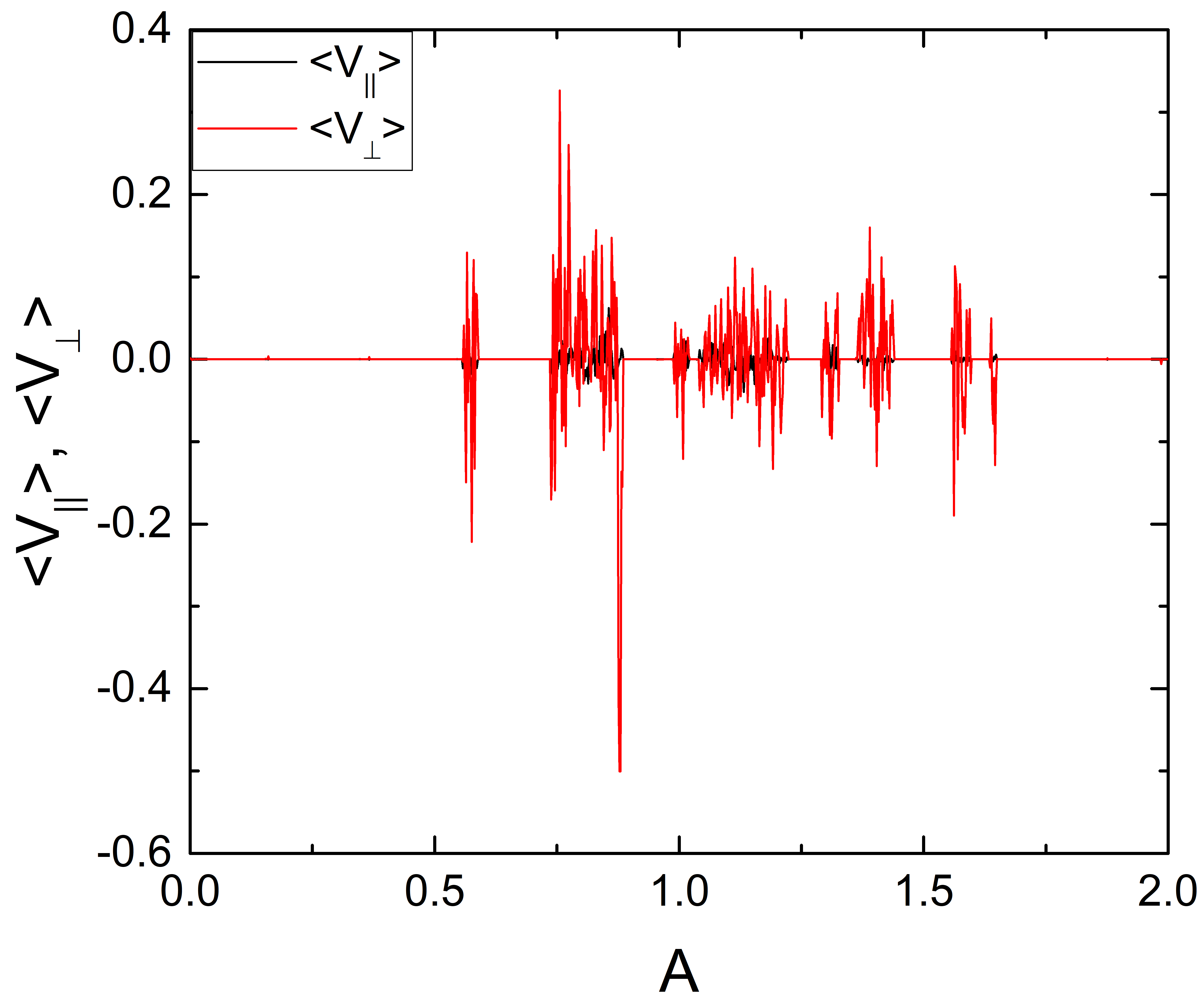}
\caption{ $\langle V_{||}\rangle$ (black)
and $\langle V_{\perp}\rangle$ (red) vs $A$ for the
system in Fig.~\ref{fig:6} with 
$\alpha_{m}/\alpha_{d} = 9.962$, $a_{0} = 0.65$,
and linear driving $F_{ac,x}$ along the $x$ direction.
The localized phases occur in the windows where both velocities are zero.
In several regions,
the orbits are localized but show
a combination of translating
and chaotic motion.    
}
\label{fig:7}
\end{figure}

For higher values of $\alpha_{m}/\alpha_{d}$,
we find
delocalized phases or 
chaotic orbits that exhibit an average drift. 
In Fig.~\ref{fig:6}
we plot
representative trajectories in a system with
$\alpha_{m}/\alpha_{d} = 9.962$.
At $A = 0.5$ in Fig.~\ref{fig:6}(a),
there is a localized 1D orbit oriented at nearly $90^\circ$ to the
driving direction.
For $A = 0.576$, the orbit is delocalized or chaotic but has a net
drift in the negative $y$ direction and a smaller net drift in the
negative $x$ direction, as shown in Fig.~\ref{fig:6}(b). 
When $A = 0.885$, as in Fig.~\ref{fig:6}(c),
we find a translating orbit with gradual
motion by one lattice
constant in the positive $y$ direction, where the skyrmion spends many
ac cycles at each location before stepping to the next location.
In Fig.~\ref{fig:6}(d) at $A = 0.936$,
a localized orbit appears that spans four plaquettes and
encircles the two obstacles at the top and bottom of the orbit.
In Fig.~\ref{fig:7} we plot $\langle V_{||}\rangle$ and
$\langle V_{\perp}\rangle$ versus $A$ to illustrate the transitions
between the different states.  The motion is localized
for $A <0.55$  and $A>1.7$, while at intermediate
values of $A$, various fluctuating and localized regions occur.

\subsection{Circular ac Drive with $A=B$}
We next consider the case of circular ac driving by applying two
ac drives that are perpendicular to each other and $90^\circ$ out of phase.
In the overdamped limit, a symmetric substrate does not produce
any directed motion
when the frequencies and amplitudes of the two ac drives are identical;
however, if the amplitudes or frequencies differ,
spatial asymmetry can appear in the particle orbit,
leading to directed motion \cite{Tierno07,Loehr16,Reichhardt03,Speer09}. 
For $A=B$ and $\omega_1=\omega_2$, if we set
$\alpha_m/\alpha_d=0$
we find no ratchet effect and the dynamics 
are the same as those found in 
vortex pinball systems,
with
transitions between localized
and delocalized orbits as the ac drive amplitude increases
\cite{Reichhardt02b}.

\begin{figure}
  \begin{minipage}{3.5in}
    \begin{minipage}{3.5in}
      \includegraphics[width=3.5in]{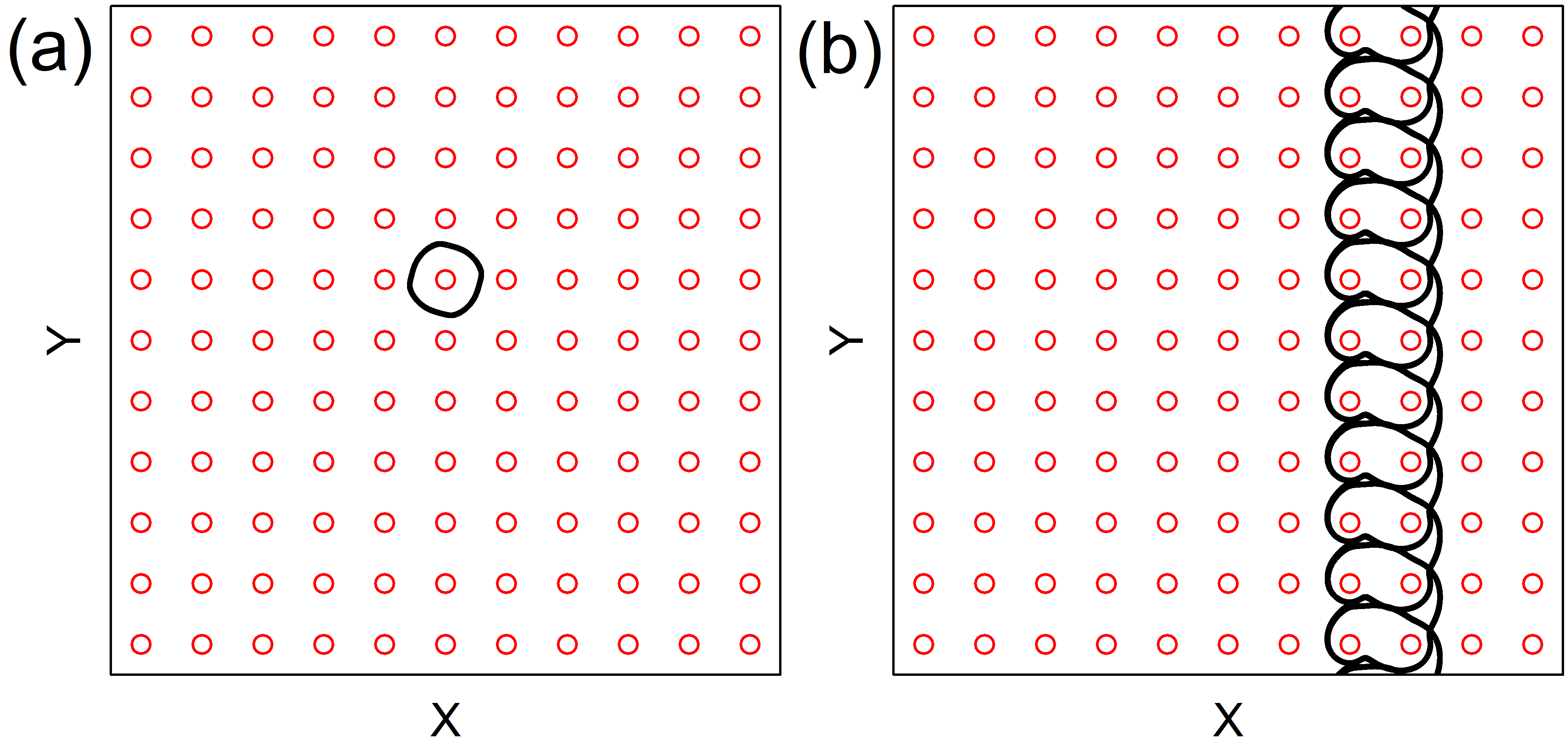}
    \end{minipage}
    \begin{minipage}{3.5in}
      \includegraphics[width=3.5in]{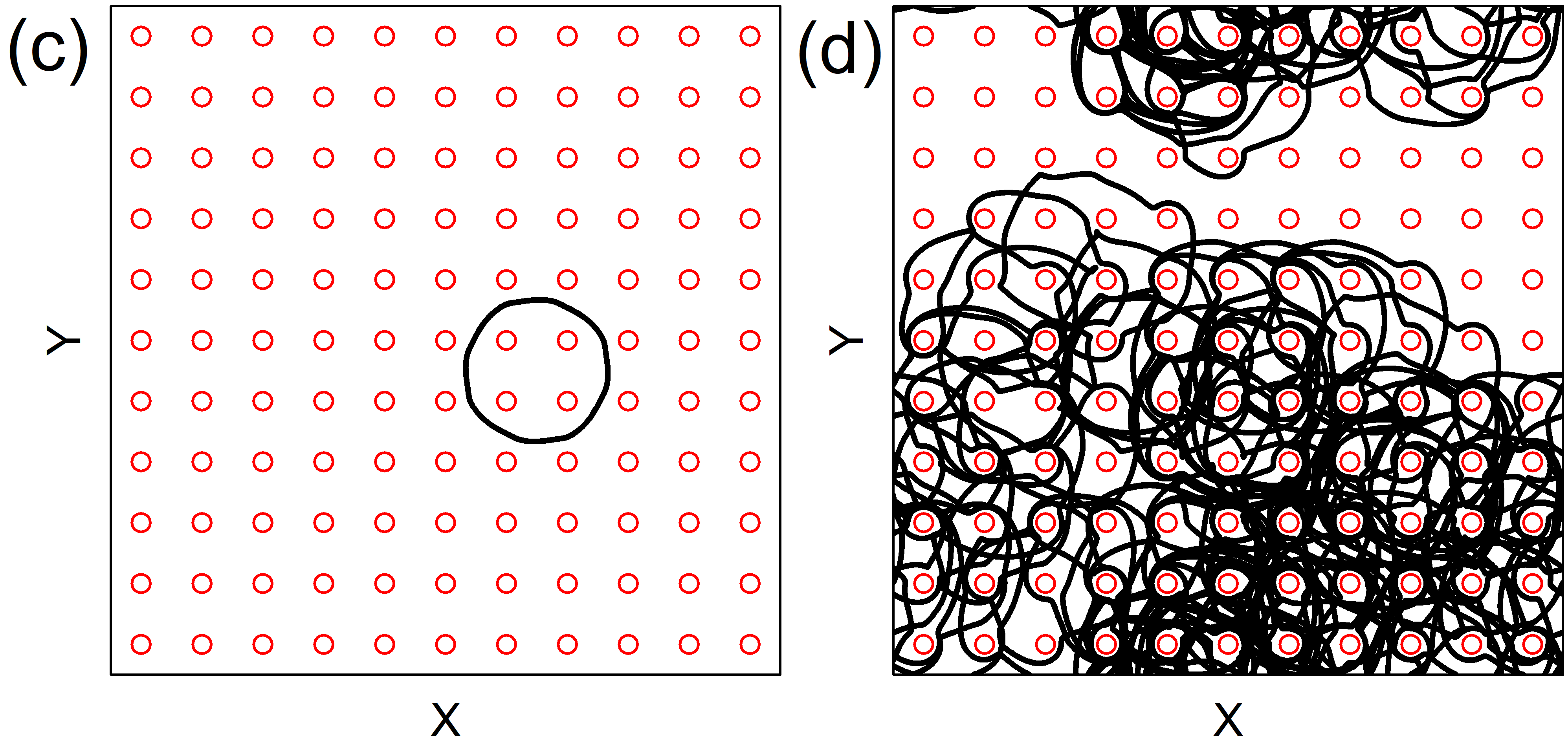}
    \end{minipage}
    \begin{minipage}{3.5in}
      \includegraphics[width=3.5in]{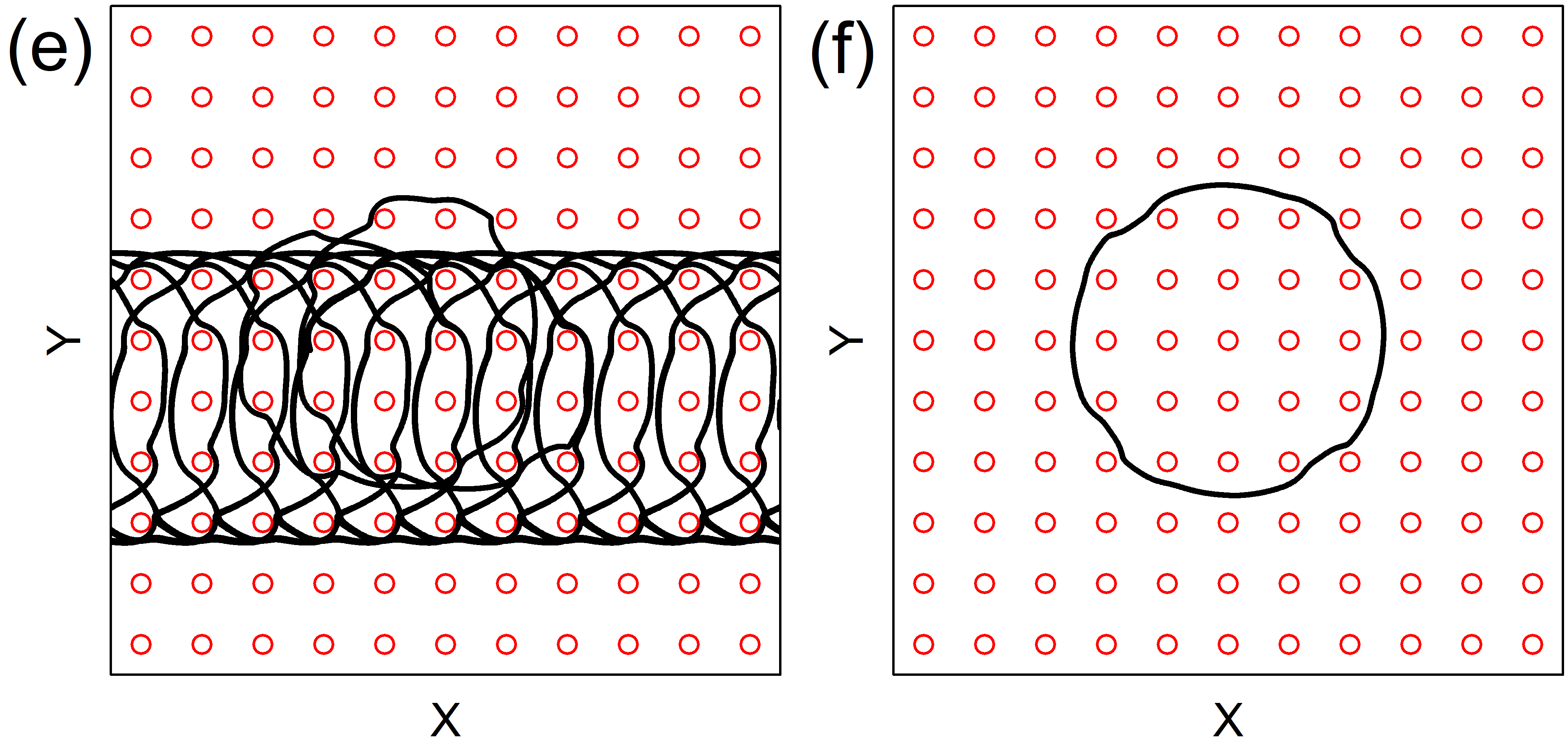}
    \end{minipage}
  \end{minipage}
\caption{Obstacles (red circles) and skyrmion trajectory (black line)
in a system with
$\alpha_{m}/\alpha_{d}  = 0.577$, $a_0=0.65$, and circular driving
with $B = A$ and $\omega_2=\omega_1=2 \times 10^{-5}$.
(a) At $A = 0.25$,
the skyrmion is in a localized orbit encircling one obstacle.  
(b) At $A = 0.375$ the orbit is translating in the positive $y$-direction.   
(c) At $A = 0.5$ there is a localized orbit encircling four obstacles.
(d) At $A = 0.672$ the orbit is diffusive.
(e) A translating orbit at $A = 0.978$.
(f) At $A = 1.082$ there is a localized orbit encircling 21 obstacles.   
} 
\label{fig:8}
\end{figure}

In Fig.~\ref{fig:8} we show skyrmion trajectories in a system with 
circular ac driving at $\alpha_{m}/\alpha_{d}  = 0.577$
for $A = B$ and $\omega_1=\omega_2=2 \times 10^{-5}$.
At $A=0.25$ in Fig.~\ref{fig:8}(a),
the skyrmion forms a localized orbit that encircles one obstacle.
In Fig.~\ref{fig:8}(b) 
at $A = 0.375$,
we find a translating orbit in which the skyrmion
spirals around two obstacles per ac drive cycle and
moves in the negative $y$-direction. 
Figure~\ref{fig:8}(c) shows a localized orbit at $A = 0.5$
where the skyrmion encircles four obstacles,
while the localized orbit at $A=0.672$ in
and Fig.~\ref{fig:8}(d)
has no net drift but exhibits diffusive motion over
long time scales.
At $A=0.978$ in Fig.~\ref{fig:8}(e),
there is
a translating orbit
in which the skyrmion
moves one lattice constant in the positive $y$ direction
during every
ac drive cycle,
while at $A = 1.082$ in Fig.~\ref{fig:8}(f), the skyrmion follows
a localized orbit encircling 21 obstacles.

\begin{figure}
\includegraphics[width=3.5in]{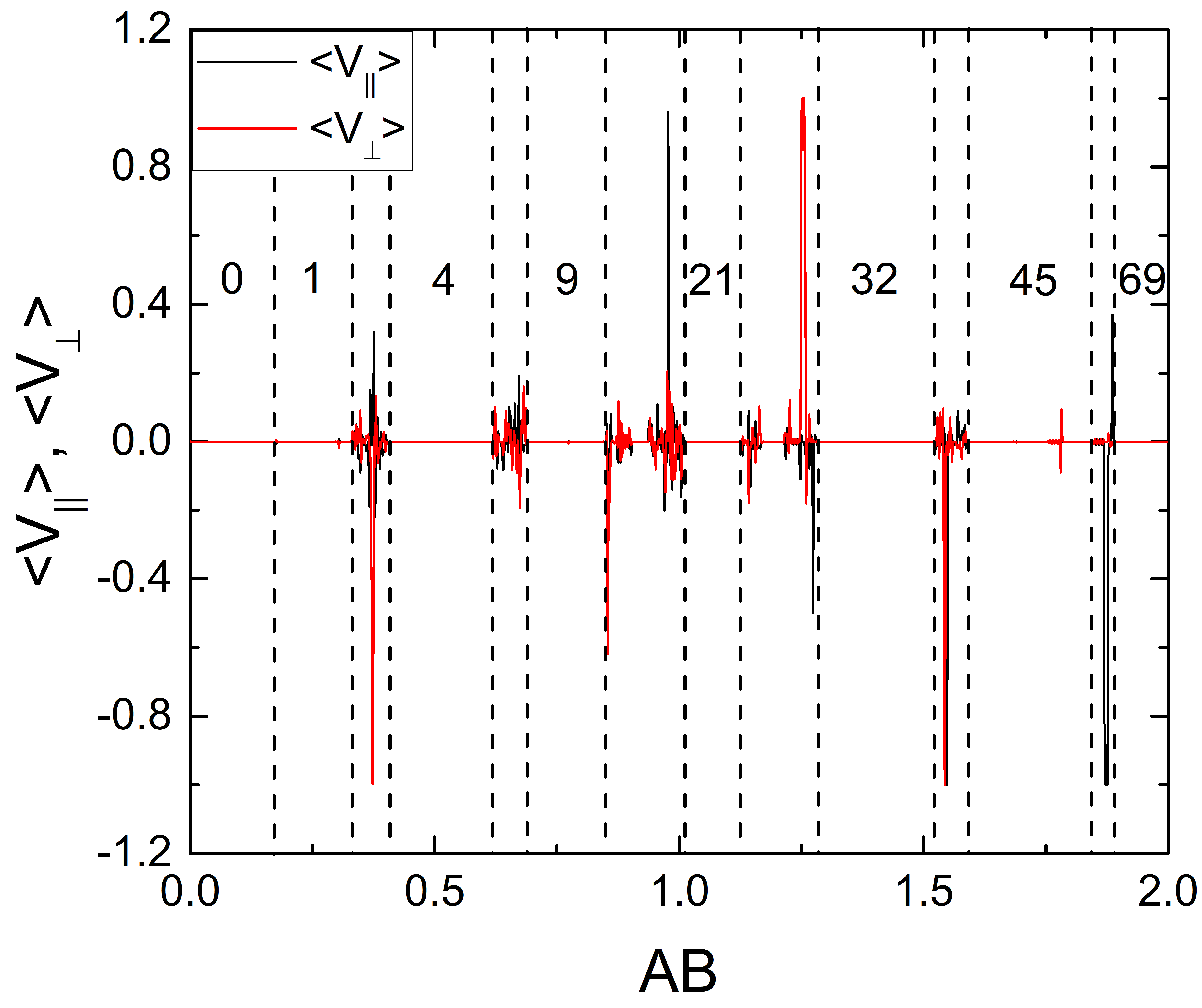}
\caption{ $\langle V_{||}\rangle$ (black) and
$\langle V_{\perp}\rangle$ (red) vs $A$ for the
system in Fig.~\ref{fig:8} with
$\alpha_{m}/\alpha_{d}  = 0.577$, $a_0=0.65$, and circular driving
with $B = A$ and $\omega_2=\omega_1=2 \times 10^{-5}$.
Dashed lines indicate the stability regions for the localized phases
where the skyrmion encircles $n = 0$, 1, 4, 9, 21, 32, 58,
and $69$ obstacles.
}
\label{fig:9}
\end{figure}

Similar orbits appear at higher ac amplitudes.  The
localized orbit of Fig.~\ref{fig:8}(f) persists over the range
$1.014 \leq A \leq 1.122$,
while an orbit encircling
26 obstacles
appears for $1.172 \leq A \leq 1.214$.
At higher $A$
we find
additional localized orbits in which the skyrmion encircles
$32$, 45, or $69$ obstacles.
In general, stable localized orbits appear
close to drives where the skyrmion
can perfectly
encircle $n^2$ obstacles; 
however, due to the square symmetry of the obstacle array, the
orbits can deviate from purely circular states so that,
for example, the
stable orbit  
encircles $26$ rather than $25$ obstacles.
In Fig.~\ref{fig:9} we plot $\langle V_{||}\rangle$
and $\langle V_{\perp}\rangle$ versus $A$
for the system in Fig.~\ref{fig:8},
highlighting the locations of some of the localized phases where 
$n = 0$, 1, 4, 9, 21, 32, 58, and $69$ obstacles are encircled.
Several delocalized regions,
including translating orbits and reversals in the translation direction,
appear in between the localized regimes.

\begin{figure}
  \begin{minipage}{3.5in}
    \begin{minipage}{3.5in}
      \includegraphics[width=3.5in]{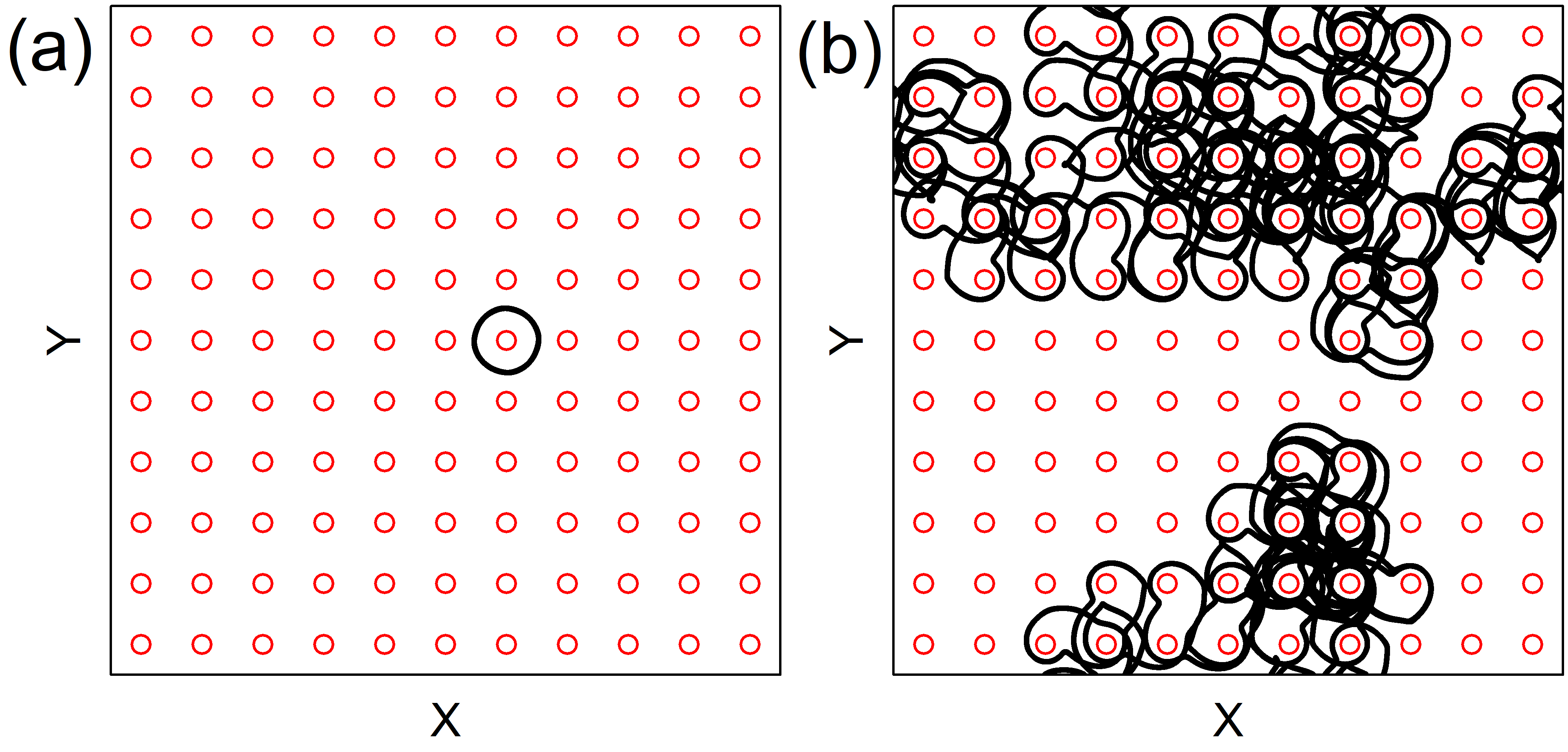}
    \end{minipage}
    \begin{minipage}{3.5in}
      \includegraphics[width=3.5in]{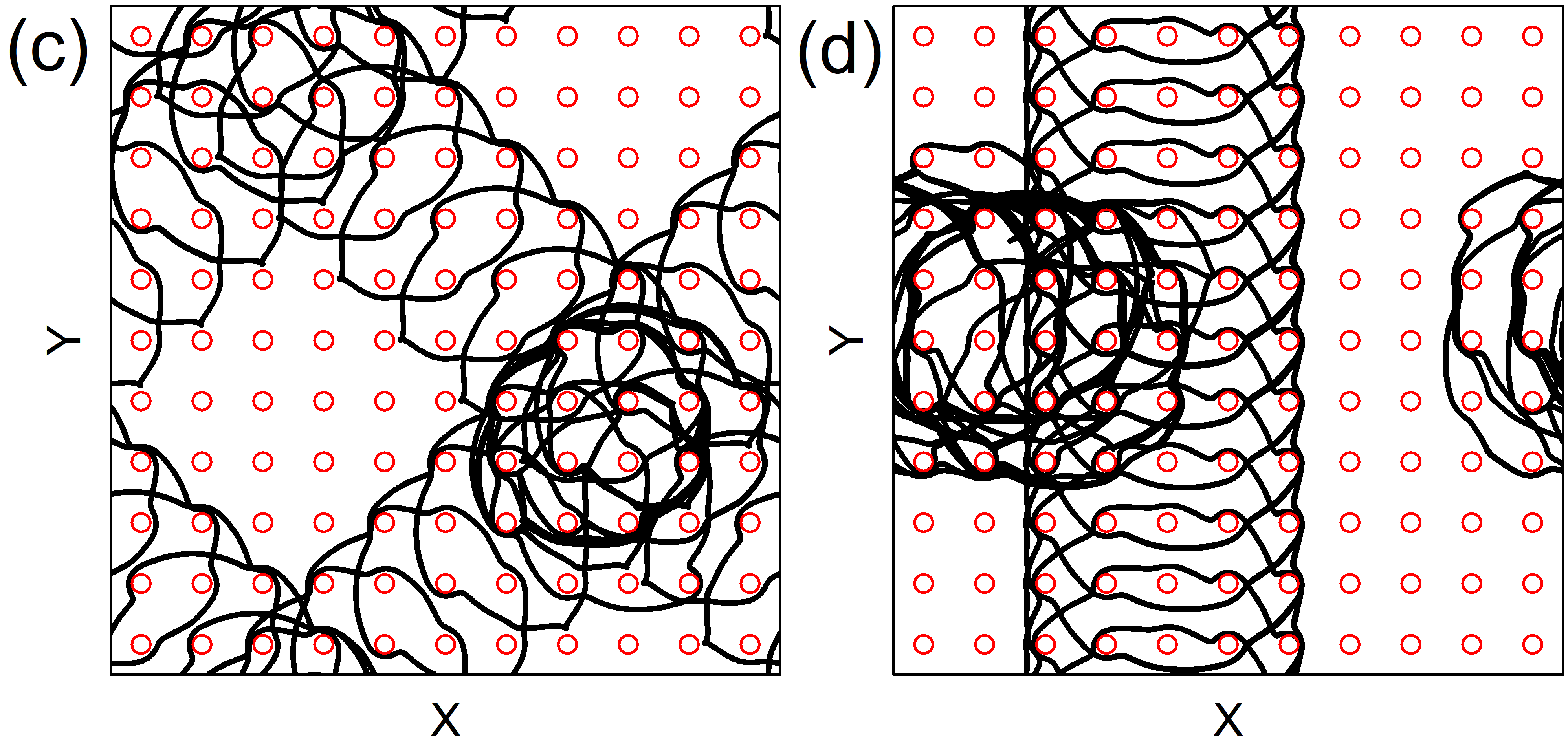}
    \end{minipage}
  \end{minipage}
\caption{Obstacles (red circles) and skyrmion trajectory (black line)
in a system with
$\alpha_{m}/\alpha_{d} = 1.732$, $a_{0} = 0.65$, and circular
driving with $B=A$ and $\omega_2=\omega_1=2\times 10^{-5}$.
(a) A localized state at $A = 0.215$.
(b) A diffusive or chaotic state at $A = 0.29$.
(c) At $A = 0.712$, there is a translating orbit
jumping between two different directions of travel at
$+45^\circ$ and $-45^\circ$.
(d) At $A = 0.836$, the skyrmion locks into an orbit that
translates in the $+y$ direction.   
}
\label{fig:10}
\end{figure}

\begin{figure}
\includegraphics[width=3.5in]{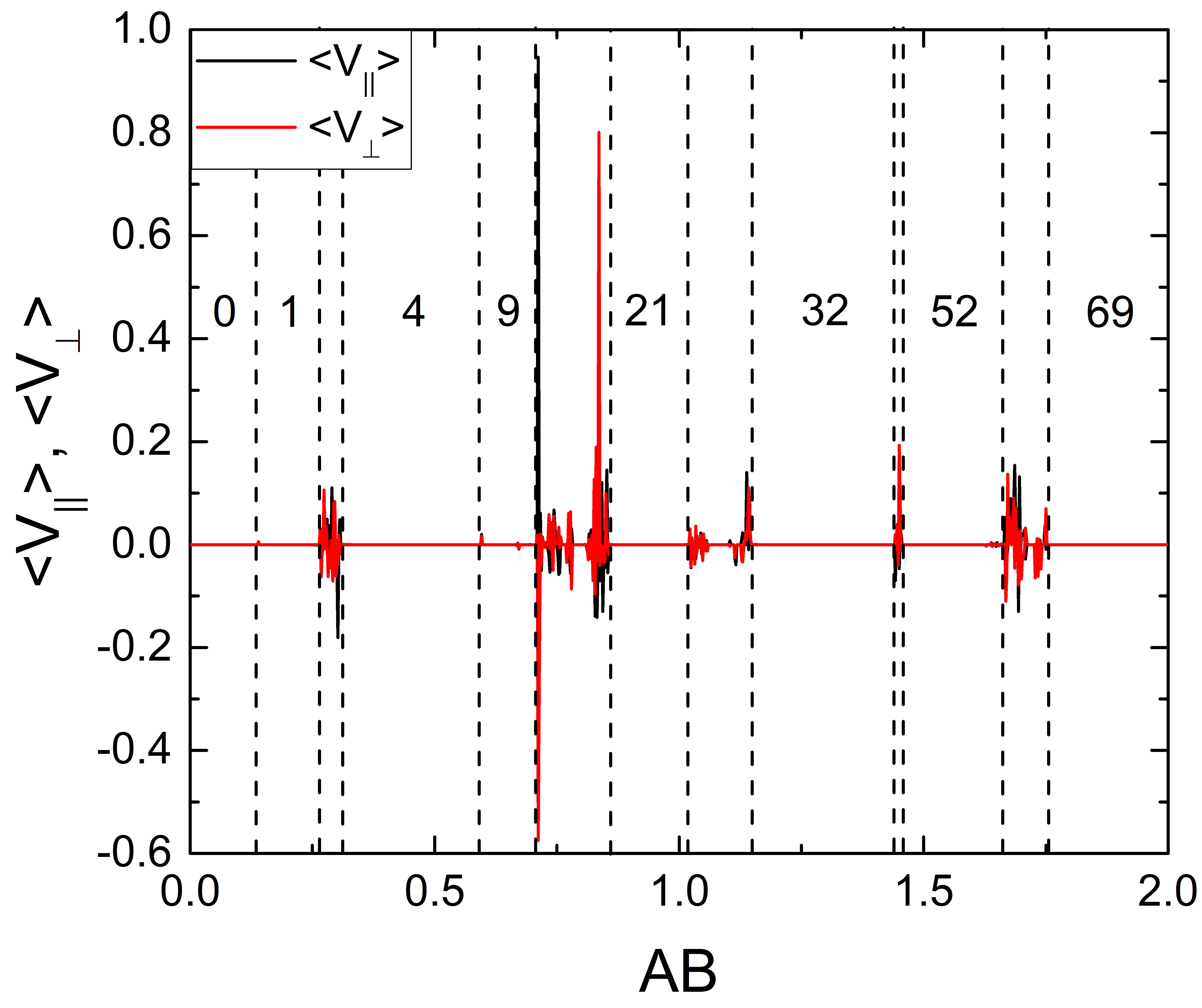}
\caption{ $\langle V_{||}\rangle$
and $\langle V_{\perp}\rangle$ vs $A$ for the
system in Fig.~\ref{fig:10} with
$\alpha_{m}/\alpha_{d} = 1.732$, $a_{0} = 0.65$, and circular
driving with $B=A$ and $\omega_2=\omega_1=2\times 10^{-5}$.
Vertical lines indicate the regions where localized phases
encircling 0, 1, 4, 9, 21, 32, 52, and 69 obstacles appear.
}
\label{fig:11}
\end{figure}

In Fig.~\ref{fig:10} we plot the
skyrmion trajectories and obstacle locations for
a system with $\alpha_{m}/\alpha_{d} = 1.732$ and circular driving with
$B=A$ and $\omega_1=\omega_2$.
For $A = 0.215$ in Fig.~\ref{fig:10}(a), the skyrmion forms a
localized orbit encircling one obstacle,
while at $A = 0.29$ in Fig.~\ref{fig:10}(b), the motion is
diffusive or chaotic.
At $A = 0.712$, Fig.~\ref{fig:10}(c) shows that there is
a translating orbit that jumps between motion along $+45^\circ$ and
$-45^\circ$.  This orbit produces
no net directed motion,
but at short times the behavior is superdiffusive.
In Fig.~\ref{fig:10}(d)
at $A = 0.836$,
the skyrmion 
locks into an orbit that translates along
the $+y$ direction.
In Fig.~\ref{fig:11} we plot
$\langle V_{||}\rangle$ and $\langle V_{\perp}\rangle$
versus $A$ for the system
in Fig.~\ref{fig:10}, with vertical lines indicating the
regions where localized orbits
encircle $n=0$, 1, 4, 9, 21, 32, 52, and $69$ obstacles.
In between the $n=9$ and $n=21$ regimes,
we find two regions of directed 
motion. Just above the $n=9$ regime,
the skyrmion translates in the positive $x$ and negative $y$ directions
along $-45^\circ$, while just below the $n=21$ regime,
the skyrmion translates in the positive $y$-direction.
There is also a
smaller region between these two translating phases
where the system forms a localized state
encircling $16$ obstacles.

\begin{figure}
  \begin{minipage}{3.5in}
    \begin{minipage}{3.5in}
      \includegraphics[width=3.5in]{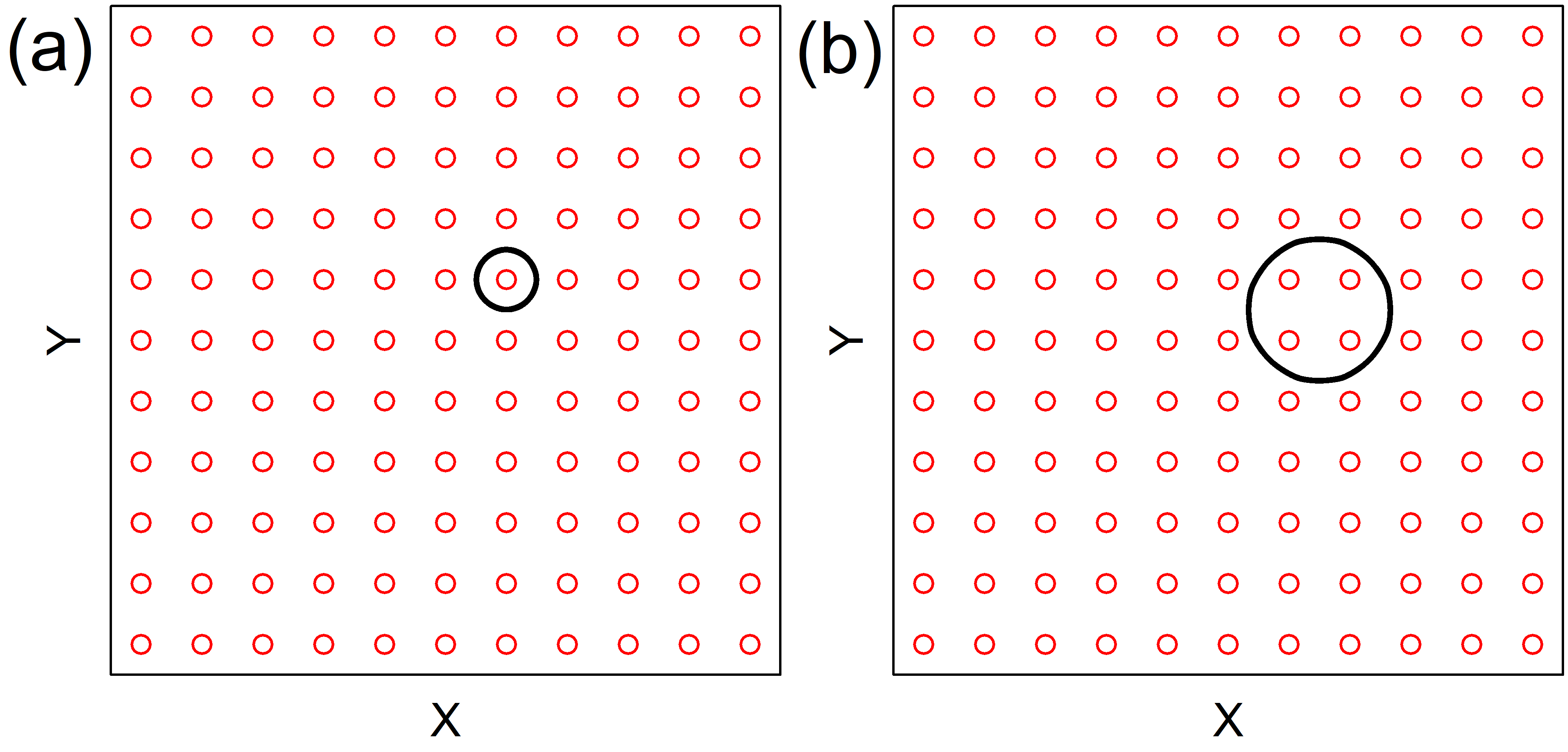}
    \end{minipage}
    \begin{minipage}{3.5in}
      \includegraphics[width=3.5in]{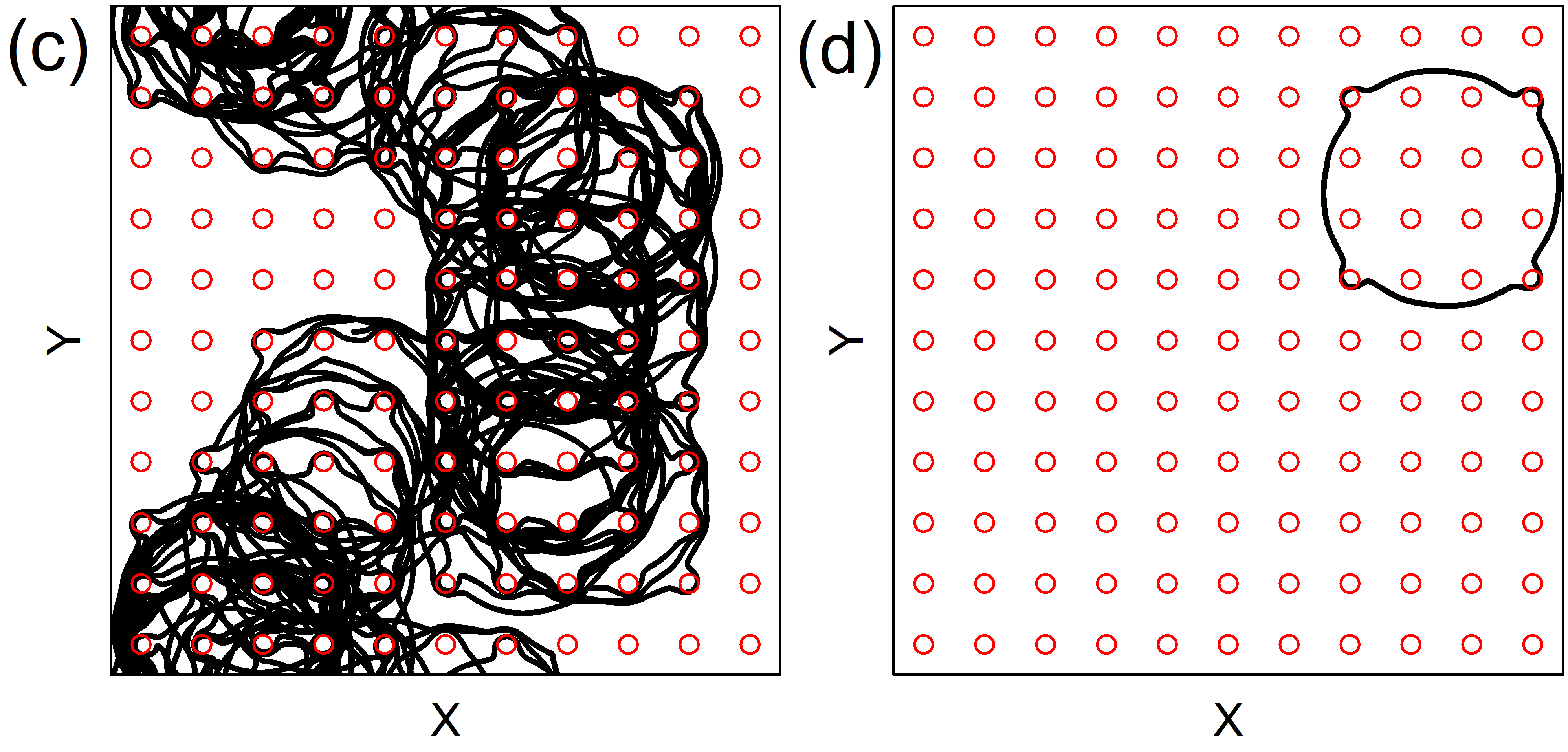}
    \end{minipage}
  \end{minipage}
\caption{Obstacles (red circles) and skyrmion trajectory (black line)
in a system with
$\alpha_{m}/\alpha_{d} = 9.962$, $a_{0} = 0.65$,
and circular driving with $B=A$ and $\omega_2=\omega_1=2 \times 10^{-5}$.
(a) A localized orbit encircling 1 obstacle at $A = 0.188$.
(b) A localized orbit encircling 4 obstacles at $A = 0.25$.
(c) A delocalized orbit at $A = 0.596$. 
(d) A localized orbit encircling 16 obstacles at
$A = 0.75$.
}
\label{fig:12}
\end{figure}

\begin{figure}
\includegraphics[width=3.5in]{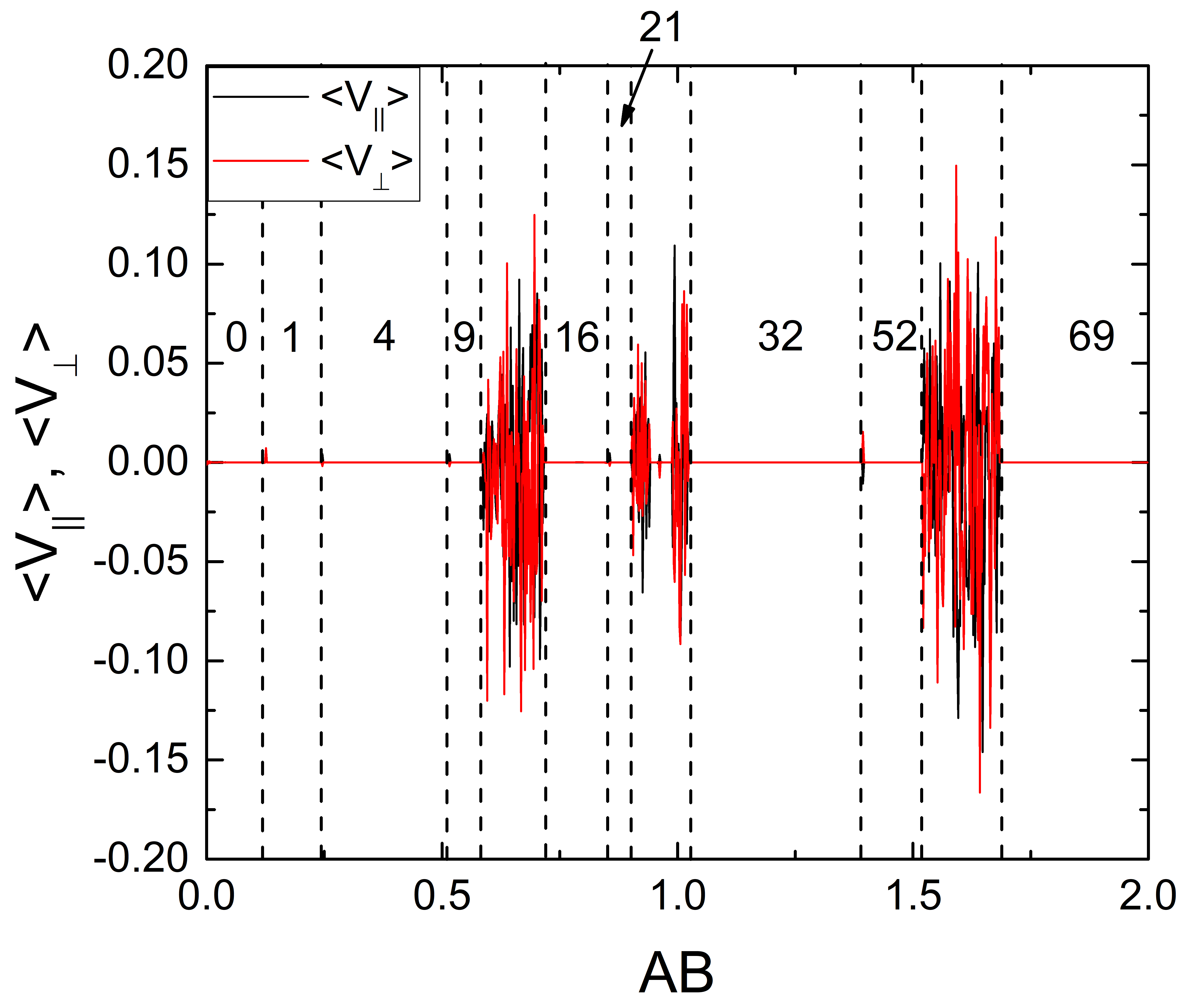}
\caption{ $\langle V_{||}\rangle$ (black) and $\langle V_{\perp}\rangle$ (red)
vs $A$ for the system in Fig.~\ref{fig:12} with
$\alpha_{m}/\alpha_{d} = 9.962$, $a_{0} = 0.65$,
and circular driving with $B=A$ and $\omega_2=\omega_1=2 \times 10^{-5}$.
The vertical lines indicate the regions where localized phases occur.
There are chaotic regions but no windows of directed flow. 
}
\label{fig:13}
\end{figure}

As we increase
$\alpha_{m}/\alpha_d$ in the circular drive system
with $A = B$, we find fewer
translating orbits and wider regions of
delocalized orbits.
In Fig.~\ref{fig:12} we illustrate skyrmion trajectories in
a system with $\alpha_{m}/\alpha_d = 9.962$.
At $A = 0.188$ in Fig.~\ref{fig:12}(a),
there is a localized orbit encircling one obstacle.
In Fig.~\ref{fig:12}(b)
at $A = 0.25$, the localized orbit encircles four obstacles.
The delocalized orbit at $A=0.596$ appears in
Fig.~\ref{fig:12}(c).
At $A = 0.75$
in Fig.~\ref{fig:12}(d),
the localized orbit
encircles 16 obstacles.
In Fig.~\ref{fig:13} we plot $\langle V_{||}\rangle$
and $\langle V_{\perp}\rangle$ versus $A$ for 
the system in Fig.~\ref{fig:12},
where the vertical lines indicate the windows of localized phases
in which the skyrmion encircles
0, 1, 4, 9, 16, 32, 52, or $69$ obstacles.
There are several chaotic regimes
but no regions of directed flow.                        

\subsection{Circular ac Drives with $A\neq B$}

\begin{figure}
  \begin{minipage}{3.5in}
    \begin{minipage}{3.5in}
      \includegraphics[width=3.5in]{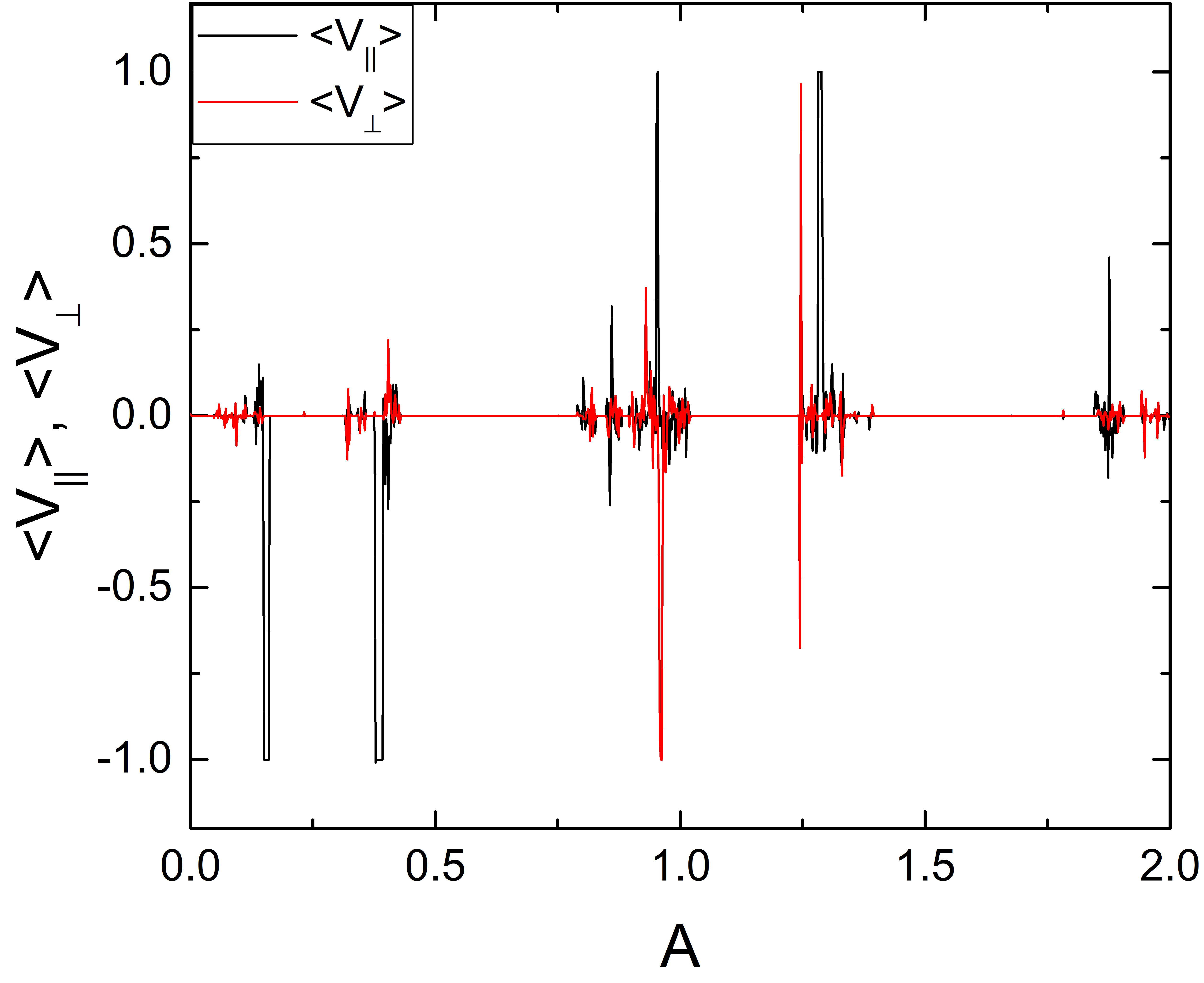}
    \end{minipage}
    \begin{minipage}{3.5in}
      \includegraphics[width=3.5in]{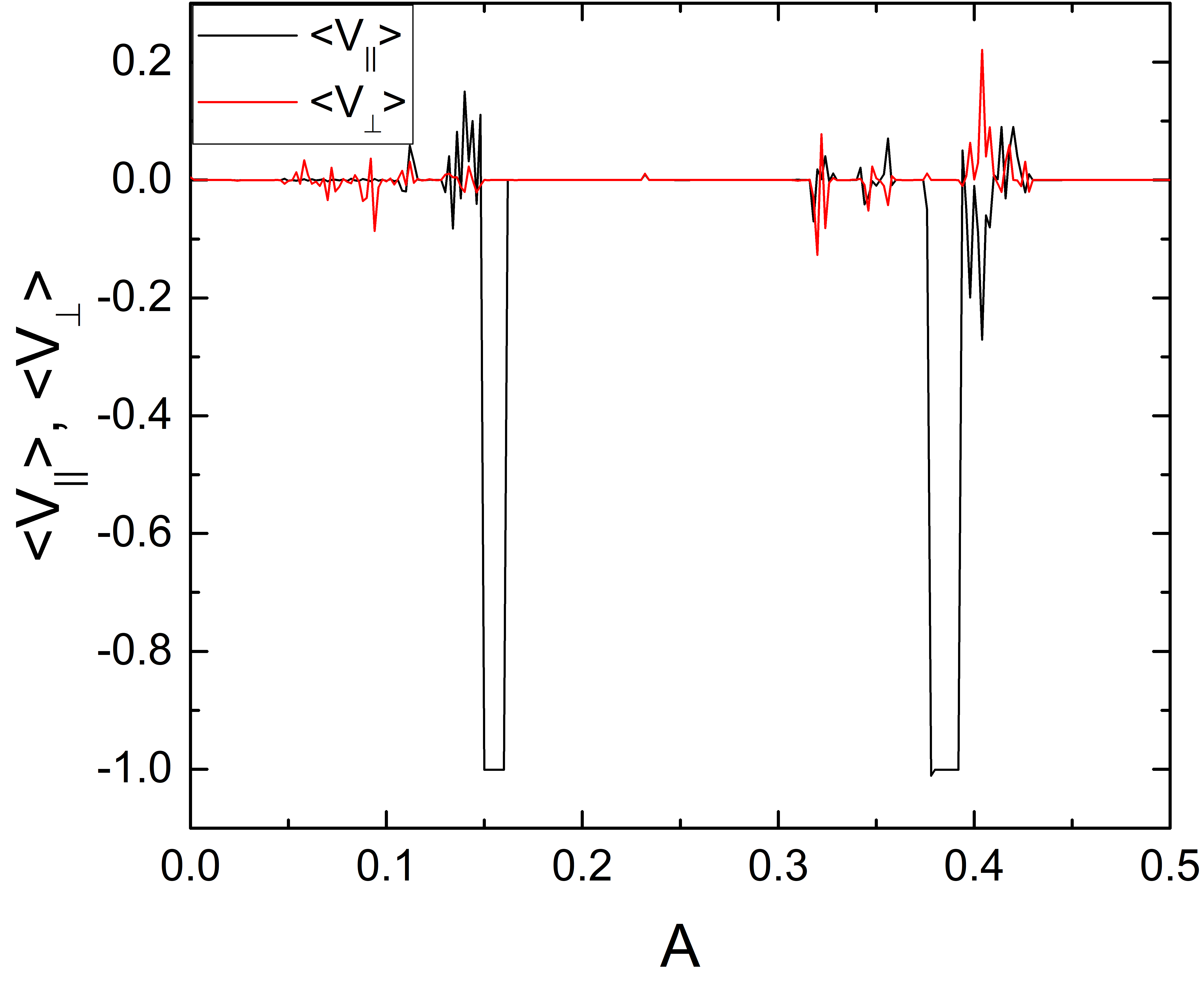}
    \end{minipage}
  \end{minipage}
\caption{(a) $\langle V_{||}\rangle$ (black)
and $\langle V_{\perp}\rangle$ vs $A$ for a system with
circular ac driving, 
$\alpha_{m}/\alpha_{d}  = 0.577$, $a_{0} = 0.65$, and 
fixed $B = 1.0$.
Several regions of translating orbits appear. (b) A blowup
of panel (a) over the range
$0 < A < 0.5$ showing two extended regions in which the skyrmion
translates in the $-x$ direction.   
}
\label{fig:14}
\end{figure}

The number of translating orbits
produced by a circular ac drive can be increased by setting $A \neq B$
so that the drive amplitude is different in the $x$ and $y$ directions.
In Fig.~\ref{fig:14} we plot $\langle V_{||}\rangle$
and $\langle V_{\perp}\rangle$ versus $A$ in a sample with
$\alpha_m/\alpha_d=0.577$, where we have fixed
$B = 1.0$. 
Here there are there six regimes of directed motion.
Two of these regimes are highlighted 
in Fig.~\ref{fig:14}(b), which is a blow up of Fig.~\ref{fig:14}(a)
over the range
$0 < A < 0.5$.
The minima in $\langle V_{||}\rangle$ correspond to
translation of the skyrmion in the negative $x$ direction by one
lattice constant every ac drive cycle.
Near
$A=0.8$, we find fractional translation
in which the skyrmion moves one lattice constant
in the negative $x$ direction every four ac drive cycles,
closely followed by a second fractional translation phase
where the skyrmion moves one lattice constant in the positive $x$ 
direction
every three ac drive cycles.
Near $A=0.95$, 
there are two additional 
phases
in which
the skyrmion moves one lattice constant in
either the negative $y$ direction
or the positive $x$ direction every
ac drive cycle.
There are several
additional translating orbits
near $A = 1.25$.

\begin{figure}
  \begin{minipage}{3.5in}
    \begin{minipage}{3.5in}
      \includegraphics[width=3.5in]{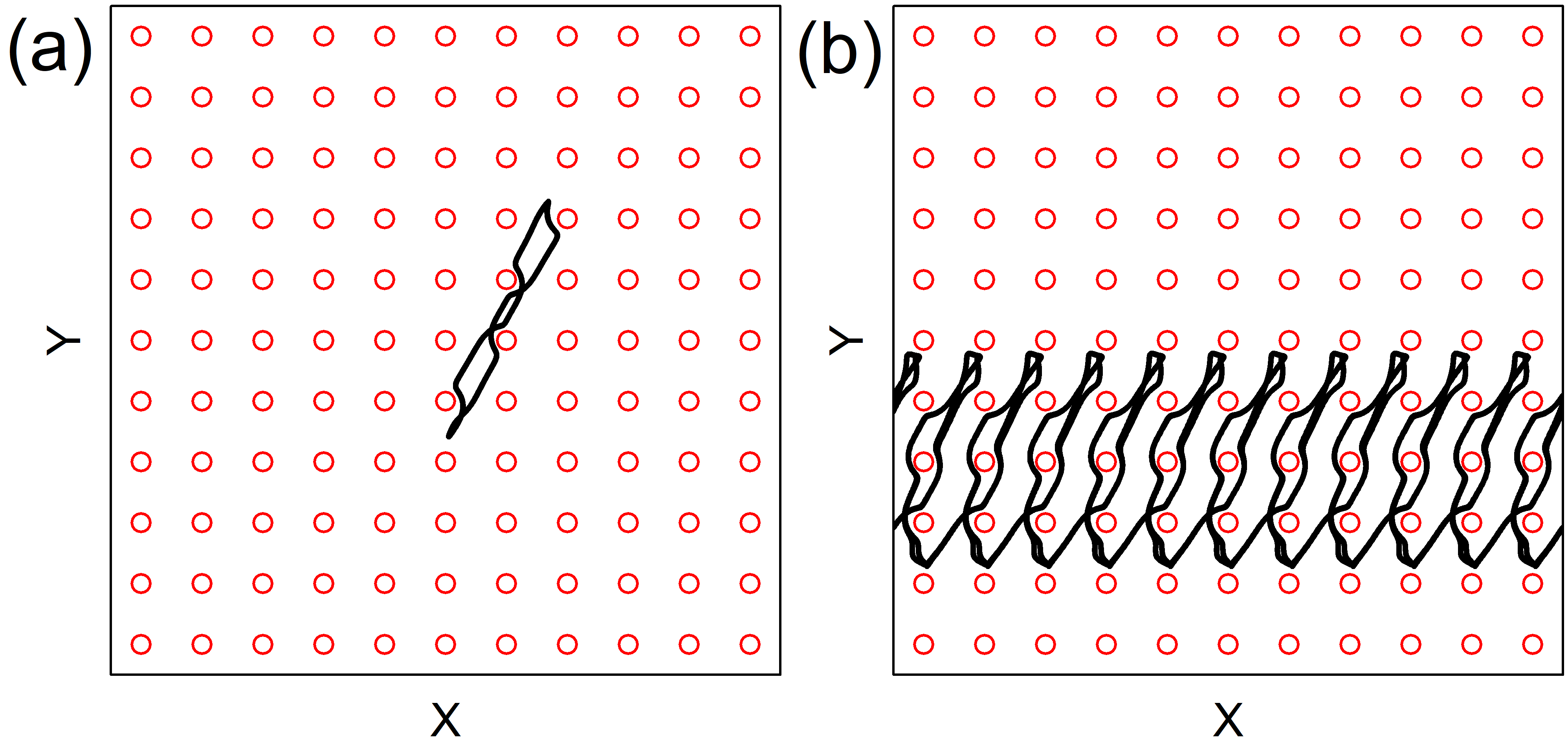}
    \end{minipage}
    \begin{minipage}{3.5in}
      \includegraphics[width=3.5in]{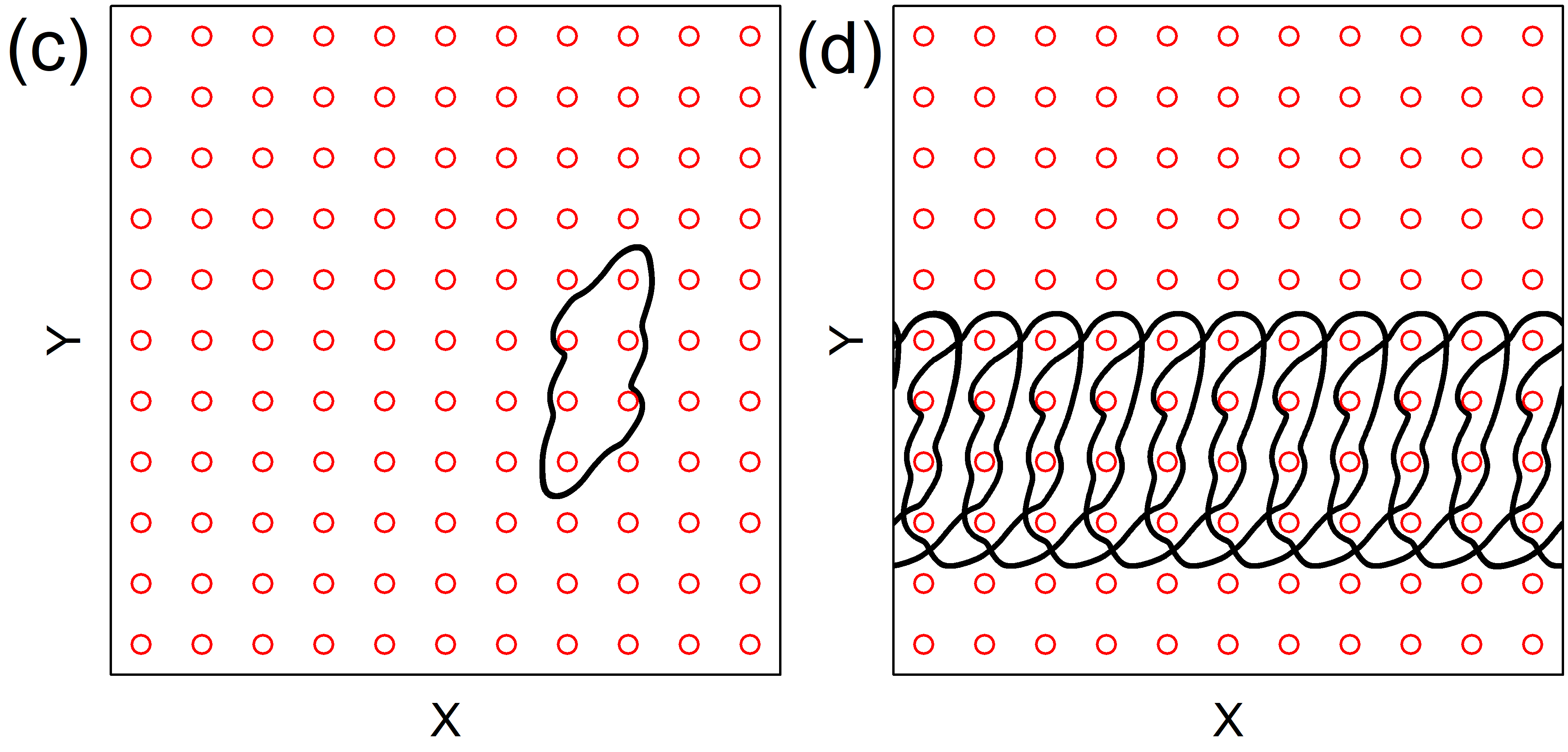}
    \end{minipage}
    \begin{minipage}{3.5in}
      \includegraphics[width=3.5in]{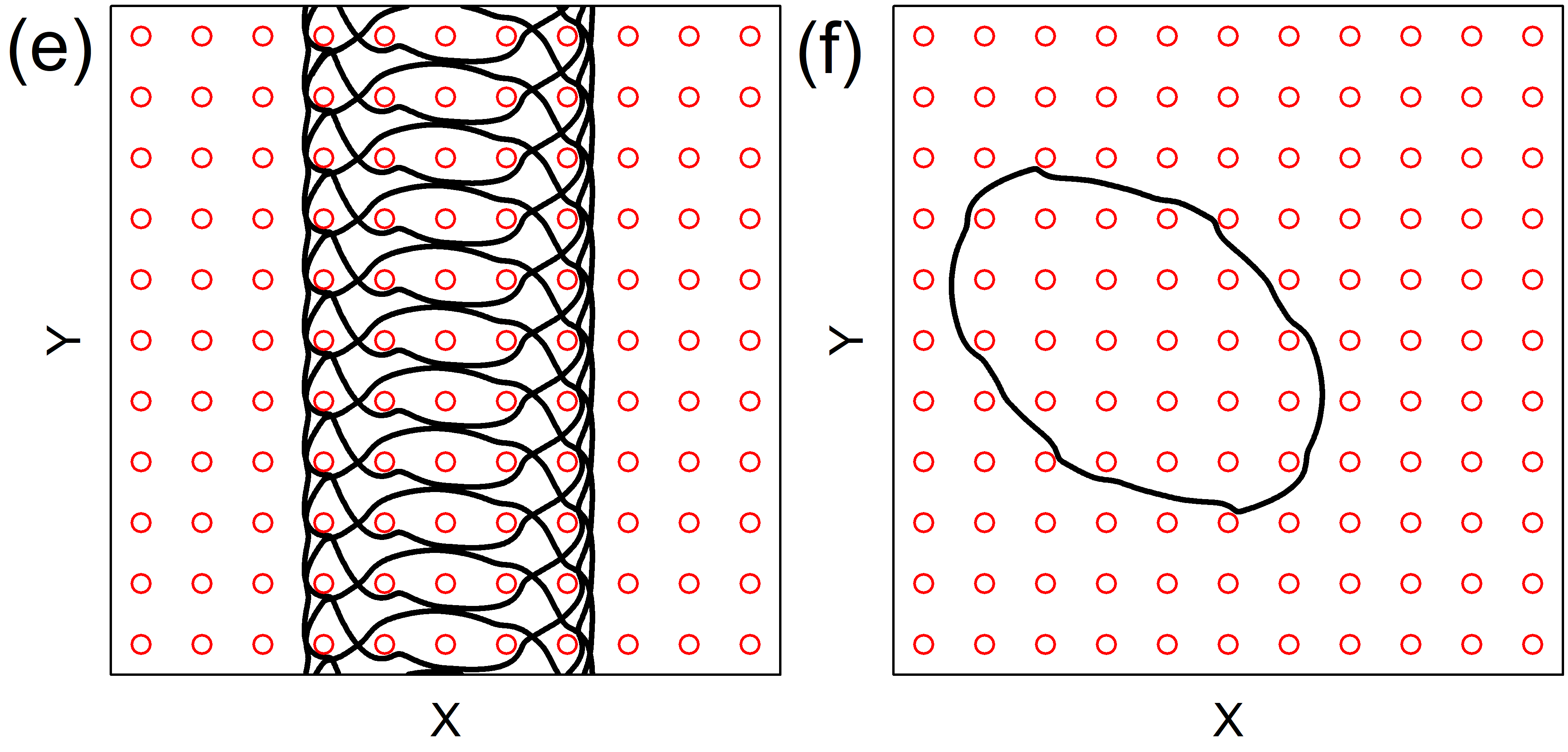}
    \end{minipage}
  \end{minipage}
\caption{Obstacles (red circles) and skyrmion trajectory (black line)
for the system in Fig.~\ref{fig:14} with 
circular ac driving, 
$\alpha_{m}/\alpha_{d}  = 0.577$, $a_{0} = 0.65$, and 
fixed $B = 1.0$.
(a) A localized phase at $A = 0.03$.
(b) At $A = 0.15$, the skyrmion translates in the negative $x$ direction.
This is the first translating phase shown in Fig.~\ref{fig:14}(b).
(c) A localized orbit at $A = 0.25$.
(d) The second phase of negative $x$ direction translation from
Fig.~\ref{fig:14}(b) at $A=0.384$.
(e) $A = 0.96$, where the skyrmion translates in the negative $y$-direction. 
(f) A localized phase encircling 23 obstacles
at $A= 1.5$.
}
\label{fig:15}
\end{figure}

In Fig.~\ref{fig:15} we illustrate the skyrmion trajectories
for the system in Fig.~\ref{fig:14}. 
Figure~\ref{fig:15}(a) shows the localized state at
$A = 0.03$, where the skyrmion
moves between three obstacles.
At $A=0.15$ in
Fig.~\ref{fig:15}(b), there is a translating orbit
in which the skyrmion
moves one lattice constant in the negative $x$ direction
during every ac drive cycle.
This corresponds to the first 
translating regime shown in Fig.~\ref{fig:14}(b). 
In Fig.~\ref{fig:15}(c) we plot the localized orbit
at $A = 0.25$, where the skyrmion orbit is asymmetric and encircles
six obstacles.
Figure~\ref{fig:15}(d) shows the second ratcheting orbit
from Fig.~\ref{fig:14}(b)
at $A = 0.384$, where the skyrmion moves in 
the negative $x$-direction.
At $A=0.96$ in
Fig.~\ref{fig:15}(e),
we find a translating orbit
where the skyrmion
moves in the negative $y$ direction.
In Fig.~\ref{fig:15}(f),
the orbit at $A  = 1.5$ is localized and encircles 23 obstacles.

\begin{figure}
\includegraphics[width=3.5in]{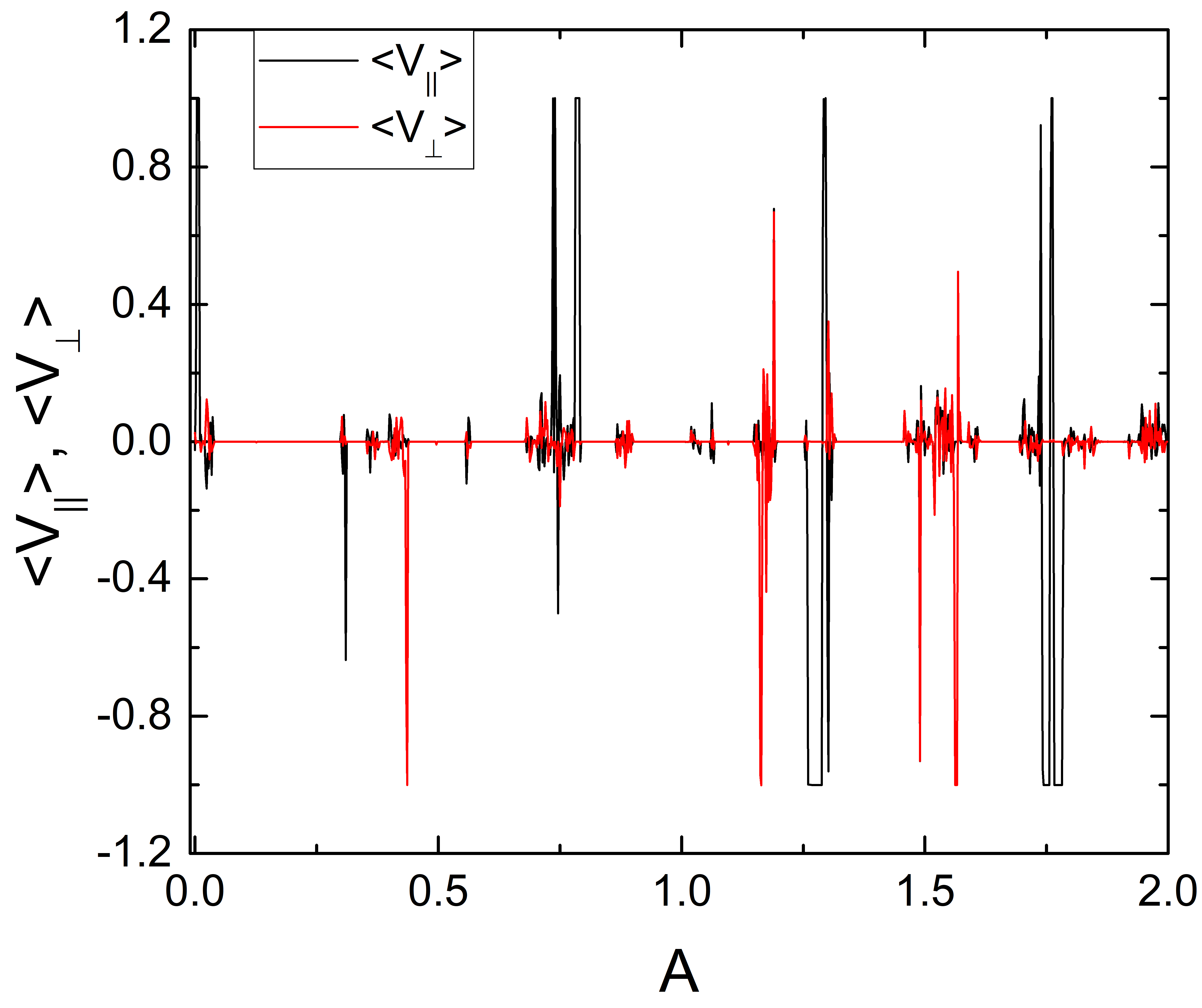}
\caption{ $\langle V_{||}\rangle$ (black) and $\langle V_{\perp}\rangle$ (red)
vs $A$ for system with
circular ac driving, 
$\alpha_{m}/\alpha_{d}  = 1.732$, $a_{0} = 0.65$, and 
fixed $B = 1.0$.
A larger number of translating phases appear.
}
\label{fig:16}
\end{figure}

When we increase $\alpha_{m}/\alpha_d$ in the
system with fixed $B = 1.0$ and varied
$A$, we find that the number of translating phases increases.
In Fig.~\ref{fig:16} we plot $\langle V_{||}\rangle$ and
$\langle V_{\perp}\rangle$ versus $A$ for 
a system with $B = 1.0$, $\alpha_{m}/\alpha_{d} = 1.732$,
and $a_{0} = 0.65$, highlighting the different translating phases
in which the skyrmion moves in either the $x$ or $y$ direction by one lattice
constant per ac drive cycle. 

\section{Dependence on Obstacle Size} 

\subsection{Linear ac Drive}

\begin{figure}
\includegraphics[width=3.5in]{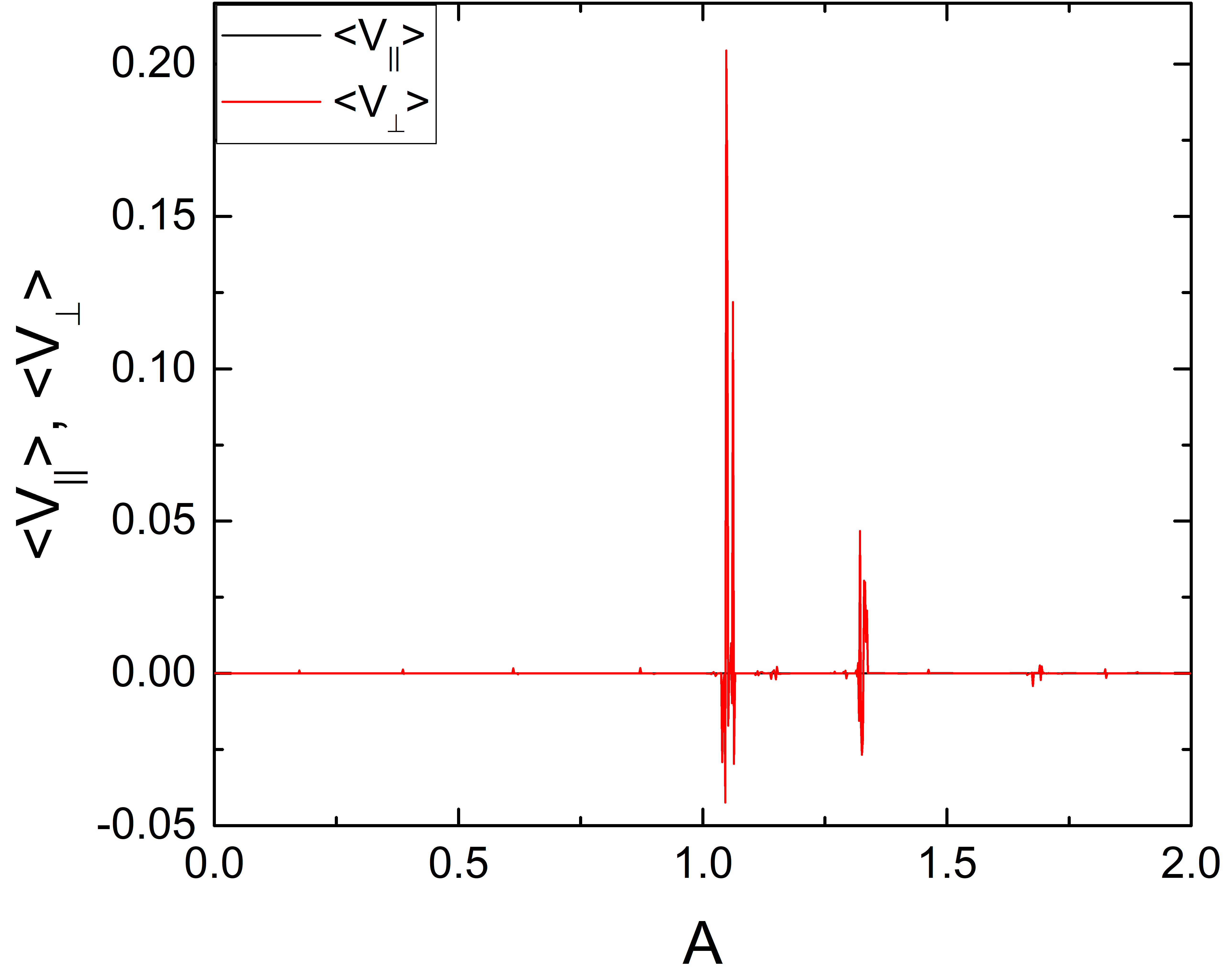}
\caption{ $\langle V_{||}\rangle$ (black) and $\langle V_{\perp}\rangle$
vs $A$ for a sample
with $a_0=0.85$,  
$\alpha_{m}/\alpha_{d} = 0.577$, and
linear ac driving with $B=0$.
The skyrmion orbit is localized over most of the range of $A$.
}
\label{fig:17}
\end{figure}

\begin{figure}
  \begin{minipage}{3.5in}
    \begin{minipage}{3.5in}
      \includegraphics[width=3.5in]{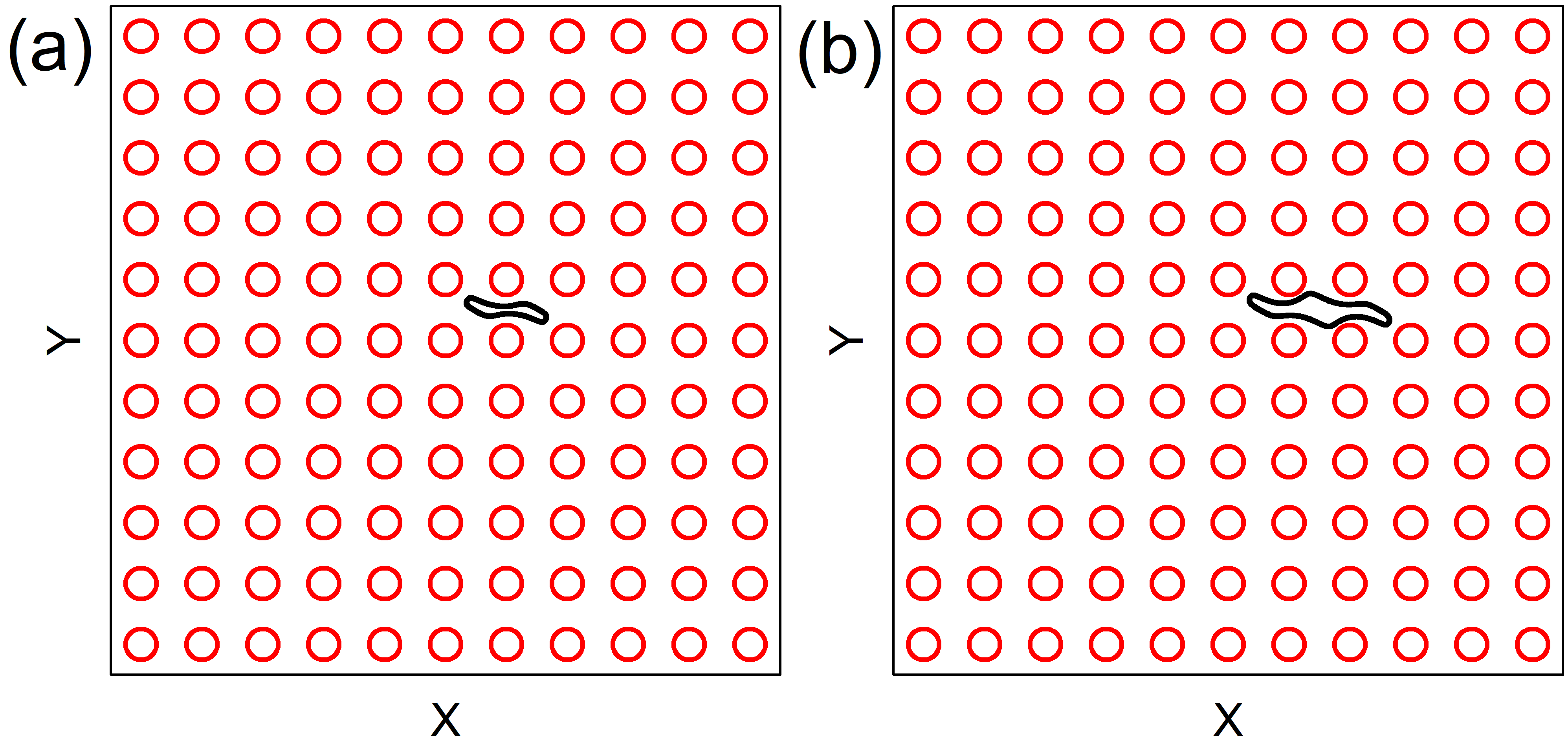}
    \end{minipage}
    \begin{minipage}{3.5in}
      \includegraphics[width=3.5in]{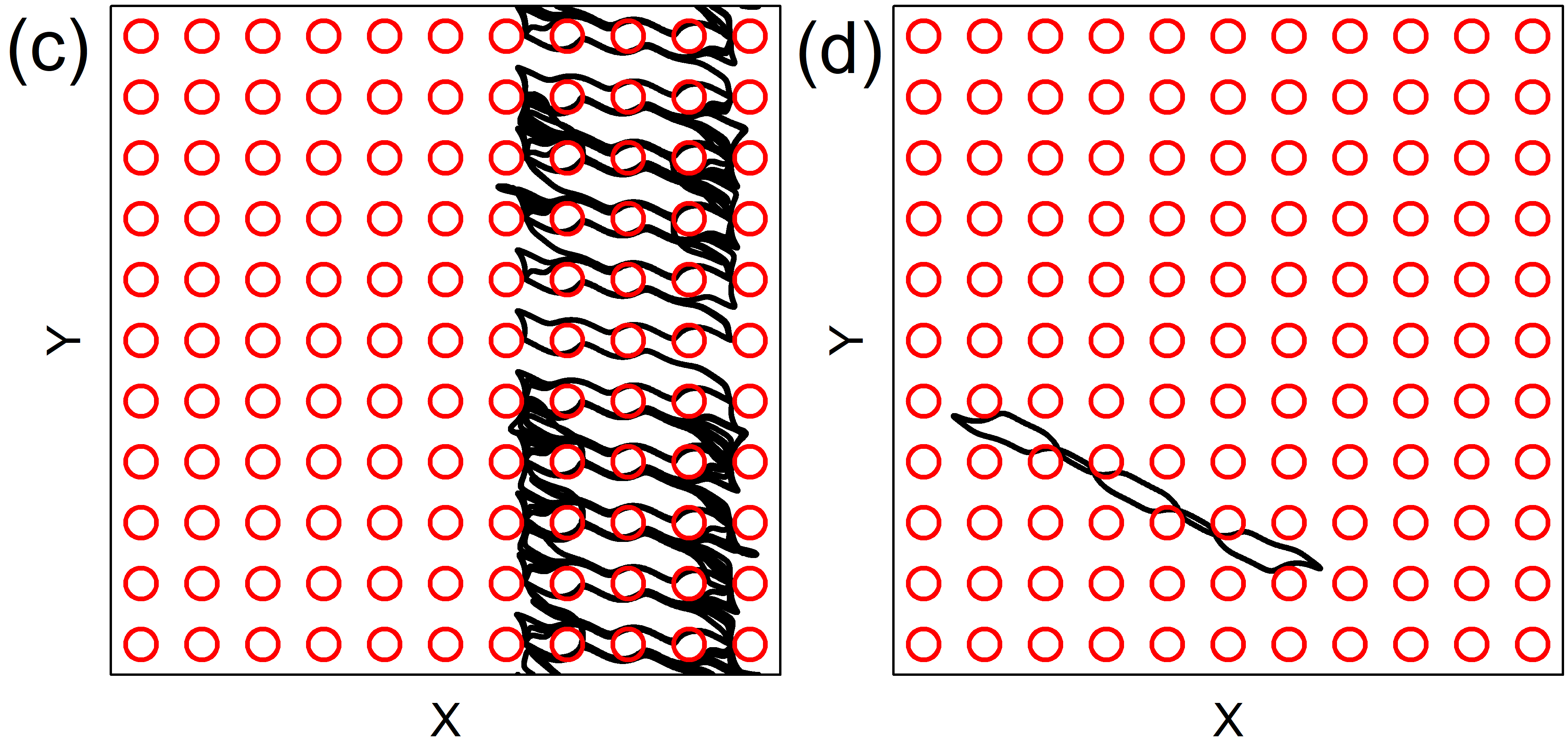}
    \end{minipage}
  \end{minipage}
\caption{Obstacles (red circles) and skyrmion trajectory (black line)
for the system in Fig.~\ref{fig:17} 
with $a_0=0.85$,  
$\alpha_{m}/\alpha_{d} = 0.577$, and
linear ac driving with $B=0$.
(a) A localized orbit at $A = 0.272$.
(b) A more extended orbit at $A = 0.5$
is delocalized.
(c) At $A = 1.047$, the skyrmion translates one lattice
constant in the $y$-direction 
every five ac drive cycles.    
(d) A localized orbit at $A = 1.5$.
}
\label{fig:18}
\end{figure}

We next consider the effects of varying the obstacle size.
We first apply a linear ac drive with $B=0$.
In general, we find that as the obstacle size increases,
the range of drives over which localized states appear increases.
In Fig.~\ref{fig:17} we plot $\langle V_{||}\rangle$
and $\langle V_{\perp}\rangle$ versus $A$ for
a system with $\alpha_{m}/\alpha_{d} = 0.577$ and $a_{0} = 0.85$
under linear ac driving.
Over the range $0<A<1.0$, all of the skyrmion orbits are
localized.  At higher $A$, there are two smaller regions of
delocalized orbits 
and one directed motion phase.
In Fig.~\ref{fig:18}(a) we illustrate the skyrmion trajectory for the 
system in Fig.~\ref{fig:17} at $A = 0.282$,
where the skyrmion forms a localized orbit moving 
between a pair of obstacles.
As $A$ increases,
the system passes
though a series of
localized states in which
the skyrmion moves in an elliptical orbit aligned in the
$x$-direction, as shown in Fig.~\ref{fig:18}(b) for $A = 0.5$.
In Fig.~\ref{fig:18}(c) at $A=1.048$,
there is a translating orbit
where the skyrmion moves one lattice constant in the positive
$y$-direction
every five ac drive cycles.
Figure~\ref{fig:18}(d) shows another localized orbit at $A = 1.5$
where the skyrmion moves between three plaquettes
at an angle to the $x$ axis. 
As $A$ increases further,
the localized orbits have a structure
similar to the orbit shown in Fig.~\ref{fig:18}(d)
but pass between an increasing number of obstacles.
Since the localized orbits
do not encircle any obstacles,
we do not highlight the different phases in Fig.~\ref{fig:17}.

\begin{figure}
  \begin{minipage}{3.5in}
    \begin{minipage}{3.5in}
      \includegraphics[width=3.5in]{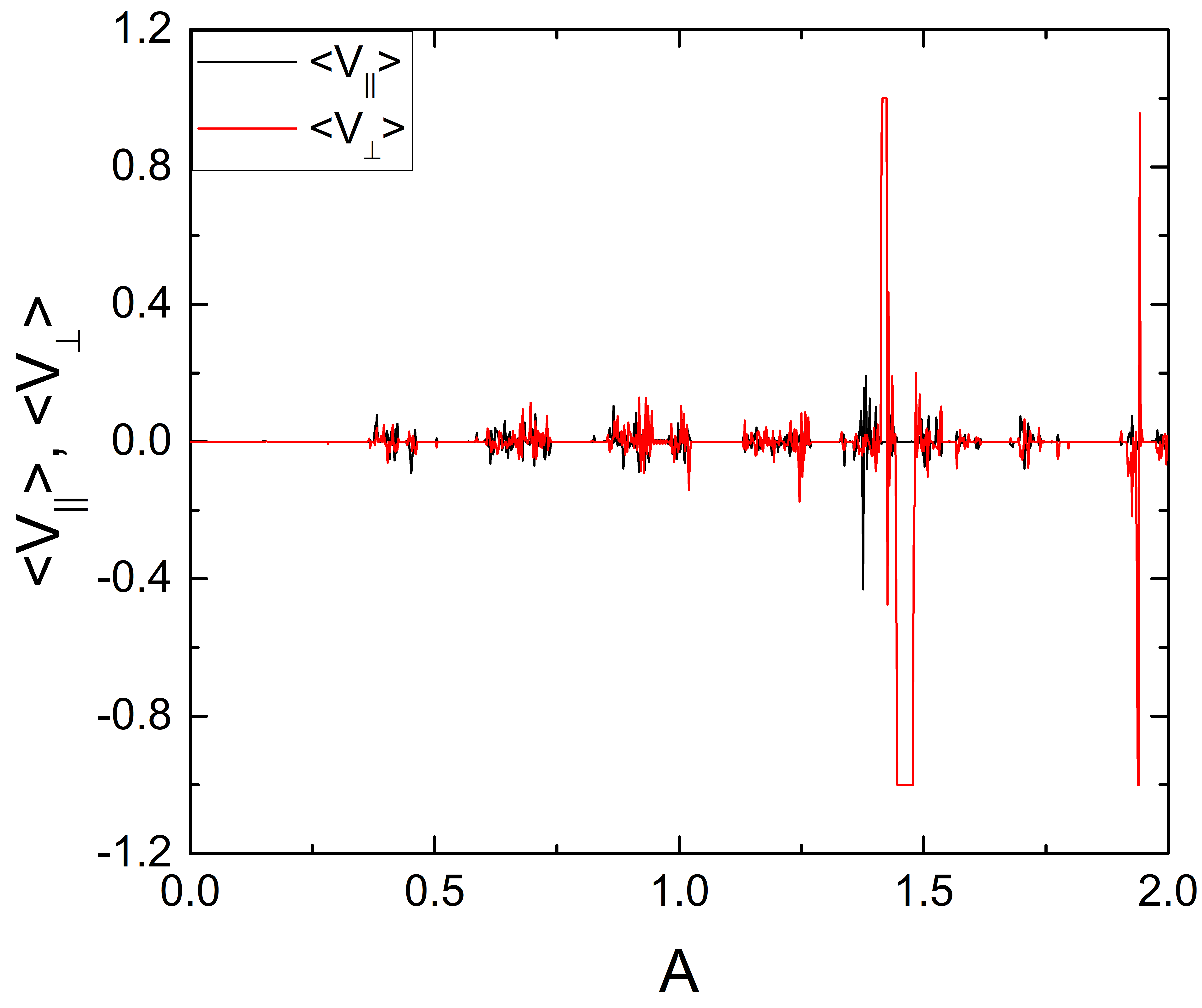}
    \end{minipage}
    \begin{minipage}{3.5in}
      \includegraphics[width=3.5in]{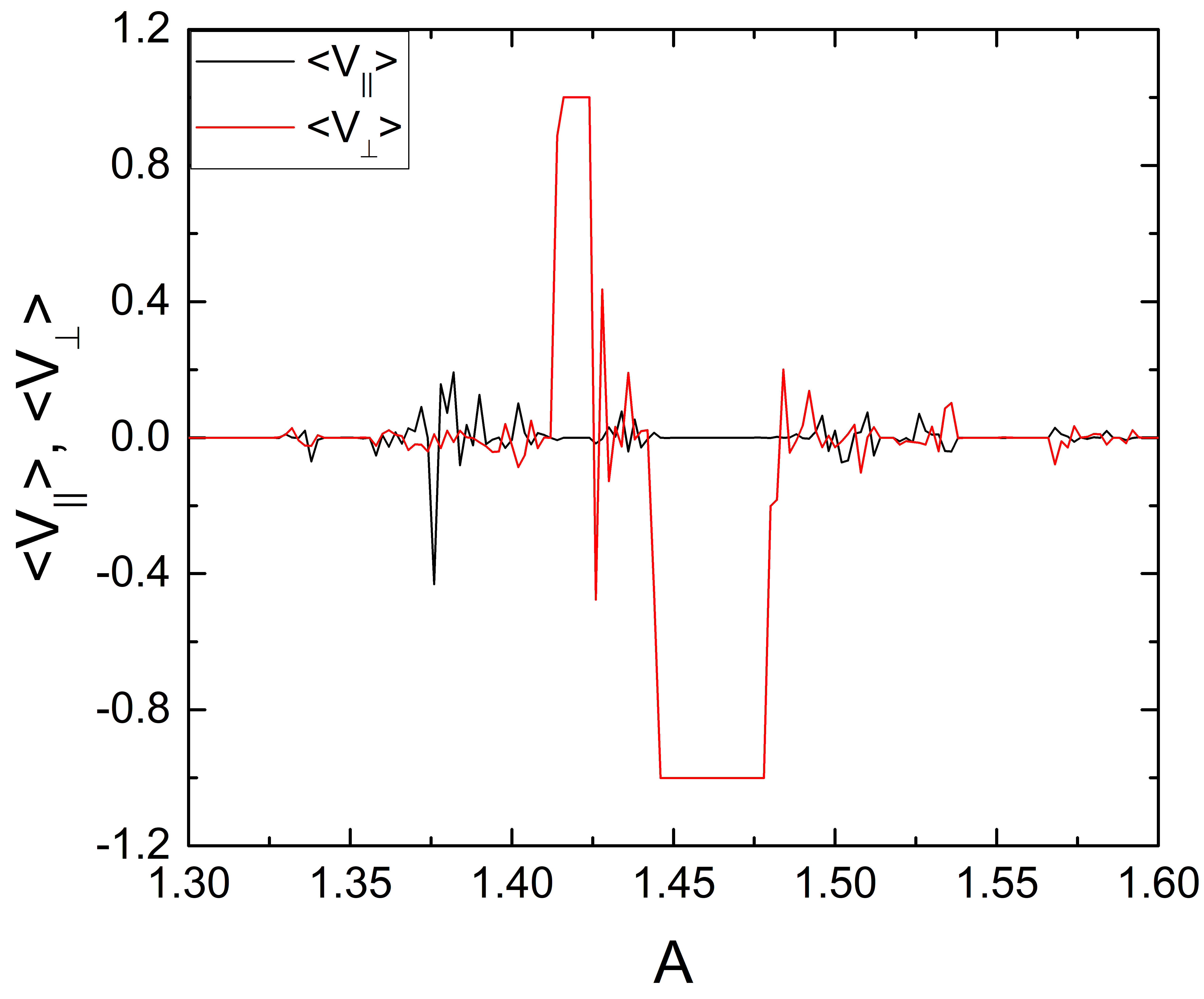}
    \end{minipage}
  \end{minipage}
\caption{(a) $\langle V_{||}\rangle$ (black) and $\langle V_{\perp}\rangle$
  (red) vs $A$ for
a sample with $\alpha_{m}/\alpha_{d} = 1.732$, 
$a_0=0.85$,  
and linear ac driving with $B=0$.
(b) A blowup of panel (a) over the range $1.3 < A < 1.6$.  
}
\label{fig:19}
\end{figure}

\begin{figure}
  \begin{minipage}{3.5in}
    \begin{minipage}{3.5in}
      \includegraphics[width=3.5in]{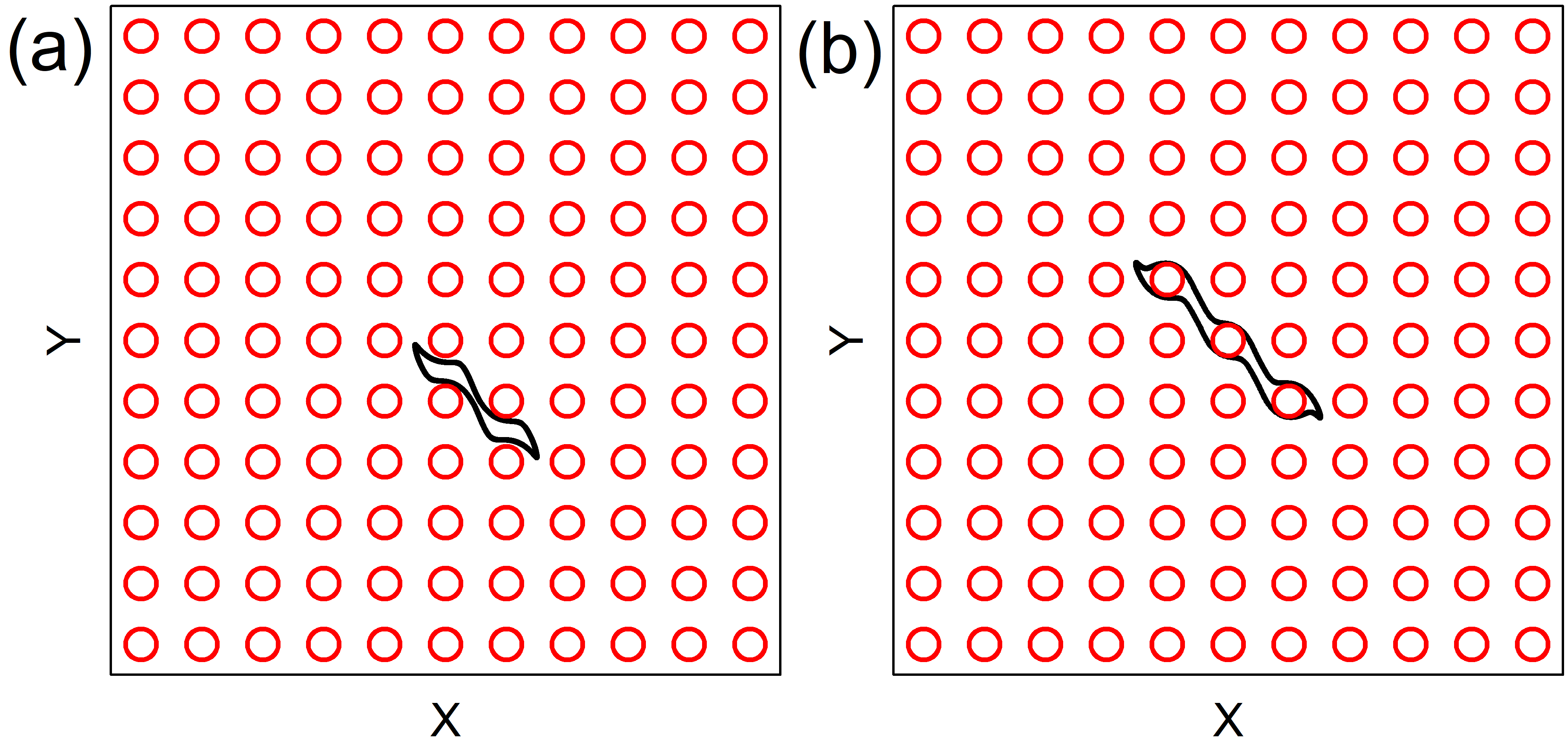}
    \end{minipage}
    \begin{minipage}{3.5in}
      \includegraphics[width=3.5in]{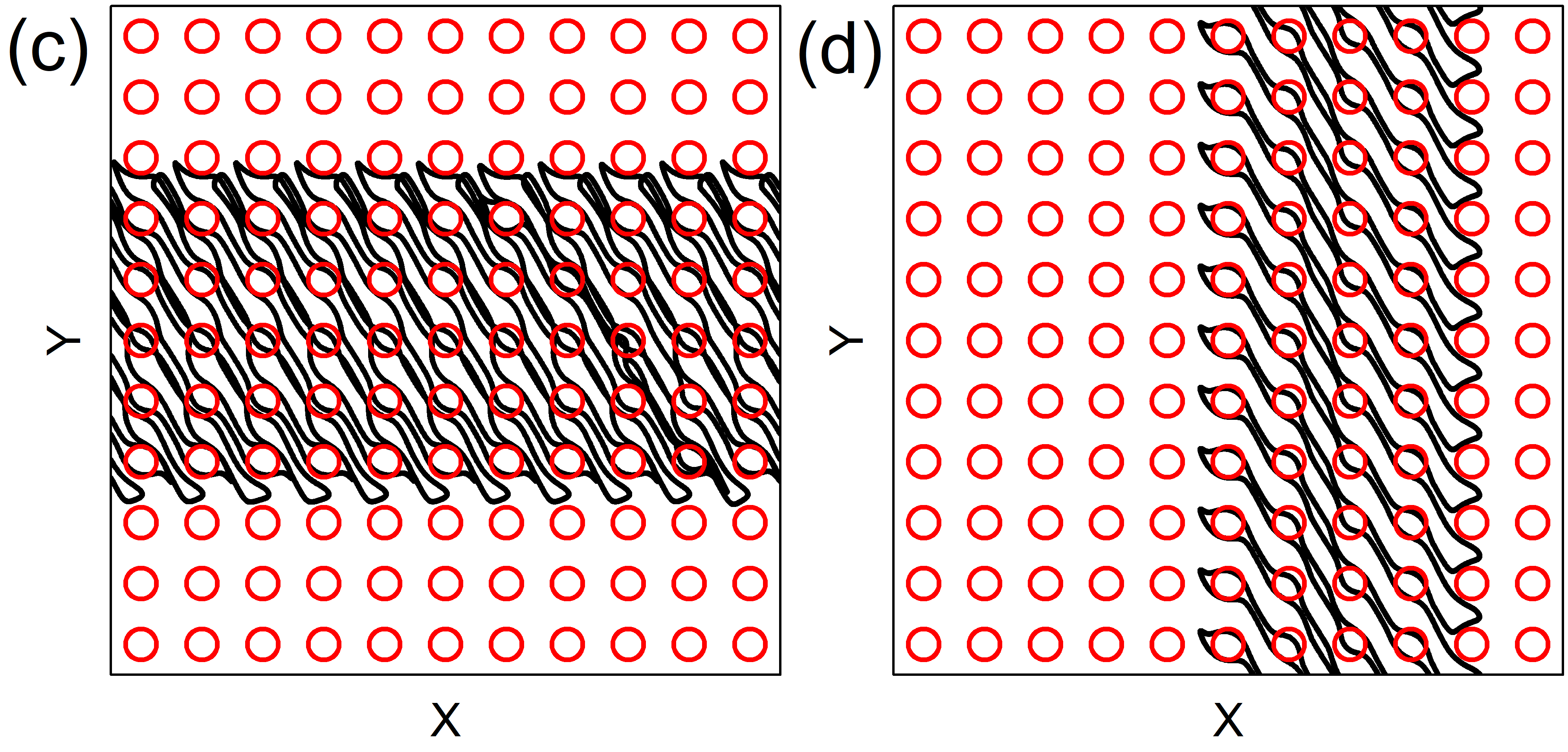}
    \end{minipage}
  \end{minipage}
\caption{Obstacles (red circles) and skyrmion trajectory (black line)
for the system
in Fig.~\ref{fig:19} 
with $\alpha_{m}/\alpha_{d} = 1.732$, 
$a_0=0.85$,  
and linear ac driving with $B=0$.
(a) A localized orbit at $A = 0.5$.
(b) A localized orbit at $A = 0.75$ where the skyrmion
encircles two obstacles per ac drive cycle.  
(c) A translating orbit at $A = 1.376$.
(d) A translating orbit at $A = 1.42$.
}
\label{fig:20}
\end{figure}

As we increase the Magnus term $\alpha_m$ in samples with larger
obstacles, 
we
find an increase in the number of regions of delocalized and
translating orbits.
In Fig.~\ref{fig:19}(a) we plot $\langle V_{||}\rangle$ and
$\langle V_{\perp}\rangle$ versus $A$ for
a sample with $a_0=0.85$, 
$\alpha_{m}/\alpha_{d} = 1.732$,
and linear ac driving.
There are several regions of chaotic flow,
several regions of translating orbits that are mostly
along the positive $y$ direction, and several reversals of the translation
direction.
In Fig.~\ref{fig:20}(a) we illustrate the localized orbit at $A = 0.5$
where the skyrmion moves between two obstacles, while
in Fig.~\ref{fig:20}(b) we show another localized orbit at $A = 0.75$
where the skyrmion encircles two obstacles
per ac drive cycle.  
Figure~\ref{fig:19}(b) displays a blowup of Fig.~\ref{fig:19}(a)
over the range $1.3 < A < 1.6$.
We find a small
region near $A = 1.376$ where the skyrmion
translates in the negative $x$ direction by one 
lattice constant every two ac drive cycles, as shown in
Fig.~\ref{fig:20}(c).
This is followed by a region of delocalized orbits. 
For $1.415 < A < 1.42$, there is a translating orbit
where the skyrmion moves one lattice constant
in the positive $y$ direction 
every ac drive cycle, as shown in Fig.~\ref{fig:20}(d).
This is followed by
a reversal to a state where the skyrmion translates in the negative
$y$ direction
for $1.425 < A < 1.48$.      

\subsection{Circular ac Drive with $A=B$}

\begin{figure}
\includegraphics[width=3.5in]{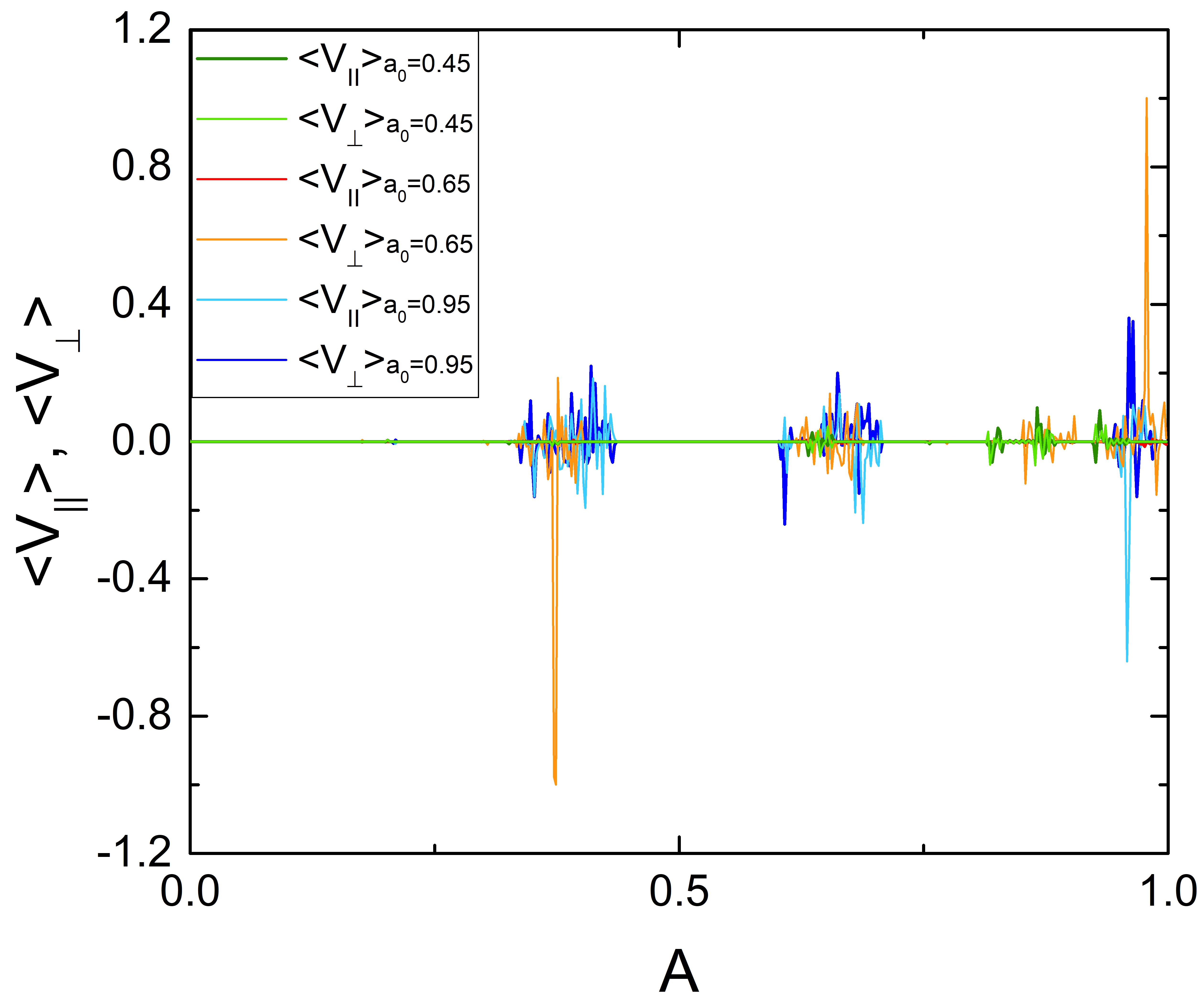}
\caption{ $\langle V_{||}\rangle$ and $\langle V_{\perp}\rangle$
vs $A$ for a system with circular ac driving, $A=B$,
$\omega_1=\omega_2$,
and $\alpha_{m}/\alpha_{d} = 0.577$ for 
$a_{0} = 0.45$ (dark and light green), $0.65$ (red and orange),
and $0.95$ (light and dark blue).  
}
\label{fig:21}
\end{figure}

\begin{figure}
\includegraphics[width=3.5in]{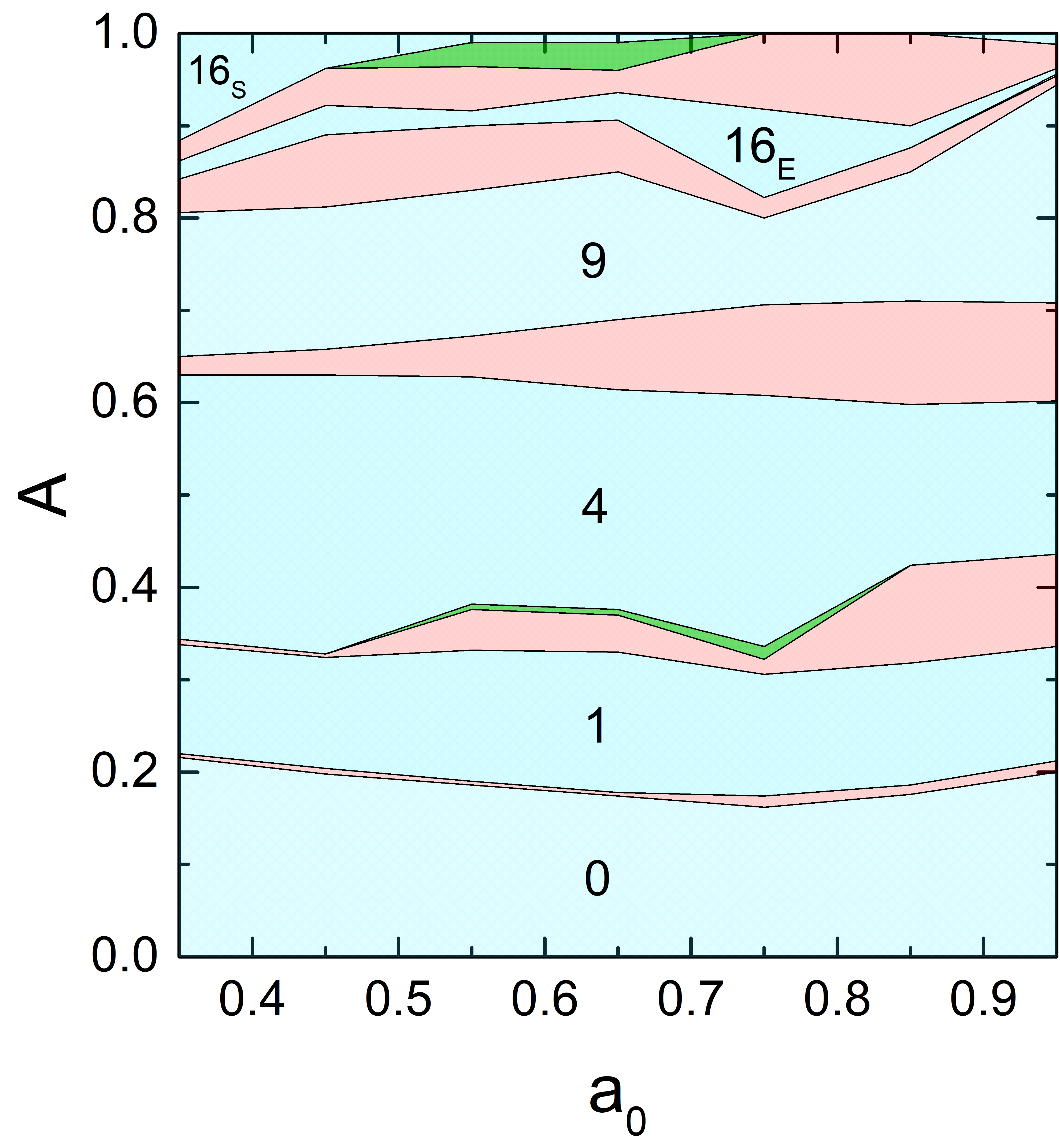}
\caption{
Dynamic phase diagram as a function of
$A$ vs $a_{0}$ for the system in Fig.~\ref{fig:21}
with $\alpha_m/\alpha_d=0.577$ and circular ac driving with
$A=B$ and  $\omega_1=\omega_2$.
Blue indicates localized phases, pink regions are delocalized states,
and green regions correspond to ordered translating orbits.
The localized phases are labeled according to whether the orbit
encircles 0, 1, 4, or 9 obstacles.  Phase 16$_E$ is a localized phase with an
elliptical orbit encircling 16 obstacles, while phase 16$_S$ is a localized
phase with a circular orbit encircling 16 obstacles.
}
\label{fig:22}
\end{figure}

\begin{figure}
  \begin{minipage}{3.5in}
    \begin{minipage}{3.5in}
      \includegraphics[width=3.5in]{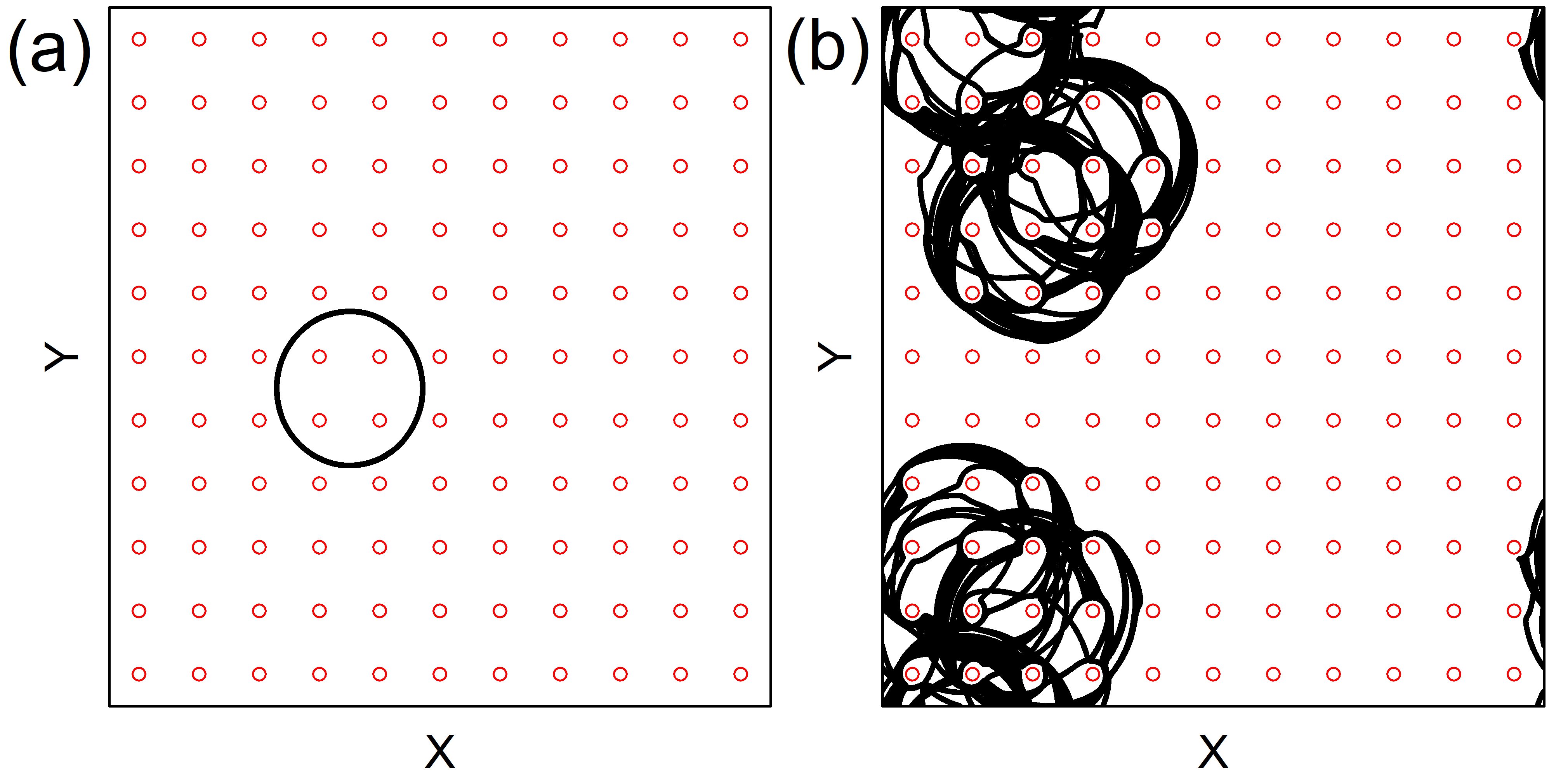}
    \end{minipage}
    \begin{minipage}{3.5in}
      \includegraphics[width=3.5in]{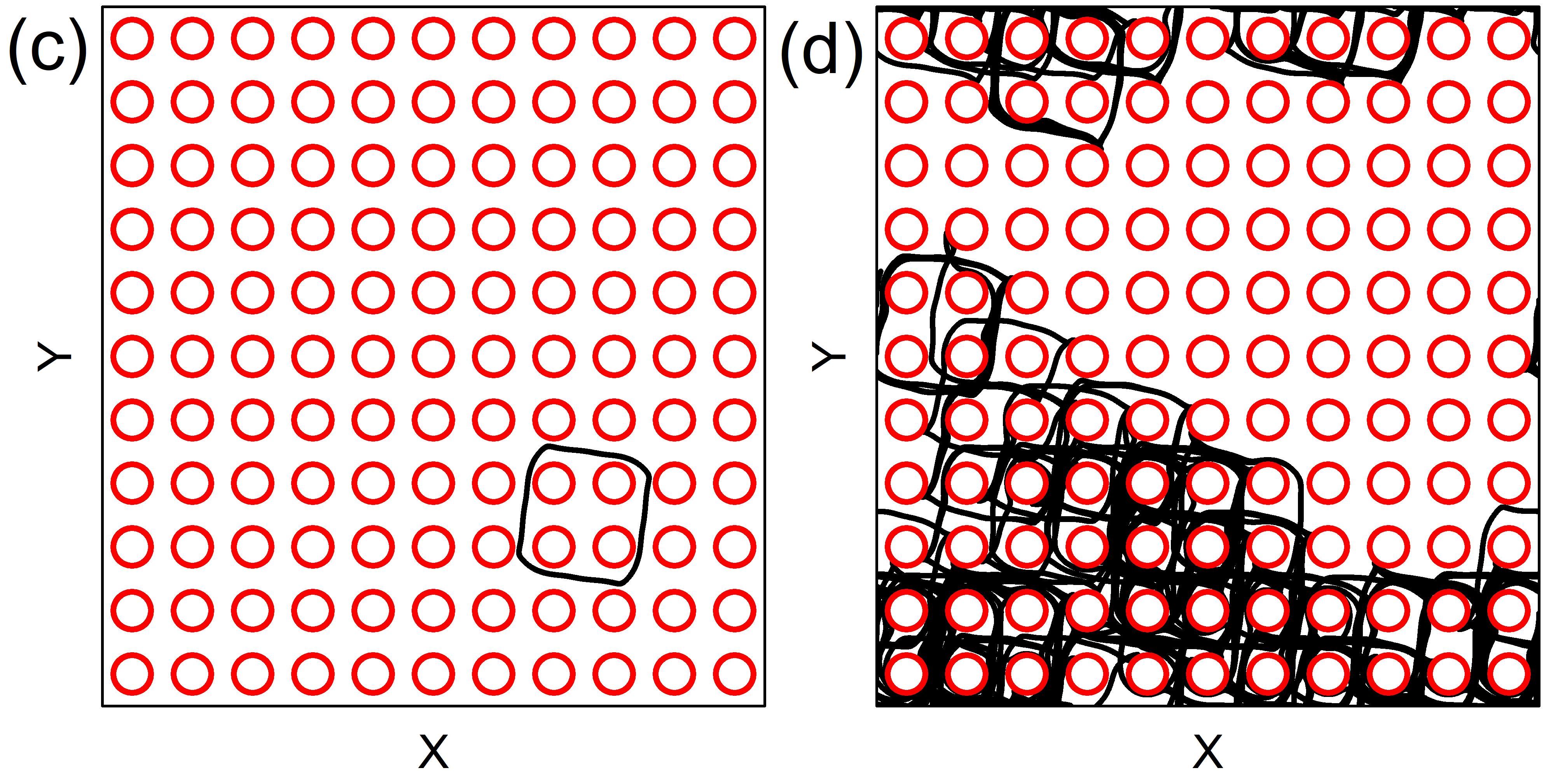}
    \end{minipage}
  \end{minipage}
\caption{Obstacles (red circles) and skyrmion trajectory (black line)
for the system in Fig.~\ref{fig:22} 
with $\alpha_m/\alpha_d=0.577$ and circular ac driving with
$A=B$ and  $\omega_1=\omega_2$.
(a) $A = 0.5$ and $a_{0} = 0.45$, where the skyrmion is in a localized orbit.
(b) $A = 0.652$ and $a_{0} = 0.45$ where the skyrmion motion
is delocalized.
(c) $A = 0.5$ and $a_{0} = 0.95$, showing a square orbit
encircling four obstacles. 
(d) $A = 0.608$ and $a = 0.95$, showing a localized phase.
}
\label{fig:23}
\end{figure}

We next consider the obstacle 
size dependence in samples with circular ac drives for $A=B$.
In Fig.~\ref{fig:21} we plot $\langle V_{||}\rangle$ and
$\langle V_{\perp}\rangle$ versus $A$
for a system with $\alpha_{m}/\alpha_{d} = 0.577$ at
$a_{0} = 0.45$, $0.65$ and $0.95$.
For small $a_{0}$, most of the orbits are localized.
By conducting a series of simulations, we
construct a dynamic phase diagram as a function of $A$ versus $a_0$
as shown in Fig.~\ref{fig:22}, where the locations of the
localized, delocalized or chaotic, and translating phases are indicated.
The localized orbits are labeled according to whether the orbit
encircles 0, 1, 4, 9, or 16 obstacles.
The localized orbits can also
have different shapes.
In Fig.~\ref{fig:23}(a) 
we illustrate the
circular $n=4$ orbit at $A = B = 0.5$ and $a_{0} = 0.45$.
In contrast, the $n=4$ orbit in Fig.~\ref{fig:23}(c)
at $A = B = 0.5$ with $a_{0} = 0.95$ 
is square.
In general, as the obstacle radius increases,
the localized orbits become more square in shape
due to the symmetry of the obstacle lattice. 
In Fig.~\ref{fig:23}(b) we show a delocalized orbit at
$A = B= 0.652$ and $a_{0} = 0.45$, 
where the skyrmion collides with various obstacles and undergoes
long time diffusion.
A  delocalized orbit
at $A = B = 0.608$ and $a_{0} = 0.95$
appears in Fig.~\ref{fig:23}(d).
For larger obstacles, the chaotic 
phases develop a stronger diffusive behavior since there
is an increased probability that the skyrmion will randomly scatter
off of the obstacles.
The phase diagram 
in Fig.~\ref{fig:22} also demonstrates
that for larger $a_0$, the localized orbits become elliptical,
such as in the 16$_E$ state which encircles 16 obstacles,
while at smaller $a_0$ the orbits are 
more circular, as in the 16$_S$ phase.
For larger $\alpha_{m}/\alpha_{d}$, 
there are a reduced number of delocalized or translating orbits.

\section{Changing ac frequency $\omega_1\neq \omega_2$}.

\begin{figure}
\includegraphics[width=3.5in]{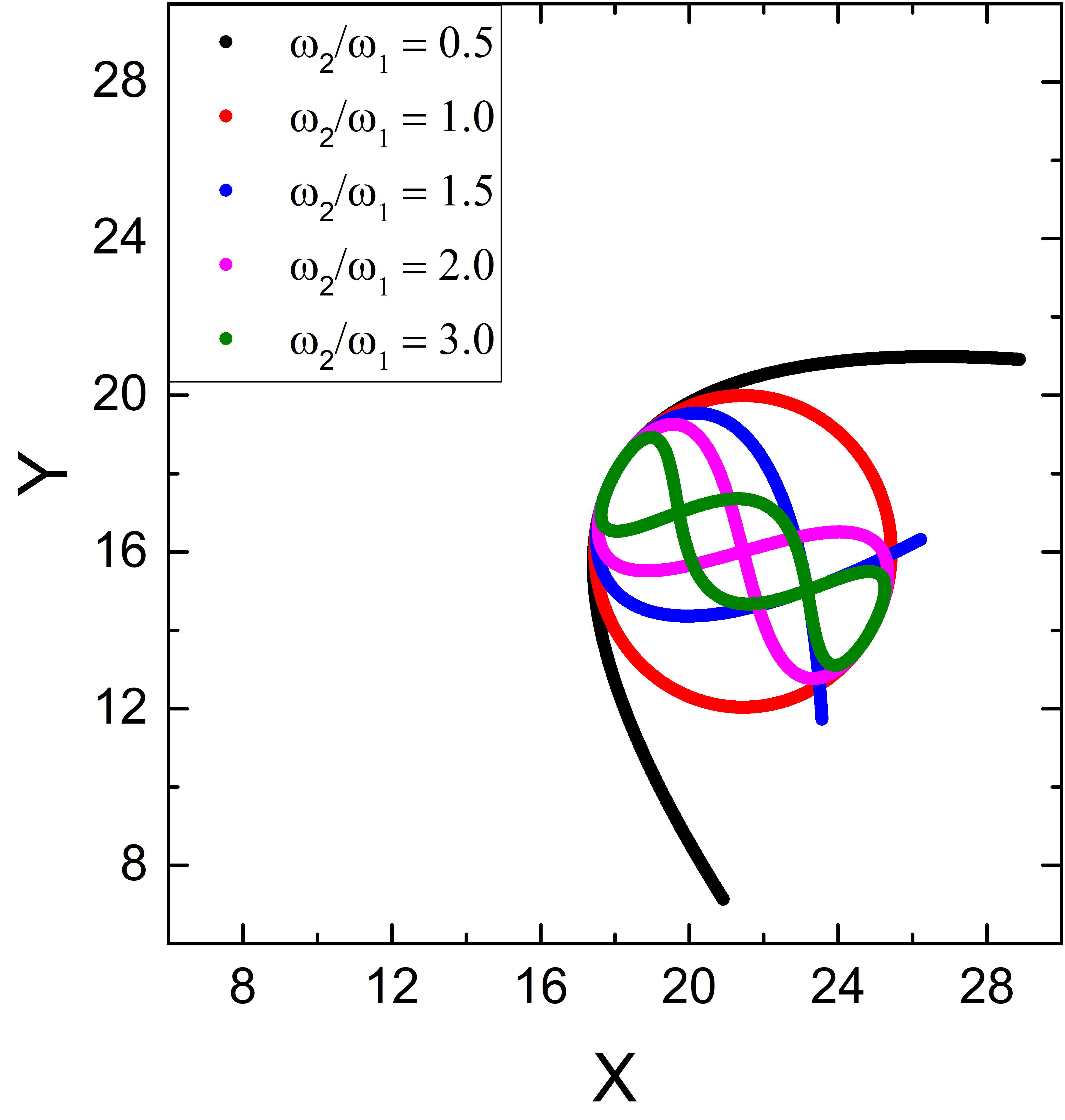}
\caption{The skyrmion trajectory in the absence of a substrate
for a system with $\alpha_{m}/\alpha_{d} = 0.45$,
$A=B$, and  
$\omega_{2}/\omega_{1} =0.5$ (black), $1.0$ (red), $1.5$ (blue), $2.0$ (pink),
and $3.0$ (green).  
}
\label{fig:24}
\end{figure}

We next consider circular driving
with $A = B$ but 
with two different frequencies,
$\omega_{1} \neq \omega_{2}$.
In this case
we find
much more extensive regions of
ratcheting or directed motion due to the additional 
asymmetry in the skyrmion orbits.
In Fig.~\ref{fig:24} we illustrate the skyrmion orbits
in the absence of a substrate
for a system with $A = B = 1.0$ 
and $\alpha_{m}/\alpha_{d} = 0.45$ 
at $\omega_{2}/\omega_{1} = 0.5$,
$1.0$, $1.5$, $2.0$, and $3.0$.
For $\omega_{1}/\omega_{2} = 1.0$, the orbit
is circular; however, for the other ratios, the orbits are asymmetric.

\begin{figure}
\includegraphics[width=3.5in]{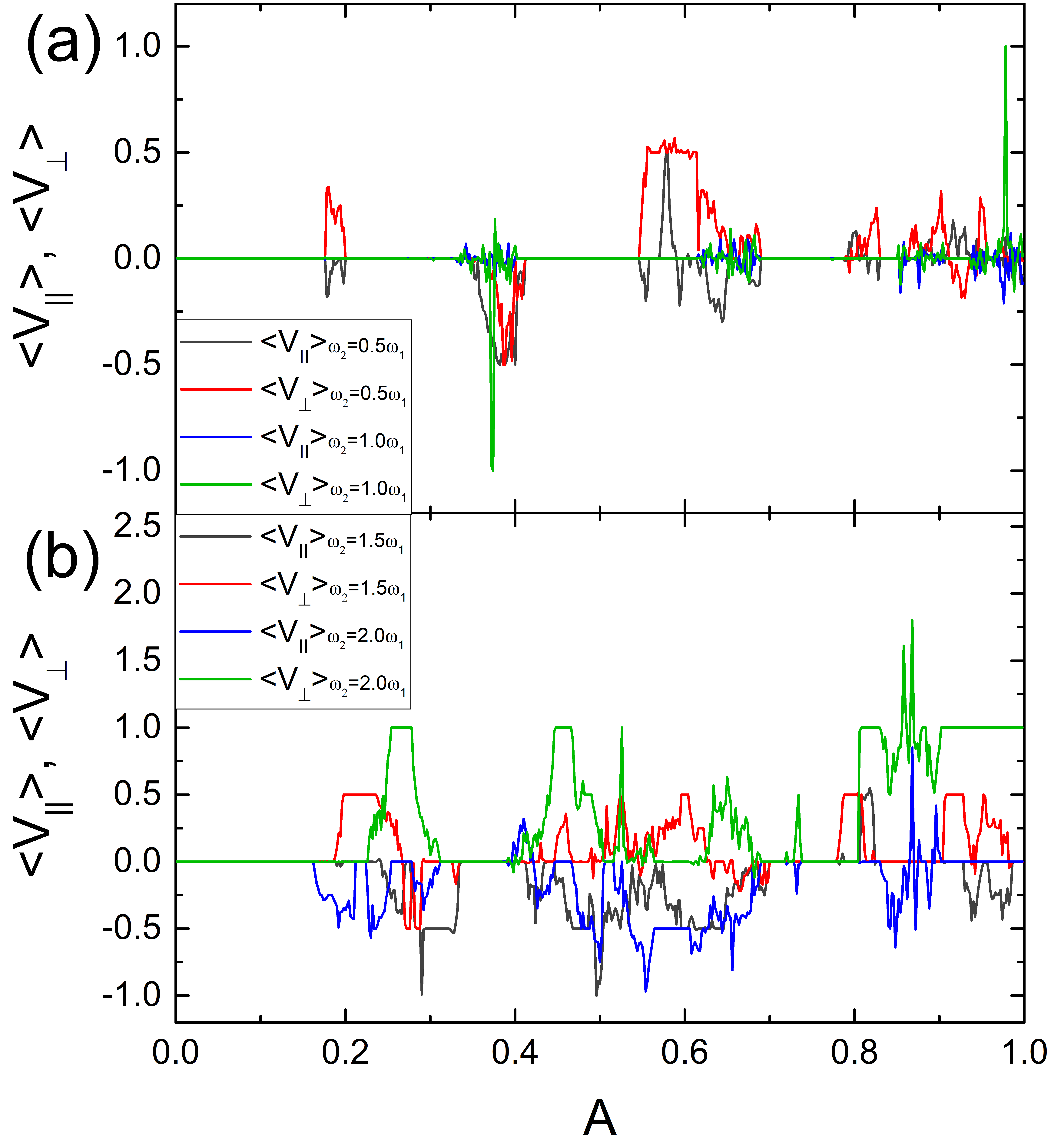}
\caption{$\langle V_{||}\rangle$ and $\langle V_{\perp}\rangle$
vs $A$ in samples with circular ac driving
for $\alpha_{m}/\alpha_{d} = 0.577$, $A=B$, and $a_0=0.65$.
(a) $\omega_{2}/\omega_{1} = 0.5$ and $\omega_2/\omega_1=1.0$.  
(b) $\omega_{2}/\omega_{1} = 1.5$ and
$\omega_{2}/\omega_{1} = 2.0$.  
}
\label{fig:25}
\end{figure}

In Fig.~\ref{fig:25}(a) we plot $\langle V_{||}\rangle$ and
$\langle V_{\perp}\rangle$ versus $A$ for
a system with circular ac driving at $\alpha_{m}/\alpha_{d} = 0.577$,
$a_0=0.65$, and $A=B$
for
$\omega_{2}/\omega_{1} = 0.5$ and $1.0$.
When $\omega_{2}/\omega_{1} = 0.5$, there are large regions
of translating orbits along with multiple reversals
in the direction of translation.
Figure~\ref{fig:25}(b) shows the same quantities
for samples with $\omega_{1}/\omega_{2} = 1.5$ and
$\omega_{1}/\omega_{s} = 2.0$.
When $\omega_{1}/\omega_{2} = 1.5$, there are several regimes
in which $\langle V_{||}\rangle = 0.5$ or $\langle V_{\perp}\rangle = 0.5$,
indicating that the skyrmion translates by one lattice constant every two
ac drive cycles,
while for $\omega_{1}/\omega_{2} = 2.0$, there are regions where
$\langle V_{||}\rangle=1.0$ or $\langle V_{\perp}\rangle = 1.0$, indicating
that the skyrmion is translating by one lattice
constant per ac drive cycle.
Near $A = 0.89$, there is a small interval over which
the skyrmion translates two lattice constants per ac drive cycle.

\begin{figure}
  \begin{minipage}{3.5in}
    \begin{minipage}{3.5in}
      \includegraphics[width=3.5in]{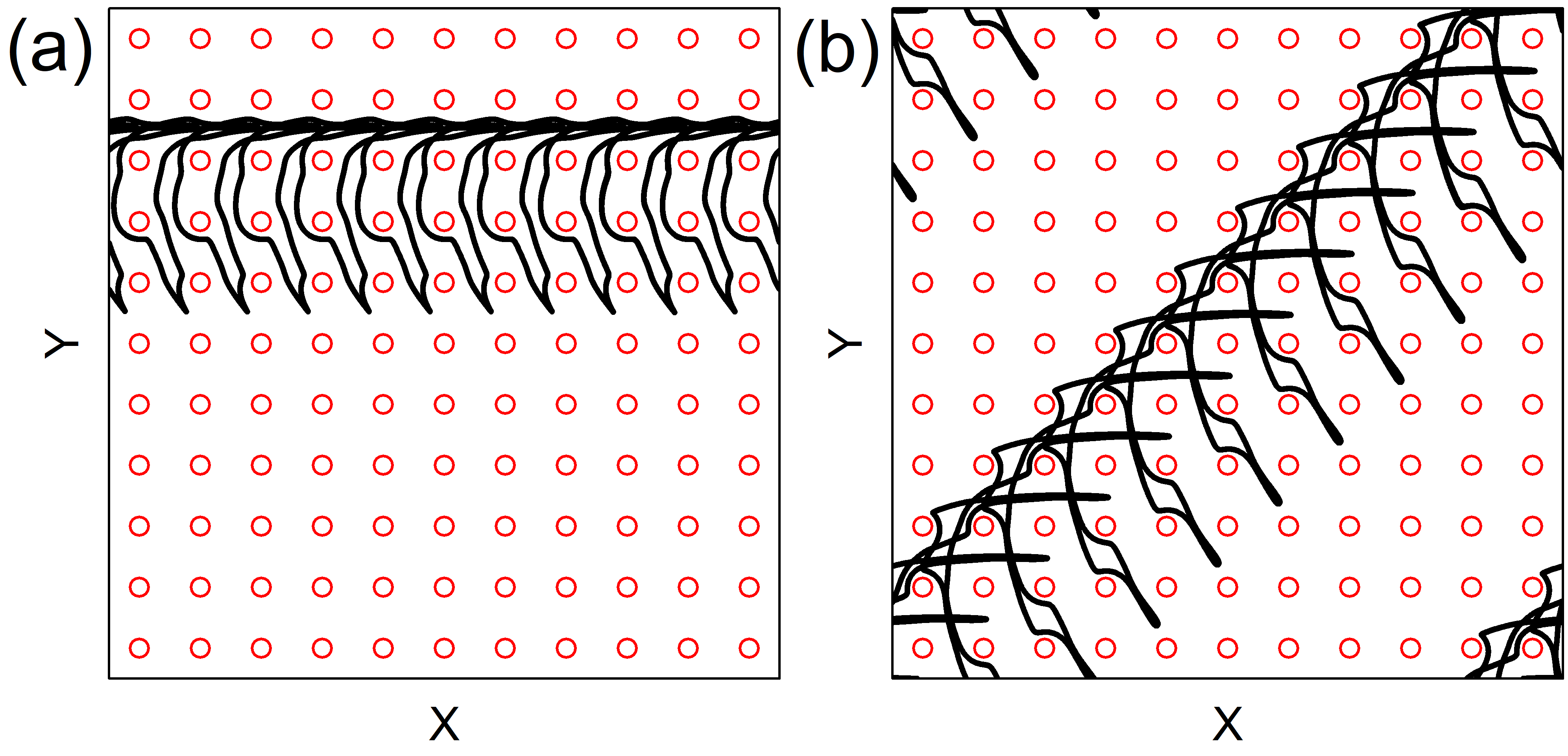}
    \end{minipage}
    \begin{minipage}{3.5in}
      \includegraphics[width=3.5in]{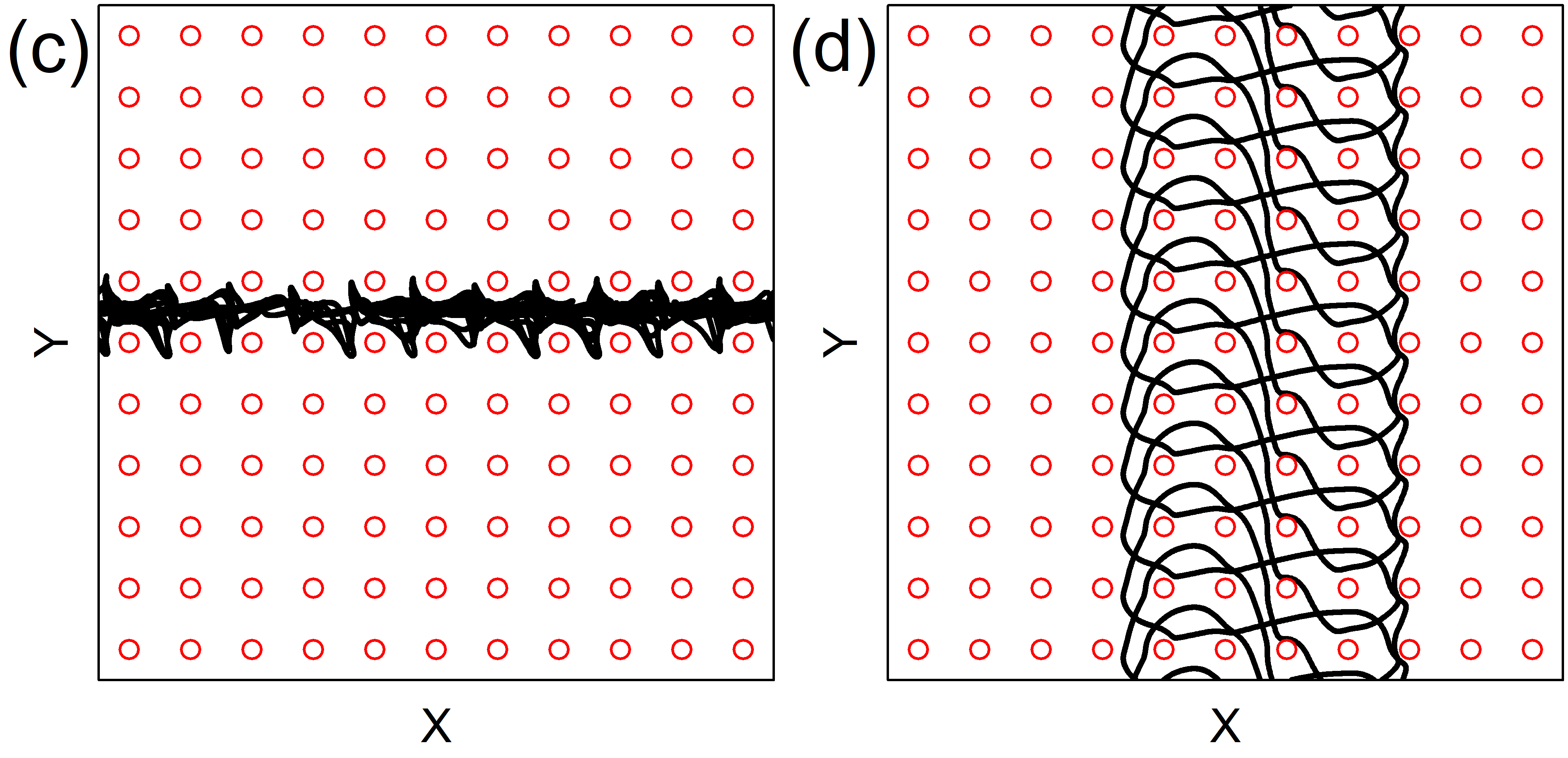}
    \end{minipage}
  \end{minipage}
\caption{
Obstacles (red circles) and skyrmion trajectory (black line)
for the system in Fig.~\ref{fig:25} with 
circular ac driving,
$\alpha_{m}/\alpha_{d} = 0.577$, $A=B$, and $a_0=0.65$.
(a) At $\omega_{2}/\omega_{1} = 0.5$ and $A = 0.4$,
the orbit translates in the $x$ direction.
(b) At $\omega_{2}/\omega_{1} = 0.5$ and $A = 0.58$,
the skyrmion translates at $45^\circ$. 
(c) At $\omega_{2}/\omega_{1} = 2.0$ and $A = 0.3$,
the orbit translates in the $x$-direction
but is also partially disordered.
(d)  At $\omega_{2}/\omega_{1} = 2.0$ and $A = 1.0$,
the
skyrmion is moving in the positive $y$-direction.   
}
\label{fig:26} 
\end{figure}

In Fig.~\ref{fig:26}(a) we illustrate the skyrmion trajectory
for the system in Fig.~\ref{fig:25}(a) 
at $\omega_{2}/\omega_{1} = 0.5$ 
and $A = 0.4$, where the skyrmion is translating in the
positive $x$-direction.
At the same ratio
$\omega_2/\omega_1=0.5$ but at
$A = 0.58$, Fig.~\ref{fig:26}(b) shows that the skyrmion
translates along
$45^\circ$. 
Figure~\ref{fig:26}(c) indicates that at
$\omega_{2}/\omega_{1} = 2.0$ and $A = 0.2$, the orbit translates
in the $x$-direction but is also disordered.
In Fig.~\ref{fig:26}(d), at the same ratio
$\omega_{2}/\omega_{1} = 2.0$ but with $A=1.0$,
the skyrmion
translates in the positive $y$-direction.

\begin{figure}
\includegraphics[width=3.5in]{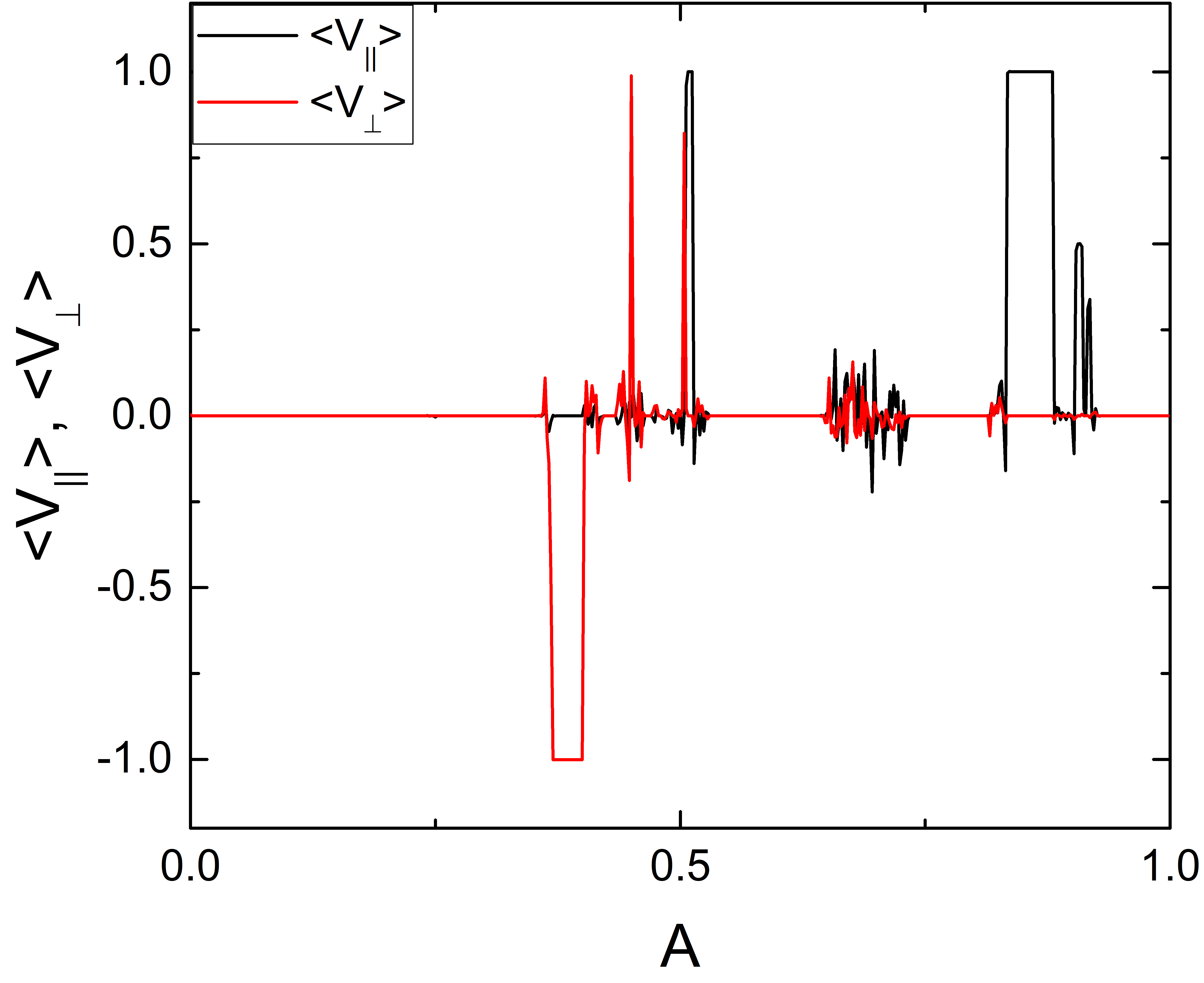}
\caption{$\langle V_{||}\rangle$ (black) and $\langle V_{\perp}\rangle$ (red)
vs $A$ for a system
with a circular ac drive at $A=B$, $a_0=0.65$, and
$\omega_{1}/\omega_{2} = 3.0$ for $\alpha_{m}/\alpha_{d} = 0.45$.
There are five regions of translating orbits.   
}
\label{fig:27}
\end{figure}

\begin{figure}
\includegraphics[width=3.5in]{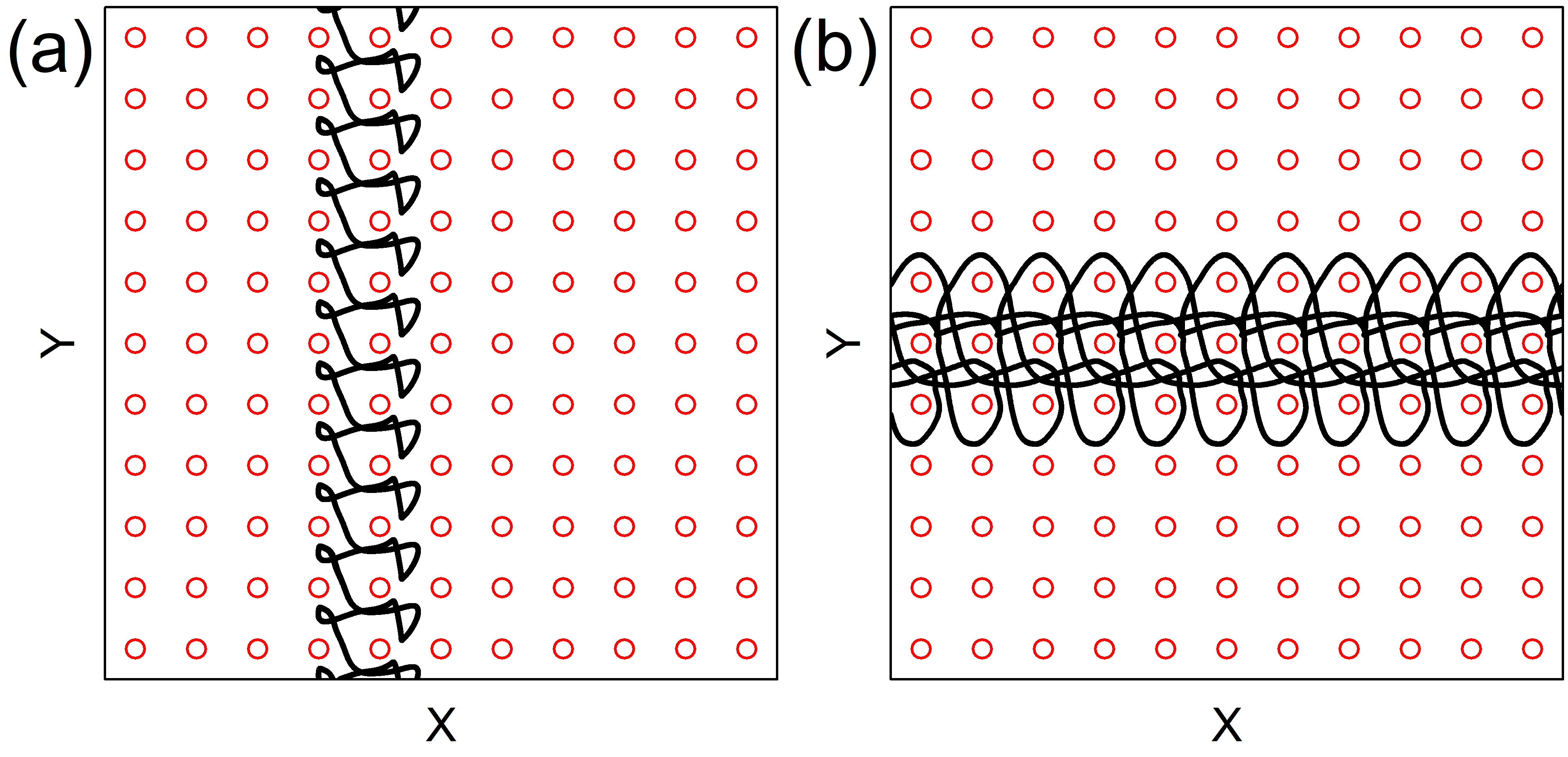}
\caption{
Obstacles (red circles) and skyrmion trajectory (black line)
for the system from Fig.~\ref{fig:27}
with a circular ac drive at $A=B$, $a_0=0.65$, 
$\omega_{1}/\omega_{2} = 3.0$ and $\alpha_{m}/\alpha_{d} = 0.45$.
(a) The translating orbit near $A = 0.35$
where the skyrmion moves in the negative $y$-direction.
(b) The translating orbit near $A = 0.85$
where the skyrmion moves in the $x$-direction.   
}
\label{fig:28}
\end{figure}

In Fig.~\ref{fig:27} we plot $\langle V_{||}\rangle$
and $\langle V_{\perp}\rangle$ versus $A$ for a system with
$A = B= 0.5$ and 
$\alpha_{m}/\alpha_{d} = 0.45$ at $\omega_{2}/\omega_{1} = 3.0$. 
We find five translating orbit phases along with
one localized phase and several
localized regions.
In Fig.~\ref{fig:28}(a) we illustrate the skyrmion
trajectory in the first translating orbit near
$A = 0.35$,
where the skyrmion moves one lattice constant
in the negative $y$-direction during each
ac drive cycle.
Figure~\ref{fig:28}(b) shows an orbit translating
in the positive $x$ direction near $A = 0.85$.
In both       
cases,
the orbit is highly complex, and the skyrmion
executes three loops before translating.

\section{Discussion}
Our results could be tested experimentally
in a system containing a periodic array of antidots
with one or two applied ac drives.
The different orbits could be measured
using
direct imaging or through
electrical detection.
Another method for observing the orbits is to analyze
the noise fluctuations, since the localized states or
ordered translating orbits should produce
low noise along with a narrow band signal at a specific
frequency.  In contrast,
the delocalized phases would exhibit broadband noise signals
or multiple frequencies due to the jumping of the skyrmion
between different orbit shapes.
It would also be interesting to consider 
a finite number of skyrmions instead of an individual skyrmion.
At low densities where the skyrmions do not interact,
we expect that the results would be similar to what we describe above;
however, for multiple interacting 
skyrmions, there could be an enhancement of the
disordered regime or even new types of ordered phases. 

In our model, the skyrmions are treated as rigid particles;
however, actual skyrmions can exhibit internal modes
or shape distortions which could induce additional phases.
This also suggests that another method for
driving skyrmions in circular orbits would be
to use oscillating fields, since
continuum studies have shown that this technique can
generate skyrmion motion even without a substrate \cite{Chen19}.
In a sample where a dc drive is superimposed on an ac drive,
various types of phase locking phenomena should appear
in the velocity force curves as
has been studied in previous work \cite{Vizarim20b}.

Our model neglects thermal effects, but we expect such effects would
become important near the transition between two different localized orbits,
and could induce a creep motion 
for certain translating orbits.
Our results also suggest that by controlling the obstacle geometry
and the ac driving,
it should be possible to cause the skyrmion to translate at a
designated
skyrmion Hall angle over a specific number of lattice constants
per ac drive cycle.
The behavior should also depend on the type of skyrmion considered. For 
antiferromagnetic skyrmions \cite{Barker16,Legrand20}
and liquid crystal skyrmions \cite{Duzgun20},  
where the Hall angle is absent,
the dynamics would be similar to those found in the vortex pinball systems.
In  other skyrmion systems where internal modes are important,
there could be complex trochoidal motion of the skyrmions
\cite{Ritzmann18}.
For antiskyrmions \cite{Nayak17}, 
additional dynamical phases could appear
since the closed orbits could be more complex.             

\section{Summary}
We have examined skyrmions interacting with a square array of repulsive
obstacles or antidots under linear or circular ac driving.
For linear ac driving, an overdamped particle follows a 1D orbit that 
does not translate; however, due to the combination of the Magnus force
and interactions with the obstacles,
skyrmions can execute a 2D orbit.
As the ac drive amplitude increases,
the skyrmion displays
a series of localized orbits in which it moves
between several different obstacle plaquettes or encircles
one or more obstacles.
In between the localized orbits,
we find that
there can be chaotic or disordered
orbits.
The transitions between the localized and delocalized orbits
resemble what is found in
electron pinball systems.
In some cases, a linear ac drive 
can produce a translating or skipping skyrmion orbit.
This ratchet effect is similar to what is found
for colloids undergoing complex closed orbits on a periodic
substrate, where the combination of the drive and the orbit asymmetry
produces enough symmetry breaking to allow directed transport to emerge.
In the skyrmion case, much simpler ac driving can be used to achieve
the same effect.
In systems with a circular ac drive,
the skyrmion follows a circular orbit in the absence of a substrate,
but when a substrate is present
we find localized stable phases, where the skyrmion encircles
an integer number of obstacles, as well as delocalized
or chaotic orbits and translating orbits.
The translating orbits produce motion along $x$ or $y$ or at a $45^\circ$
angle, since these are the major symmetry directions of the square obstacle
lattice.
The distance moved by the skyrmion in the translating orbits
ranges from 1, 1/2, 1/3, 1/4, or 1/5 of a lattice constant per ac drive
cycle.
There are also chaotic phases that show gradual translation.
As the Magnus force increases,
additional translating phases appear; however, for the largest
Magnus forces, disordered phases begin to dominate the behavior.
We also also examine the different phases
as a function of the obstacle size.
The extent of the translating orbits can be enhanced by
varying the amplitude or frequency of the circular ac drive in the $x$
and $y$ direction to create
asymmetric orbits.     
Our results show that skyrmions interacting with periodic obstacle arrays
under ac driving
provide another example of 
pinball dynamics where the Magnus force
induces additional features such as translating orbits. 
These results also suggest new ways to control skyrmion motion
for device applications by using strictly ac driving.

\acknowledgments
This work was supported by the US Department of Energy through
the Los Alamos National Laboratory.  Los Alamos National Laboratory is
operated by Triad National Security, LLC, for the National Nuclear Security
Administration of the U. S. Department of Energy (Contract No. 892333218NCA000001).
N.P.V. acknowledges funding from Funda\c{c}\~{a}o de Amparo {\' a}
Pesquisa do Estado de S\~{a}o Paulo - FAPESP (Grant 2018/13198-7).

\bibliography{mybib}

\end{document}